\documentclass[12pt,a4paper,final]{iopart}
\usepackage{iopams}  
\usepackage{graphicx}
\usepackage{hyperref}
\usepackage{cleveref}
\usepackage{pst-node}
\begin{document}
\title[Einstein Euler Flow]{The nonlinear stability of n+1 dimensional FLRW spacetimes}
\author{Puskar Mondal \footnote{e-mail: puskar$\_$mondal@fas.harvard.edu}}

\begin{abstract}
We prove nonlinear Lyapunov stability of a family of `$n+1$'-dimensional cosmological models of general relativity locally isometric to the Friedman Lema\^itre Robertson Walker (FLRW) spacetimes including a positive cosmological constant. In particular, we show that the perturbed solutions to the Einstein-Euler field equations around a class of spatially compact FLRW metrics (for which the spatial slices are compact negative Einstein spaces in general and hyperbolic for the physically relevant $n=3$ case) arising from regular Cauchy data remain uniformly bounded and decay to a family of metrics with constant negative spatial scalar curvature. To accomplish this result, we employ an energy method for the coupled Einstein-Euler field equations in constant mean extrinsic curvature spatial harmonic gauge (CMCSH). In order to handle Euler's equations, we construct energy from a current that is similar to the one derived by Christodoulou \cite{christodoulou} (and which coincides with Christodoulou's current on the Minkowski space) and show that this energy controls the desired norm of the fluid degrees of freedom. The use of a fluid energy current together with the CMCSH gauge condition casts the Einstein-Euler field equations into a coupled elliptic-hyperbolic system. Utilizing the estimates derived from the elliptic equations, we first show that the gravity-fluid energy functional remains uniformly bounded in the expanding direction. Using this uniform boundedness property, we later obtain sharp decay estimates if a positive cosmological constant $\Lambda$ is included, which suggests that the accelerated expansion of the physical universe that is induced by the positive cosmological constant is sufficient to control the non-linearities in the case of small data. A few physical consequences of this stability result are discussed.   
 
\end{abstract}

\section{Introduction and previous studies}
Since its conception, the FLRW cosmological solution has been used extensively to model the dynamics of our universe. With the advent of accelerated expansion of the physical universe, a cosmological constant (the simplest form of the dark energy) is often included in the model and such a model is designated as the $\Lambda$CDM model \cite{peebles2003cosmological,carroll2001cosmological}. This model is developed based on the assumption of \textit{global} homogeneity and isotropy of the physical universe, the so-called \textit{cosmological principle}. However, the astronomical observations that motivate the cosmological principle are restricted to a fraction (possibly quite small) of the universe and as such do not provide a hint towards global properties such as the topology of the universe. Therefore, the assumption of global homogeneity and isotropy seems to be rather restrictive and the corresponding topology turns out to be rather limited: $\mathbb{S}^{3}$ (or its two-fold quotient $\mathbb{RP}^{3}=\mathbb{S}^{3}/\mathbb{Z}_{2}$), $\mathbb{H}^{3}$ or $\mathbb{E}^{3}$. Each of these spaces is equipped with its standard metric of constant positive, zero, or negative sectional curvature, respectively. If one removes the global assumption and rather imposes a \textit{local} condition, then there exist numerous closed manifolds that admit locally homogeneous and isotropic metrics and that can be constructed as quotients of $\mathbb{E}^{3}$, $\mathbb{S}^{3}$ and $\mathbb{H}^{3}$ by the discrete, proper and torsion-free subgroups of their respective isometry groups. However, in the current article, we will only focus on the universe models with spatial slices that are closed hyperbolic manifolds. More generally, if one considers an $n+1$ ($n\geq 3$) dimensional problem, then one may consider spatial manifolds that are negative Einstein (for the $n=3$ case these negative Einstein spaces correspond to hyperbolic manifolds due to Mostow rigidity \cite{lebrun1994einstein}). Another motivation to consider the models with negative spatial sectional curvature is that the recent astronomical observations \cite{aghanim2018planck, ryan2018constraints} seem to indicate a possibility that the spatial universe may have slightly negative curvature. These compact variants of the most general FLRW solutions may be written as follows 
\begin{eqnarray}
\label{eq:model}
^{n+1}\hat{g}=-dt\otimes dt+a(t)^{2}\gamma_{ij}dx^{i}\otimes dx^{j},
\end{eqnarray}
where $R[\gamma]_{ij}=-\frac{1}{n}\gamma_{ij},~t\in (0,\infty),~a(t)$ is the scale factor satisfying $\partial_{t}a(t)>0$.

One important question that arises in the context of FLRW cosmology is whether these models are predicted by general relativity? To conclude that they are would seem to hinge on a proof that the purely theoretical variants of FLRW models (spatially compact models foliated by compact hyperbolic spaces) are dynamically stable. In other words, if one were to perturb these solutions in a suitable function space setting, then the natural question would be whether the perturbed spacetimes exist for all \textit{time} as solutions of Einstein's field equations. Moreover one would like to understand whether all small data solutions (in suitable function space settings after subtracting the background solutions given by (\ref{eq:model})) of the Einstein's field equations remain bounded for all time in terms of a suitable norm of the initial data and whether they exhibit asymptotic decay. There are other important physical motivations such as large-scale structure formation in the universe through cosmological perturbations \cite{percival2005cosmological, macpherson2019einstein,east2018comparing}. So far, the state-of-the-art work on the structure formation is limited to linearized perturbation theory, numerical relativity on flat space background ($\mathbb{E}^{3}$), or $N-$ body simulations where a Newtonian approximation is applied \cite{bertschinger1991cosmological,springel2005simulations,kitaura2013cosmological, lucie2018machine}. Each of these studies reveals a plethora of interesting physics. Moving one step further, one would want to study the problem of including the complete non-linearities of Einstein's field equations, and deduce whether the large-scale structure formation is hinted at by the Einstein dynamics. A completely satisfactory study would be one that considers a large data problem, which is far from being currently solved. In addition, recently several inconsistencies of the FLRW models have been proposed by a few recent studies \cite{buchert2012backreaction, buchert2018observational,heinesen2020solving} which require a fully non-linear treatment of the field equations. We will not address such issues here since they require other information in conjunction with the nonlinear stability result and therefore remain as a matter for future study. Overall, it is crucial to study the fully non-linear stability of these cosmological models in order to draw any physical conclusions whatsoever. 

The story of cosmological perturbation theory is not new. A substantial study has been carried out on the topic by several authors \cite{hawking1966perturbations, weinberg1972gravitation, bardeen1980gauge, kodama1984cosmological}) since the earliest investigation by Lifsitz \cite{lifshitz1992gravitational}. However, these studies are formal mode stability analyses and as such are unable to yield sufficient conditions for a `true' linear stability where one is interested in the asymptotic behavior of all solutions to the linearized Einstein field equations, not simply the fixed modes. Even though such true linear stability has been rigorously established \cite{mondal2021linear}, it lacks the nonlinear effects which can lead to the potential formation of naked singularities or shocks (or both). The most interesting dynamical behaviors are expected to occur in the fully non-linear regime. In a sense, the feature that makes Einstein's field equations (in the presence or absence of sources) tremendously rich is the non-linearities present. In the regime of fully nonlinear stability of Einstein's equations (may that be vacuum or otherwise), there has been considerable progress in the last three decades. Global non-linear stability proof of de-Sitter spacetime by \cite{friedrich1986existence} marked the initiation of this progress. This was followed by the proof of the stability of Minkowski space by \cite{christodoulou1993global}. Later \cite{lindblad2010global} provided a simpler proof utilizing the so-called `weak null condition' of Einstein's equations in the spacetime harmonic gauge. In the context of cosmological spacetimes, Andersson and Moncrief \cite{andersson2004future} established the stability of the $3+1$ dimensional Milne universe in constant mean extrinsic curvature spatial harmonic gauge (CMCSH) and later generalized this result to arbitrary dimensions ($n+1$, $n\geq 3$) \cite{andersson2011einstein}. Following these fundamental studies, numerous studies have been performed regarding small data global existence and stability problems associated with Einstein's equations including various sources \cite{ringstrom2008future, oliynyk2016future, fajman2021attractors, fajman2017nonvacuum, rodnianski2018regime, lefloch2021nonlinear}. Two of the studies that are most relevant to us are the ones by Rodnianski and Speck \cite{rodnianski2009stability} and Speck \cite{speck2012nonlinear}. Rodnianski and Speck \cite{rodnianski2009stability} studied the small data perturbations of the $\mathbb{R}\times \mathbb{T}^{3}$ type FLRW model with a positive cosmological constant in spacetime harmonic gauge, where an irrotational perfect fluid model was used as the source term. This essentially reduced the problem to Einstein's equations coupled to a nonlinear scalar field. Later Speck \cite{speck2012nonlinear} extended this analysis on the same topology and removed the irrotational condition. Both of these studies utilize the rapid expansion induced by the positive cosmological constant to avoid pathology such as shock formation associated with the perfect fluid source. The issue of shock formation is tremendously important in the non-linear setting (this is another important difference between linear and non-linear stability analysis of FLRW spacetimes). \cite{christodoulou2007formation} proved the formation of shocks in relativistic perfect fluids evolving on a Minkowski background for arbitrarily small initial data. This may be attributed to the physical fact that the Minkwoski space does not posses the expansion property and therefore the purely geometric dispersion alone is not sufficient to suppress the shock formation. On a compact spatial topology, the energy concentration (and subsequent formation of singularities) is more favorable since there are no non-compact directions through which energy may dissipate (decay). Therefore, the expansion becomes the only factor that may prevent energy concentration by non-linearities and the problem essentially turns out to be a competition between the dissipation caused by expansion and energy concentration by non-linearities. The global existence or finite-time blow-up then depends on which of these two opposing forces dominate. For accelerated expansion, it turns out that the expansion dominates the non-linearities for small data perturbations yielding global existence.

Despite the large amount of studies that are present in the literature, there are no non-linear stability results for the FLRW spacetimes to our knowledge that contain spatially compact slices with constant negative sectional curvature. In addition, the proof of the stability of $\mathbb{R}\times \mathbb{T}^{3}$ model by \cite{speck2012nonlinear} uses the spacetime harmonic gauge while we work in the constant mean extrinsic curvature spatial harmonic gauge (CMCSH). Both of our proofs use a form of energy current (an $n+1$ vector field on the spacetime) that was derived by Christodoulou \cite{christodoulou2007formation}. The reason behind using the energy current is explained below. We construct the following current vector field bilinear in perturbations and their higher order derivatives
\begin{eqnarray}
\label{eq:energycurrent_rough}
\mathcal{C}^{\mu}\partial_{\mu}:=\left\{\frac{(\gamma_{ad}-1)}{\gamma_{ad}\rho N}[1+g(v,v)]^{1/2}\dot{\rho}^{2}\nonumber+2(\gamma_{ad}-1)\frac{g_{kl}v^{k}\dot{v}^{l}}{N[1+g(v,v)]^{1/2}}\dot{\rho}\right.\\\nonumber 
\left.+\frac{\gamma_{ad}\rho[1+g(v,v)]^{1/2}}{N}\left(-\frac{g(v,\dot{v})^{2}}{1+g(v,v)}+g_{kl}\dot{v}^{k}\dot{v}^{l}\right)\right\}\partial_{t}\\ 
+\left(\frac{(\gamma_{ad}-1)v^{i}\dot{\rho}^{2}}{\gamma_{ad}\rho}+2(\gamma_{ad}-1)\dot{v}^{i}\dot{\rho}+\gamma_{ad}\rho v^{i}(-\frac{g(v,\dot{v})^{2}}{1+g(v,v)}+g_{kl}\dot{v}^{k}\dot{v}^{l})\right)\partial_{i}
\end{eqnarray}
where $v^{i}\partial_{i}$ is the $n-$velocity of the fluid (i.e., the projection of the $n+1$-velocity on to a constant $t$ hypersurface), $\rho$ is the mass energy density, $N$ is the lapse function, $g$ is the induced Riemannian metric on a $t=$constant hypersurface, and $\gamma_{ad}$ is the adiabatic index lying in the range $\gamma_{ad}\in (1,\frac{n+1}{n})$. Here $\dot{v}:=\nabla_{\vec{\alpha}}v$ and $\dot{\rho}=\nabla_{\vec{\alpha}}(\rho-\bar{\rho})$ are the multi spatial derivatives of the fields $\rho-\bar{\rho}$ ($\bar{\rho}$ is the background mass energy density) and $v$. $\nabla_{\vec{\alpha}}$ is roughly defined as $\nabla_{\vec{\alpha}}=\nabla[\gamma]_{I_{1}}\nabla[\gamma]_{I_{2}}\nabla[\gamma]_{I_{3}}....\nabla[\gamma]_{I_{\alpha}}$ where $\nabla[\gamma]$ is the covariant derivative with respect to a background Riemannian metric $\gamma$. we show, if $\gamma_{ad}\in (1,\frac{n+1}{n})$, then $-\int_{\Sigma}\mathcal{C}^{\mu}\mathbf{n}_{\mu}$ is positive definite and controls the required norm of the matter density (or perturbed matter density to be precise) and the fluid's $n$-velocity field for sufficiently small data, $\mathbf{n}$ is the unit future pointing time-like vector field orthogonal to a constant $t$ hypersurface $\Sigma$. While evaluating the time derivative of the energy through a divergence identity, one needs to compute $\nabla_{\mu}\mathcal{C}^{\mu}$. In order for the energy argument to close, $\nabla_{\mu}\mathcal{C}^{\mu}$ should not contain any derivative terms of the fields $\dot{\rho}$ and $\dot{u}$ (or even if there are derivative terms they should be of total divergence form so that after integration they disappear). This cancellation of top order derivatives of the physical fields is obtained if one uses this special energy current (\ref{eq:energycurrent_rough}). This 
is the main purpose of constructing this special energy current since the relativistic Euler's equations are not of the diagonal form and therefore an ordinary definition of energy fails to close the energy argument. Once the energy argument closes, the remaining lower order terms are estimated using available Sobolev and multiplication inequalities.

Another technicality that arises in our study is that the trivial solutions (or background solutions (\ref{eq:model})) form a finite dimensional sub-manifold (center manifold) of the infinite dimensional configuration space if the dimension is strictly larger than $3+1$. This occurs simply because the trivial solutions are described by $R[\gamma]_{ij}=-\frac{1}{n}\gamma_{ij}$ which form the Einstein moduli space after a gauge fixing (CMCSH) is performed. The stability analysis is slightly tricky in this situation since the perturbations that are parallel to the moduli space do not have detectable dynamics in the sense of Einstein's evolution equations since each perturbed metric $\gamma^{*}$ will satisfy the criteria $R[\gamma^{*}]_{ij}=-\frac{1}{n}\gamma^{*}_{ij}$ and therefore will be a trivial solution of the Einstein's equations. In order to handle this additional technicality, we invoke the shadow gauge condition introduced by \cite{andersson2011einstein}. Implementation of shadow gauge enables us to explicitly handle the perturbations that are parallel and perpendicular to the moduli space.       

Our analysis indicates that if the expansion factor $a(t)$ appearing in the background solutions (\ref{eq:model}) obeys  suitable integrability conditions and the adiabatic index $\gamma_{ad}$ lies in the range $(1,\frac{n+1}{n})$ ($(1,4/3)$ in the physically relevant `3+1' case; $\gamma_{ad}=1$ and $\gamma_{ad}=4/3$ correspond to a pressure-less fluid or `dust' and a `radiation' fluid, respectively), then the energy functional remains uniformly bounded by its initial value and moreover the perturbations decay in the expanding direction. This is in a sense equivalent to the fact that if the perturbations to the matter part are restricted within a smaller cone contained in the `sound' cone (in the tangent space; to be defined later), that is, they do not propagate at a speed higher than $\sqrt{\frac{1}{n}}$ (in the unit of light speed), then the stability (Lyapunov or asymptotic to be clarified later) holds. On the other hand, there exists a seemingly contradictory result in the purely non-relativistic case of a self-gravitating fluid system. For a self-gravitating non-relativistic 3-dimensional fluid body (Euler-Poisson system), Chandrasekhar \cite{chandrasekhar2013hydrodynamic} used a virial identity argument to show that the static isolated compact solution of the Euler-Poisson system is stable for $\gamma_{ad}>\frac{4}{3}$. However, if one observes carefully, our result and Chandrasekhar's result are not contradictory since the notions of stability are different in the two contexts. In the context of Chandrasekhar's argument, stability is defined by the negativity of total energy i.e., the domination of gravitational energy (stable $\iff$ gravitationally bound system i.e., the fluid body does not disperse). Now, of course, a system with strong gravity is stable in the sense of Chandrasekhar (since total energy is dominated by gravitational energy and therefore negative) which is nonsensical in our fully relativistic context (ultra-high gravity could focus and blow up forming a singularity, indicating instability in our context).

The stability criteria are satisfied by the universe model in which one includes a positive cosmological constant $\Lambda$ (inclusion of $\Lambda$ forces the integrability conditions on $a(t)$ to hold) and in such case, one obtains a uniform asymptotic decay (in a suitable sense) which matches with that of the linear theory. However, turning off the cosmological constant results in losing the uniform boundedness property simply because one does not have the desired integrability properties for the scale factor. This may be attributed to the fact that a positive cosmological constant induced accelerated expansion wins over the gravitational effect at the level of small data. However, we do not claim that turning off the cosmological constant leads to instability but simply we are unable to reach a definite conclusion with the currently available method. It is worth pointing out that \cite{fajman2021slowly} proved the non-linear asymptotic stability of the Milne universe (devoid of background matter density) including a dust source (slowly expanding spacetime). In such a case, the individual dust particles travel along their respective geodesics and the sound speed vanishes. This intuitively suggests that one may avoid shock formation in such a scenario with the help of ordinary polynomial expansion. However, we do not know whether such a result hold for FLRW spacetimes where the background mass-energy density does not vanish. We will investigate the borderline cases of $\gamma_{ad}=0$ and $\gamma_{ad}=\frac{n+1}{n}$ in the future using special techniques.

\subsection{Notations}
The `$n+1$' dimensional $C^{\infty}$ spacetime manifold is denoted by $\hat{M}$. We are interested in spacetime manifolds $\hat{M}$ of globally hyperbolic type i.e., the topology is the product topology $\mathbb{R}\times \Sigma$, where $\Sigma$ denotes an $n$-dimensional closed spatial manifold diffeomorphic to a Cauchy hypersurface (i.e., every inextendible causal curve intersects $\Sigma$ exactly once). We denote the space of Riemannian metrics on $\Sigma$ by $\mathcal{M}_{\Sigma}$. We explicitly work in the $L^{2}$ (with respect to a background metric) Sobolev space $H^{s}$ for $s>\frac{n}{2}+2$. The general $L^{p}$ Sobolev space of order $s$ is denoted by $W^{s,p}$. We will consider tensors as sections of a suitable vector bundle $\mathcal{B}$ over $(\Sigma,g)$, where $g\in \mathcal{M}_{\Sigma}$ is the dynamical metric that solves the Einstein's equations written in $n+1$ form. The $L^{2}$ inner product on a fibre of such a rank-$k$ bundle is defined as
\begin{eqnarray}
\langle u|v\rangle_{L^{2}}:=\int_{\Sigma}u_{i_{1}i_{2}i_{3}....i_{k}}v_{j_{1}j_{2}j_{3}....j_{k}}\gamma^{i_{1}j_{1}}\gamma^{i_{1}j_{2}}\gamma^{i_{3}j_{3}}....\gamma^{i_{k}j_{k}}\mu_{g},
\end{eqnarray}
where $\gamma\in \mathcal{M}_{\Sigma}$ is a $C^{\infty}$ background metric lying in a small enough neighbourhood of $g$ in a suitable space and $\mu_{g}=\sqrt{\det(g_{ij})}dx^{1}\wedge dx^{2}\wedge dx^{3}\wedge.........\wedge dx^{n}$ is the volume form associated with the metric $g$. 
The standard norms are defined naturally as follows
\begin{eqnarray}
||u||_{L^{2}}:=\langle u|u\rangle^{1/2}_{L^{2}}\\
||u||_{L^{\infty}}:=\sup_{\Sigma}(u_{i_{1}i_{2}i_{3}....i_{k}}u_{j_{1}j_{2}j_{3}....j_{k}}\gamma^{i_{1}j_{1}}\gamma^{i_{1}j_{2}}\gamma^{i_{3}j_{3}}....\gamma^{i_{k}j_{k}})^{1/2}
\end{eqnarray}
and so on. We adapt the following definition of a \textit{twisted} rough Laplacian acting on sections of $\mathcal{B}^{k}$
\begin{eqnarray}
\Delta^{\gamma}_{g}u:=-\frac{1}{\mu_{g}}\nabla[\gamma]_{i}(g^{ij}\mu_{g}\nabla[\gamma]_{j}u).
\end{eqnarray}
This Laplacian is self-adjoint with respect to the following inner product on derivatives
\begin{eqnarray}
\label{eq:twistedinner}
\langle\nabla[\gamma]u|\nabla[\gamma]v\rangle_{L^{2}}:=\int_{\Sigma}g^{ij}\nabla[\gamma]_{i}u_{i_{1}i_{2}....i_{k}}\nabla[\gamma]_{j}v_{j_{1}j_{2}....j_{k}}\gamma^{i_{1}j_{1}}\gamma^{i_{1}j_{2}}....\gamma^{i_{k}j_{k}}\mu_{g},
\end{eqnarray}
where $\mu_{g}$ is the volume form $\sqrt{\det(g)}dx^{1}\wedge dx^{2}\wedge dx^{3}\wedge....\wedge dx^{n}$.
The covariant derivative with respect to the dynamical metric $g$ will be denoted simply by $\nabla$ while that with respect to $\gamma$ will be denoted by $\nabla[\gamma]$. We denote the connection coefficients of metric $g$ and $\gamma$ by $\Gamma[g]$ and $\Gamma[\gamma]$, respectively. The Riemann and Ricci curvatures are also denoted similarly.
We define the ordinary Laplacian of $g$ in the following way such that it has non-negative spectrum i.e., \begin{eqnarray}
\Delta_{g}\equiv-g^{ij}\nabla_{i}\nabla_{j}.
\end{eqnarray}
We also define a Lichnerowicz type Laplacian $\mathcal{L}_{g,\gamma}$ acting on sections of a symmetric rank-2 vector bundle, which will play a crucial role in our analysis
\begin{eqnarray}
\mathcal{L}_{g,\gamma}u_{ij}=\Delta^{\gamma}_{g}u_{ij}-2R[\gamma]_{i}~^{k}~_{j}~^{l}u_{kl}.
\end{eqnarray}
The spectra of this operator is assumed to be non-negative \cite{andersson2011einstein}. 

Lastly, for functions $f(t)\in \mathbb{R}_{\geq 0}$ and $g(t)\in \mathbb{R}_{\geq 0}$ $f(t)\lesssim g(t)$ means $f(t)\leq Cg(t)$ and $f(t)\gtrsim g(t)$ means $f(t)\geq Cg(t)$, and $f(t)\approx g(t)$ implies $C_{1}g(t)\leq f(t)\leq C_{2}g(t)$ for $0<C_{1},C_{2},C<\infty$. The involved constants may depend only on a fixed background geometry. The spaces of symmetric covariant 2-tensors and vector fields on $\Sigma$ are denoted by $S^{0}_{2}(\Sigma)$ and $\mathfrak{X}(\Sigma)$, respectively. By $A^{tr}$ and $A^{TT}$, we will denote symmetric trace-less and transverse-trace-less 2-tensors, respectively.

\subsection{\textbf{Motivation and Overview}}
In this section we describe our motivation for this study, state a rough version of the main theorem and discuss the overall structure of the article. As we have mentioned in the introduction before, the obvious motivation is to understand whether the purely theoretical FLRW solutions are predicted by general relativity. If the overall characteristics of this model still persist after adding perturbations (in a suitable sense), then we may hope that this purely theoretical model is physically realized. This question is directly related to several aspects of modern cosmology and therefore very crucial to answer. 

In addition to the above motivation, we want to study the stability properties of this particular variant of FLRW models (\ref{eq:model}) because several recent studies \cite{ashtekar2015general, moncrief2019could, moncrief2021einstein} have suggested that the observed local homogeneity and isotropy (almost) of the physical universe may be derived dynamically from the Einstein's equations (including physically reasonable matter sources). This idea is supported by the temporal behavior of a certain \textit{weak} Lyapunov functional (which takes the form of a re-scaled volume functional of the spatial manifold; the spatial manifold is assumed to be of negative Yamabe type for a technical reason, see \cite{moncrief2021einstein} for more detail) that controls the lowest norm of the gravity and matter fields. Such a Lyapunov functional monotonically decays in the expanding direction and achieves its infimum on solutions that are isometric to our background solutions (\ref{eq:model}). In that limit, the Lyapunov functional becomes a topological invariant of the manifold (the so-called $\sigma$-constant that is closely related to the Yamabe invariant). Applying a well-established theorem from geometry \cite{anderson2004geometrization}, one then observes that the value of the Lyapunov functional (and hence the volume of the physical universe) is dominated by the hyperbolic components that are present in the spatial manifold. Since the compact hyperbolic manifolds are locally homogeneous and isotropic, this result tends to indicate that there is a dynamical mechanism at work within the framework of Einstein flow that drives the physical universe towards an asymptotic state which is characterized by a locally homogeneous and isotropic spatial metric. 

However, since the Lyapunov functional only controls a very rough norm ($H^{1}\times L^{2}$ of (metric, second fundamental form)) of the data, one is unable to draw a definite conclusion without studying the long time behavior of the classical solutions (or solutions that lie in the higher Sobolev spaces) of the Einstein field equations. This is due to the fact that gravitational energy can concentrate and blow up before the weak Lyapunov functional ever achieves its infimum. In the presence of matter sources, there are additional complications since matter-energy could blow up as well. The prototypical example of such matter singularity is present in the current context. The matter source in the context of FLRW cosmology is the perfect fluid and it is known that the relativistic perfect fluid forms shock singularities in finite time regardless of the size of the initial data on a Minkowski background (\cite{christodoulou2007formation}). These are the associated issues and therefore in order to understand the extent to which the weak Lyapunov functional decays to its infimum (or maybe obstructed from doing so), it is necessary to study the global properties of the solutions. As it happens, in the current context we would only want to study a `small data' problem, which is expected to provide a level of insight into several physics questions associated with the asymptotic behavior of the physical universe. Without going into the technical details, we state the main theorem and discuss a few physical implications. Let us denote the gravitational degrees of freedom by $(g,k)$, where $g$ and $k$ are the first and second fundamental forms of the constant time hypersurfaces foliating the spacetime. The fluid's mass-energy density and the spatial velocity fields are denoted by $\rho$ and $v$, respectively. A rough version of the main stability theorem is as follows\\
\textbf{Rough sketch of the main theorem:} \textit{Let $(g_{0},k^{tr}_{0},\rho_{0},v_{0})\in H^{s}\times H^{s-1}\times H^{s-1}\times H^{s-1},~s>\frac{n}{2}+2$ be the initial data for the re-scaled Einstein-Euler-$\Lambda$ system with the cosmological constant $\Lambda>0$. Let us also consider that the Cauchy data for the re-scaled background solutions are $(g_{B.G},k^{tr}_{B.G},\rho_{B.G},v_{B.G})$ which satisfy $R[g_{B.G}]_{ij}=-\frac{1}{n}(g_{B.G})_{ij}, k^{tr}_{B.G}=0, \rho_{B.G}=C_{\rho}, v^{i}_{B.G}=0$, $C_{\rho}$ is a constant. Now assume that the following open smallness condition holds $||g_{0}-g_{B.G}||_{H^{s}}+||k^{tr}_{0}-k^{tr}_{B.G}||_{H^{s-1}}+||\rho_{0}-\rho_{B.G}||_{H^{s-1}}+||v_{0}-v_{B.G}||_{H^{s-1}}<\epsilon$, $\epsilon>0$ is chosen sufficiently small. If $t \mapsto (g(t), k^{tr}(t), \rho(t), v(t))$ is the maximal development of the Cauchy problem for the re-scaled Einstein-Euler-$\Lambda$ system in constant mean extrinsic curvature spatial harmonic gauge (CMCSH) (\ref{eq:metricevol}-\ref{eq:mc}) with initial data $(g_{0},k^{tr}_{0},\rho_{0},v_{0})$, then the following holds in the limit of infinite time
\begin{eqnarray}
\lim_{t\to\infty}(g(t),k^{tr}(t),\rho(t),v(t))=(\gamma^{\dag},0,\rho^{'},0),
\end{eqnarray}
where $\gamma^{\dag}$ satisfies $R[\gamma^{\dag}]=-1$ and $\rho^{'}$ depends on the initial re-scaled density $\rho_{0}$ and $\partial_{i}\rho^{'}\neq 0$ in general.
}

Let us now explain the main theorem putting aside the technical details. The statement essentially deals with the \textit{asymptotic} stability of the solutions that are characterized by constant negative scalar curvature ($-1$ to be precise in this occasion) and sufficiently close to the FLRW solutions. This corresponds to a \textit{Lyapunov} stability of the FLRW solutions. In other words, the solutions of the linearized Einstein-Euler-$\Lambda$ equations about the background FLRW solutions (\ref{eq:model}) decay to nearby solutions which have constant negative scalar curvature while the background FLRW spacetimes (\ref{eq:model}) are described by the property that the spatial metric $g_{B.G}$ is negative Einstein i.e., $R[g_{B.G}]_{ij}=-\frac{1}{n}(g_{B.G})_{ij}$. Now, the space of negative Einstein metrics is a subspace of the space of metrics with constant negative scalar curvature. Simple use of the triangle inequality shows that the asymptotic state should be the space of constant negative scalar curvature metrics sufficiently close to and containing the background FLRW solutions. Since the latter is a subspace of the former, one may fine-tune the initial conditions to obtain the asymptotic solution to be exactly of FLRW type (i.e., the solution has a spatial metric that is negative Einstein). But such a procedure proves to be too restrictive and physically undesirable. 

If for the moment we focus on the physical $3+1$ case, then ideally we want the asymptotic solution to have constant negative sectional curvature (i.e., be hyperbolic or negative Einstein by Mostow rigidity) not just constant negative scalar curvature and have a spatially uniform background energy density. In other words, if one perturbs an FLRW solution, it does not generically come back to an FLRW solution in infinite time. Now notice that the local spatial homogeneity and isotropy criteria required constancy of the sectional curvature and as such in 3 spatial dimensions, a negative Einstein metric automatically has constant sectional curvature. However, the space of metrics with constant negative scalar curvature is a much larger space and does not necessarily exhibit the local spatial homogeneity and isotropy criteria (only a subspace that is described by hyperbolic metrics does). This result is not completely satisfactory but on the other hand, physically expected as these inhomogeneous and anisotropic characteristics may be attributed to the structure formation (see \cite{lifshitz1992gravitational, peebles2020large} about the role of cosmological perturbation theory in structure formation). This is indeed a remarkable fact that indicates there is a dynamical mechanism at work within the Einstein-Euler-$\Lambda$ flow that naturally drives the physical universe to an anisotropic and inhomogeneous state (small deviation from homogeneity and isotropy in a suitable sense) leading to cosmological structure formation. On the other hand, a Lyapunov functional that controls a very rough norm of the data may be unable to detect any structure that is smoothed out by the expansion of the universe. However, at the moment, we do not claim that such a mechanism of structure formation is yet so compelling since we can not prove that these background solutions are \textit{not} asymptotically stable and approach a homogeneous and isotropic state.   

We note that \cite{roy2011global} reported this behavior of the FLRW spacetimes as well considering `dust' as the matter source through a rather heuristic argument. They constructed a finite-dimensional dynamical system by spatially averaging out the inhomogeneities and showed that \textit{FLRW cosmologies are unstable in some relevant cases: averaged models are driven away from the background through structure formation and accelerated expansion}. The phenomenon of accelerated expansion driving the solutions away from the background is previously noted in \cite{mondal2019attractors} which dealt with a particular case of vacuum gravity (background solutions were described by the compact hyperbolic manifolds) including a positive cosmological constant. The solutions decay asymptotically to the nearby ones described by constant negative scalar curvature. On the other hand, in the same problem if the cosmological constant is turned off (i.e., the vacuum energy is absent), then the perturbed solutions do come back to the background solution asymptotically \cite{andersson2004future, andersson2011einstein}. Similar phenomena of asymptotic stability are observed in the study of \cite{fajman2021slowly}, where the background solutions once again are simply the Milne universe that does not contain a non-zero energy density. In order to reach a definite conclusion about forming large-scale structures by non-linear perturbations of a background with non-vanishing energy density would require further analysis including radiation sources and possibly more observational inputs.  

Now we discuss the overview of the article. Section 2 provides the necessary detail about the formulation of the gauge fixed Einstein-Euler field equations including a positive cosmological constant $\Lambda$, the re-scaling of the field equations, and the background solutions. Here we state a local well-posedness theorem (local existence, uniqueness, and Cauchy stability of the solution as well as the conservation of gauge conditions and constraints along the solution curve) for the coupled Einstein-Euler system in CMCSH gauge. This yields an `in time' continuation criteria for the solutions and plays a crucial role in the final step of the global existence proof. We do not describe in detail the properties of the background solutions rather only the key features that are essential to our analysis. Since we are studying the general `$n+1$' dimensional case, the trivial solutions need to be handled carefully since they form a finite-dimensional manifold for $n>3$ (Einstein moduli space). Therefore, we briefly describe the center manifold of the re-scaled dynamics when the Einstein moduli space is non-trivial and the kinematics of the perturbations (this is crucial) under the assumption of a shadow gauge condition (introduced by \cite{andersson2011einstein}). 

Section $3$ is one of the most important sections of the article. In this section, we construct the energy current for the fluid and from which we construct a suitable energy. We prove that under the physical assumptions, this energy is positive definite. Simultaneously, we construct a wave equation type energy functional for the gravitational degrees of freedom and show that the total energy of the system controls the suitable norms of the fluid and gravity degrees of freedom for sufficiently small data (coercive property). In this section, we explicitly compute the time evolution of the fluid energy functional as well. In order to do so, we first obtain a set of commuted Euler equations for the fluid degrees of freedom. Using these commuted equations, we show that the spacetime divergence of the energy current $\mathcal{C}^{\mu}\partial_{\mu}$ does not contain any contribution from the principal terms. This is an essential feature of a hyperbolic system. In a sense, the relativistic Euler equations are examples of hyperbolic equations and as such theoretically derivable from a Lagrangian (see \cite{moncrief1977hamiltonian,bao1985hamiltonin,demaret1980hamiltonian,brown1993action}). The gravity part of the total energy is relatively easy to handle since the application of CMCSH gauge cancels the problematic terms obstructing the ellipticity of the map $g\mapsto Ricci[g]$ therefore also making the Einstein evolution equations hyperbolic (essentially we assumed this property while stating a local existence theorem).

 While computing the time derivative of the total energy, we will encounter terms involving the lapse function and the shift vector field which therefore need to be estimated by other means. However, the applications of the CMCSH gauge condition yields elliptic equations for the lapse function and the shift vector field that are sourced by the matter degrees of freedom. Therefore, we obtain the necessary elliptic estimates and use those to control the time derivative of the energy. One subtlety is that apart from estimating the lapse and the shift vector fields, we also need to estimate the time derivative of the lapse function. We complete these estimates in section $4.2$. Using the necessary estimates for the gauge variables, in section $4.3$ and $4.4$, we obtain a differential inequality for the energy, which upon integration and using the smallness of the data yields the uniform boundedness property of the energy (essentially closing a bootstrap argument). Utilizing the local existence theorem, we immediately then obtain a global existence theorem. Using the uniform boundedness property, we prove decay separately. We close this section by stating the fully rigorous version of the main theorem and sketching a proof of the geodesic completeness of the perturbed spacetimes.           

\section{Gauge fixed field equations and re-scaling}
\subsection{Complete Einstein-Euler system and gauge fixing}
The ADM formalism splits the spacetime described by an `$n+1$' dimensional Lorentzian manifold $\tilde{M}$ into $\mathbb{R}\times M$. Here each level set $\{t\}\times M$ of the time function $t$ is an orientable n-manifold diffeomorphic to a Cauchy hypersurface (assuming the spacetime admits a Cauchy hypersurface) and equipped with a Riemannian metric. Such a split may be executed by introducing a lapse function $N$ and shift vector field $X$ belonging to suitable function spaces on $\tilde{M}$ and defined such that
\begin{eqnarray}
\partial_{t}&=&N\hat{n}+X,
\end{eqnarray}
where $t$ and $\hat{n}$ are time and a hypersurface orthogonal future directed timelike unit vector i.e., $\hat{g}(\hat{n},\hat{n})=-1$, respectively. The above splitting writes the spacetime metric $\hat{g}$ in local coordinates $\{x^{\alpha}\}_{\alpha=0}^{n}=\{t,x^{1},x^{2},....,x^{n}\}$ as 
\begin{eqnarray}
\hat{g}&=&-N^{2}dt\otimes dt+g_{ij}(dx^{i}+X^{i}dt)\otimes(dx^{j}+X^{j}dt)
\end{eqnarray} 
and the stress-energy tensor as
\begin{eqnarray}
\mathbf{T}&=&E\mathbf{n}\otimes\mathbf{n}+2\mathbf{J}\odot\mathbf{n}+\mathbf{S},
\end{eqnarray}
where $\mathbf{J}\in \Omega^{1}(M)$, $\mathbf{S}\in S^{0}_{2}(M)$, and $A\odot B=\frac{1}{2}(A\otimes B+B\otimes A$). Here, $\Omega^{1}(M)$ and $S^{0}_{2}(M)$ are the space of one form fields and the space of symmetric covariant 2-tensor fields, respectively.  Here $E:=\mathbf{T}(\mathbf{n},\mathbf{n})$ is the energy density observed by a time-like observer with $n+1$-velocity $\mathbf{n}$, $\mathbf{J}_{i}=-\mathbf{T}(\partial_{i},\mathbf{n})$ is the momentum density, $\mathbf{J}_{i}=-\mathbf{T}(\mathbf{n},\partial_{i})$ is the energy flux density, and $\mathbf{S}_{ij}=\mathbf{T}(\partial_{i},\partial_{j})$ is the momentum flux density (with respect to the chosen constant $t$ hypersurface $M$). The choice of a spatial slice in the spacetime leads to consideration of the second fundamental form $k_{ij}$ which describes how the slice is curved in the spacetime. The trace of the second fundamental form ($tr_{g}k=\tau$) is the mean extrinsic curvature of the slice, which will play an important role in the analysis. Under such decomposition, the Einstein equations 
\begin{eqnarray}
R_{\mu\nu}-\frac{1}{2}Rg_{\mu\nu}+\Lambda g_{\mu\nu}&=&T_{\mu\nu}
\end{eqnarray}
take the form ($8\pi G=c=1$)
\begin{eqnarray}
\partial_{t}g_{ij}&=&-2Nk_{ij}+L_{X}g_{ij},\\\nonumber
\partial_{t}k_{ij}&=&-\nabla_{i}\nabla_{j}N+N\{R_{ij}+\tau k_{ij}-2k_{ik}k^{k}_{j}-\frac{1}{n-1}(2\Lambda-S+E)g_{ij}\\\nonumber
&&-\mathbf{S}_{ij}\}+L_{X}k_{ij}
\end{eqnarray}
along with the constraints (Gauss and Codazzi equations)
\begin{eqnarray}
\label{eq:HC}
R(g)-|k|^{2}+\tau^{2}&=&2\Lambda+2E,\\
\label{eq:MC}
\nabla_{j}k^{j}_{i}-\nabla_{i}\tau&=&-\mathbf{J}_{i},
\end{eqnarray} 
where $S=g^{ij}\mathbf{S}_{ij}$. The vanishing of the covariant divergence of the stress energy tensor i.e., $\nabla_{\nu}T^{\mu\nu}=0$ is equivalent to the continuity equation and equations of motion of the matter
\begin{eqnarray}
\frac{\partial E}{\partial t}&=&L_{X}E+NE\tau-N\nabla_{i}\mathbf{J}^{i}-2\mathbf{J}^{i}\nabla_{i}N+N\mathbf{S}^{ij}k_{ij},\\
\frac{\partial \mathbf{J}^{i}}{\partial t}&=&L_{X}\mathbf{J}^{i}+N\tau \mathbf{J}^{i}-\nabla_{j}(N\mathbf{S}^{ij})+2Nk^{i}_{j}\mathbf{J}^{j}-E\nabla^{i}N.
\end{eqnarray}
Here we want to study the Einstein-Euler coupled system. Let the `$n+1$' velocity field of a perfect fluid be denoted by $\mathbf{u}$ which satisfies the normalization condition $\hat{g}(\mathbf{u},\mathbf{u})=-1$. One may for convenience decompose the $n+1$ velocity $\mathbf{u}$ into its component parallel and perpendicular to constant $t$ hypersurface $M$ as follows 
\begin{eqnarray}
\mathbf{u}=v-\hat{g}(\mathbf{u},\mathbf{n})\mathbf{n}
\end{eqnarray}
where $v$ is a vector field parallel to the spatial manifold $M$ i.e., $v\in \mathfrak{X}(M)$. Importantly note that $\mathbf{u}_{i}=v_{i}$ but $\mathbf{u}^{i}\neq v^{i}$ unless the shift vector field $X$ vanishes. 
The stress energy tensor for a perfect fluid with $n+1$ velocity field $\mathbf{u}$ reads
\begin{eqnarray}
\mathbf{T}=(P+\rho)\mathbf{u}\otimes\mathbf{u}+P\hat{g},
\end{eqnarray}
where $P$ and $\rho$ are the pressure and the mass energy density, respectively and $\hat{g}$ is the spacetime metric. After projecting onto the spatial manifold $M$, several components of the stress energy tensor may be computed as follows
\begin{eqnarray}
E=(P+\rho)(N\mathbf{u}^{0})^{2}-P, 
\mathbf{J}^{i}=(P+\rho)\sqrt{1+g(v,v)}v^{i},\\\nonumber 
\mathbf{S}=(P+\rho)v\otimes v+P g. 
\end{eqnarray}
However, notice that the field equations do not close with the information of the stress-energy tensor alone, and therefore one needs an equation of state relating pressure $P$ and mass-energy density $\rho$. The choice of the equation of state is non-trivial and there are numerous studies about the equation of state alone in the cosmology literature (e.g., \cite{babichev2004dark, chavanis2018simple, nakamura1999determining}). In the standard model of cosmology ($\Lambda$CDM), one frequently uses a barotropic equation of state of the type $P=(\gamma_{ad}-1)\rho$, where $\gamma_{ad}$ is the adiabatic index. The speed of sound $C_{s}$ is defined as $C^{2}_{s}:=\frac{\partial P}{\partial\rho}$, where the derivative is computed at constant entropy. However, for the barotropic equation of state chosen here, the entropy equation decouples from the field equations and therefore plays no role in our analysis. The speed of sound is then a constant $\sqrt{\gamma_{ad}-1}$. Due to causality, one must have $0\leq C^{2}_{s}\leq 1$ yielding $\gamma_{ad}\in [1,2]$. This is equivalent to the fact that the sound cone is contained within the light cone in the tangent space. Here $\gamma_{ad}=1$ corresponds to pressure-less fluid or `dust' and $\gamma_{ad}=2$ corresponds to a ultra-relativistic stiff fluid. We note an important fact that choosing this equation of state, we ignore the rest energy of the fluid which may be relevant in the late epoch of the universe evolution. In other words, the equation of state $P=(\gamma_{ad}-1)(\rho-ne)$, $n$ being the baryon number density and $e$ the fluid rest energy per particle, may be more appropriate \cite{eosnew}. Here we will stick with the equation of state $P=(\gamma_{ad}-1)\rho$ for simplicity and in such case, the baryon number conservation equation $\nabla_{\mu}(nu^{\mu})=0$ is a simple consequence of the Euler's equations i.e., $\nabla_{\nu}T^{\mu\nu}=0$. With the barotropic equation of state, the components of the stress energy tensor are computed to be     
\begin{eqnarray}
E=(P+\rho)(1+g(v,v))-P=\rho+\gamma_{ad}\rho g(v,v)\\\nonumber
\mathbf{J}^{i}=\gamma_{ad}\rho\sqrt{1+g(v,v)}v^{i},
\mathbf{S}=\gamma_{ad}\rho v\otimes v+(\gamma_{ad}-1)\rho g.
\end{eqnarray} The complete system of evolution and constraint equations reads
\begin{eqnarray}
 \partial_{t}\rho+\frac{\gamma\rho N}{[1+g(v,v)]^{1/2}}\nabla_{i}v^{i}=L_{X}\rho\nonumber-\frac{\gamma_{ad}\rho g(\partial_{t}v,v)}{1+g(v,v)}+\frac{\gamma_{ad}\rho Nk(v,v)}{1+g(v,v)}-\frac{2\gamma_{ad}\rho v_{i}L_{X}v^{i}}{1+g(v,v)}\\\nonumber 
-\frac{\gamma_{ad}N\rho L_{v}N}{[1+g(v,v)]^{1/2}}+Nk^{i}_{i}\rho-\frac{NL_{v}\rho}{[1+g(v,v)]^{1/2}}.
 \end{eqnarray}
and
\begin{eqnarray}
\gamma_{ad}\rho\left[u^{0}\partial_{t}v^{i}-2Nu^{0}k^{i}_{j}v^{j}-u^{0}(L_{X}v)^{i}+v^{j}\nabla_{j}v^{i}+(1+g(v,v))\nabla^{i}N\right]\\\nonumber 
+(\gamma_{ad}-1)\left[\nabla^{i}\rho+v^{i}L_{v}\rho+u^{0}v^{i}(\partial_{t}\rho-L_{X}\rho )\right]=0
\end{eqnarray}
\begin{eqnarray}
\label{eq:evol1}
\partial_{t}g_{ij}&=&-2Nk_{ij}+L_{X}g_{ij},\\
\label{eq:evol2}
\partial_{t}k_{ij}&=&-\nabla_{i}\nabla_{j}N+N\left\{R_{ij}+\tau k_{ij}-2k_{ik}k^{k}_{j}-\frac{2\Lambda}{n-1} g_{ij}\right.\\\nonumber
&&\left.-\gamma_{ad}\rho v_{i}v_{j}+\frac{\gamma_{ad}-2}{n-1}\rho g_{ij}\right\}
+L_{X}k_{ij},
\end{eqnarray}
\begin{eqnarray}
\label{eq:HC}
R(g)-|k|^{2}+\tau^{2}&=&2\Lambda+2\rho\left\{1+\gamma_{ad} g(v,v)\right\},\\
\label{eq:MC}
\nabla_{i}k^{ij}-\nabla^{j}\tau&=&-\gamma_{ad}\rho\sqrt{1+g(v,v)}v^{i},
\end{eqnarray} 
where $u^{0}$ is explicitly computed to be $u^{0}=\frac{[1+g(v,v)]^{1/2}}{N}$.

As clearly seen from the field equations, the gauge variables $N$ and $X$ do not satisfy evolution equations as physically expected. Therefore, we need to fix the gauge choice (or choose a gauge slice) to obtain equations for $N$ and $X$. As we have explicitly mentioned before, we will use the constant mean extrinsic curvature spatial harmonic gauge (CMCSH). Setting the mean extrinsic curvature ($\tau:=\tr_{g}k$) of the hypersurface $\Sigma$ as a constant over $\Sigma$ allows it to play the role of time. However, one is obligated to address the issue related to the existence of a CMC hypersurface in the spacetime $\hat{M}$ \cite{bartnik1988remarks, galloway2018existence, rendall1996constant, andersson1999existence}. Luckily, our background solution (\ref{eq:model}) possesses constant mean extrinsic curvature slices, which is evident through explicit calculations. Therefore simply setting the mean extrinsic curvature of the perturbed spacetimes to be constant on each spatial slice seems to be a reasonable choice
\begin{eqnarray}
\label{eq:uniformmean}
\partial_{i}\tau&=&0.
\end{eqnarray}
Such an assumption can be made without loss of generality for solutions that are close to the background solution \cite{fajman2016local, rodnianski2018stable}. If one evaluates the time derivative of $\tau$ using the evolution equations, one obtains the following equation
\begin{eqnarray}
\frac{\partial \tau}{\partial t}=\Delta_{\gamma}N+\left\{|k|^{2}_{g}+[\frac{(n\gamma-2)}{n-1}+\gamma_{ad} g(v,v)]\rho-\frac{2\Lambda}{n-1}\right\}N+L_{X}\tau
\end{eqnarray}
which upon utilizing the CMC gauge condition (\ref{eq:uniformmean}) yields the elliptic equation for the lapse function 
\begin{eqnarray}
\label{eq:lapseequation}
\Delta_{g}N+\left\{|k|^{2}_{g}+[\frac{(n\gamma-2)}{n-1}+\gamma_{ad} g(v,v)]\rho-\frac{2\Lambda}{n-1}\right\}N=&\frac{\partial \tau}{\partial t}.
\end{eqnarray}
Note that we may obtain an elliptic equation for $\partial_{t}N$ by commuting the lapse equation (\ref{eq:lapseequation}) with the vector field $\partial_{t}$. Since this causes equivalent loss of one derivative of every field present in the equation, care must be taken in order to estimate $\partial_{t}N$. We will do so in a later section. 

In addition to the temporal gauge fixing, we need to fix the spatial gauge as well. We use the spatial harmonic gauge introduced by Andersson and Moncrief \cite{andersson2003elliptic} to study the local well-posedness of the vacuum Einstein equations. Let $\chi:(\Sigma,g)\to (\Sigma,\gamma)$ be a harmonic map with Dirichlet energy functional $\frac{1}{2}\int_{\Sigma}g^{ij}\frac{\partial\chi^{k}}{\partial x^{i}}\frac{\partial\chi^{l}}{\partial x^{j}}\gamma_{jl}\mu_{g}$. Since the harmonic maps are, by definition, critical points of this Dirichlet energy functional, $\chi$ satisfies the following formal Euler-Lagrange equations
\begin{eqnarray}
g^{ij}\left(\partial_{i}\partial_{j}\chi^{k}-\Gamma[g]_{ij}^{l}\partial_{l} \chi^{k}+\Gamma[\gamma]_{\alpha\beta}^{k}\partial_{i} \chi^{\alpha}\partial_{j} \chi^{\beta}\right)=0.
\end{eqnarray}
Now the spatial harmonic gauge fixing amounts to setting $\chi=id$, which yields a vanishing tension field 
\begin{eqnarray}
\label{eq:sh}
-g^{ij}\left(\Gamma[g]^{k}_{ij}-\Gamma[\gamma]^{k}_{ij}\right)=0.
\end{eqnarray}
A subtle issue that arises in the context of spatial gauge fixing is the existence of a harmonic map between $(\Sigma,g)$ and $(\Sigma,\gamma)$. There are a plethora of studies that address such questions \cite{schoen1997lectures}. The one study that is most relevant to the current context is the one by Eells and Sampson \cite{eells1964harmonic} who proved if $M$ is a compact manifold with non-positive Riemann curvature, then any continuous map from a compact manifold into $M$ is homotopic to a harmonic map. Andersson and Moncrief \cite{andersson2003elliptic} applied this gauge to prove a local existence theorem for the vacuum Einstein's equations in arbitrary dimensions. Later they have successfully implemented this gauge to prove a small data nonlinear stability result of the `Milne' model and its higher-dimensional generalizations (so-called `Lorentz cone spacetimes') \cite{andersson2004future, andersson2011einstein}. In addition to the study by \cite{andersson2004future}, there have been several other studies where the spatial harmonic gauge is utilized to study the long-time existence issues of the Einstein field equations coupled to matter sources. The implementation of the spatial harmonic gauge condition (\ref{eq:sh}) yields the following elliptic equation for the shift vector field
\begin{eqnarray}
\Delta_{g}X^{i}-R^{i}_{j}X^{j}&=&-2k^{ij}\nabla_{j}N-2N\nabla_{j}k^{ij}+\tau\nabla^{i}N\\\nonumber
&&+(2Nk^{jk}-2\nabla^{j}X^{k})(\Gamma[g]^{i}_{jk}-\Gamma[\gamma]^{i}_{jk})-g^{ij}\partial_{t}\Gamma[\gamma]_{ij}^{k}.
\end{eqnarray}
We designate this gauge fixed system as the CMCSH-Einstein-Euler-$\Lambda$ system.  

\subsection{Re-scaled equations and local well-posedness}
The background solution we are interested in is the constant negative sectional spatial curvature FLRW model in `$3+1$' dimensions. In higher dimensions ($n>3$), however, the spatial metrics are negative Einstein (not necessarily hyperbolic). The solution is expressible in the following warped product form   
 \begin{eqnarray}
 \label{eq:fixed_space}
 ^{n+1}g&=&-dt\otimes dt+a(t)^{2}\gamma_{ij}dx^{i}\otimes dx^{j},
 \end{eqnarray}
 where $R_{ij}(\gamma)=-\frac{1}{n}\gamma_{ij}$ and $t\in (0,\infty)$. Clearly we have $N=1,~X^{i}=0,~g_{ij}=a(t)^{2}\gamma_{ij}$. These spacetimes are globally foliated by constant mean curvature slices which is evident from the following simple calculations 
 \begin{eqnarray}
 k_{ij}&=&-\frac{1}{2N}\partial_{t}g_{ij}=-a\dot{a}\gamma_{ij},\\\nonumber
 \tau(t)&=&k_{ij}g^{ij}=-a\dot{a}\gamma_{ij}\frac{1}{a(t)^{2}}\gamma^{ij}=-\frac{n\dot{a}}{a}.
 \end{eqnarray}
A standard linear algebra decomposition $k_{ij}=k^{tr}_{ij}+\frac{\tr_{g}k}{n}g_{ij}$, where $k^{tr}_{ij}$ is the trace-less part of $k_{ij}$ and $\tr_{g}k=\tau$ is its trace, yields $k^{tr}=0$ for the background solution (\ref{eq:fixed_space}).   
Note that in order to study the dynamics of the perturbations, we need to re-scale the field equations. We want to re-scale in such a way as to make the background time independent and then analyze the associated re-scaled dynamical system. For this particular purpose, we will need the background density and $n-$velocity. Let us now explicitly calculate the scale factor $a(t)$, the background density, and the velocity. Note that we are only interested in the asymptotic behaviour of the scale factor at $t\to\infty$ instead of its precise expression for our stability analysis. The evolution equation for  energy density of the background solutions reads
 \begin{eqnarray}
 \frac{\partial\rho}{\partial t}&=&-n\gamma_{ad}\frac{\dot{a}}{a}\rho,
 \end{eqnarray} 
 integration of which yields
 \begin{eqnarray}
\rho a(t)^{n\gamma_{ad}}=C_{\rho},
\end{eqnarray}
for some finite time independent positive constant $C_{\rho}$. The momentum constraint (\ref{eq:MC}) equation yields 
\begin{eqnarray}
v^{i}=0
\end{eqnarray}
since $k^{tr}=0$ and $\partial_{i}\tau=0$ in CMCSH gauge.
Utilizing the Hamiltonian constraint, we obtain the following ODE for the scale factor
\begin{eqnarray}
\label{eq:backint}
\frac{1}{a}\frac{da}{dt}=\sqrt{\frac{1}{n(n-1)}\left(2\Lambda+\frac{1}{a^{2}}+\frac{2C_{\rho}}{a^{n\gamma_{ad}}}\right)},
\end{eqnarray}
integration of which yields at $t\to \infty$
\begin{eqnarray}
\label{eq:expansion}
a\sim e^{\alpha t}
\end{eqnarray}
where $\alpha:=\sqrt{\frac{2\Lambda}{n(n-1)}}$. Taking $\Lambda= 0$, integration of (\ref{eq:backint}) 
yields
\begin{eqnarray}
\label{eq:asymp}
a\sim a(0)+\frac{t}{\sqrt{n(n-1)}}
\end{eqnarray}
as $t\to\infty$.

Now let us focus on obtaining a re-scaled system of evolution and constraint equations. Recall that the metric in local coordinates $(t,x^{i})$ may be written as 
 \begin{eqnarray}
 \hat{g}&=&-N^{2}dt\otimes dt+g_{ij}(dx^{i}+X^{i}dt)\otimes(dx^{j}+X^{j}dt)
 \end{eqnarray}
 and the background solutions are written as 
 \begin{eqnarray}
 ^{n+1}g_{background}=-dt\otimes dt+a^{2}\gamma_{ij}dx^{i}\otimes dx^{j},
 \end{eqnarray}
where $R[\gamma]_{ij}=-\frac{1}{n}\gamma_{ij}$ and $t\in (0,\infty)$. If we assign $a$ the dimension of length (natural) (similar to the re-scaling used by \cite{andersson2011einstein}), the dimensions of the rest of the entities follow 
$t\sim length,~x^{i}\sim (length)^{0},~g_{ij}\sim (length)^{2},~N\sim (length)^{0},~X^{i}\sim (length)^{-1}, ~\tau \sim(length)^{-1}$,~$\kappa^{tr}_{ij}\sim length$. Therefore, the following scaling follows naturally 
\begin{eqnarray}
\tilde{g}_{ij}=a^{2} g_{ij},
\tilde{N}=N,
\tilde{X}^{i}=\frac{1}{a}X^{i},
\tilde{k^{tr}}_{ij}=ak^{tr}_{ij}.
\end{eqnarray}
Here, we denote the dimension-full entities with a tilde sign and dimensionless entities are left alone for simplicity. The Hamiltonian constraint (\ref{eq:HC}) and the momentum constraint (\ref{eq:MC}) take the following forms 
 \begin{eqnarray}
 R(g)-|k^{tr}|^{2}+1+2C_{\rho}a^{2-n\gamma}&=&2a(t)^{2}\tilde{\rho}\left(1+\gamma_{ad} \tilde{g}(\tilde{v},\tilde{v})\right),\\\nonumber
 \nabla_{j}k^{trij}&=&a(t)^{3}\gamma_{ad}\tilde{\rho}\sqrt{1+\tilde{g}(\tilde{v},\tilde{v})}\tilde{v}^{i}.
 \end{eqnarray}
 Notice that we are yet to re-scale the matter fields i.e., $(\rho,v)$. We have already noticed that the background energy density satisfies $\rho(t)=\frac{C_{\rho}}{a^{n\gamma_{ad}}}$ for some constant $C_{\rho}>0$ depending on the initial density. Therefore, we naturally scale the energy density by $\frac{1}{a^{n\gamma_{ad}}}$ i.e.,
 \begin{eqnarray}
 \tilde{\rho}&=&\frac{1}{a(t)^{n\gamma_{ad}}}\rho
 \end{eqnarray}
 so that the background density becomes time independent and an $O(1)$ entity. Notice that $C_{\rho}$ is a constant with dimension of $(length)^{n\gamma_{ad}-2}$.
Now, since, the background $n-$velocity vanishes for the background solution, we do not have a straightforward scaling. However, we have the normalization condition for the $n+1-$velocity field  $\textbf{u}=-^{n+1}g(u,n)n+v$
 \begin{eqnarray}
 -(\tilde{N}\tilde{\textbf{u}}^{0})^{2}+\tilde{g}_{ij}\tilde{v}^{i}\tilde{v}^{j}&=&-1,
 \end{eqnarray}
 and therefore scaling of the metric $\tilde{g}_{ij}=a^{2}g_{ij}$ yields the following natural scaling of $v^{i}$
 \begin{eqnarray}
 \tilde{v}^{i}&=&\frac{1}{a}v^{i},
 \end{eqnarray}
 since the right hand side is simply a constant. 
 The complete evolution and constraints of the gravity coupled fluid system may be expressed in the following re-scaled form 
 \begin{eqnarray}
 \label{eq:metricevol}
 \frac{\partial g_{ij}}{\partial t}&=&2\frac{\dot{a}}{a}\left(N-1\right)g_{ij}-\frac{2}{a}Nk^{tr}_{ij}+\frac{1}{a}(L_{X}g)_{ij},
 \end{eqnarray}
 \begin{eqnarray}
\frac{\partial k^{tr}_{ij}}{\partial t}&=&-\underbrace{\frac{\dot{a}}{a}\left((N-1)(n-2)+n-1\right)}_{should~be~<0~to~generate~decay}k^{tr}_{ij}-\frac{2}{a}Nk^{tr}_{ik}k^{trk}_{j} +\underbrace{\frac{1}{a}NR_{ij}}_{gauge~fixing~is ~required}\\\nonumber 
 &&-\left(\frac{a}{n^{2}}(\tau^{2}-\frac{2n\Lambda}{n-1})+\frac{n\gamma_{ad}-2}{n(n-1)}a^{1-n\gamma_{ad}}C_{\rho}-\frac{N(1+2C_{\rho}a^{2-n\gamma})}{a(n-1)}\right.\\\nonumber
 &&\left.+\frac{\gamma_{ad}-2}{n-1}a^{1-n\gamma_{ad}}N\rho g_{ij}\right)g_{ij}-\nonumber \frac{1}{a}\nabla_{i}\nabla_{j}N-a^{1-n\gamma_{ad}}N\gamma\rho v_{i}v_{j} +\frac{1}{a}(L_{X}k^{tr})_{ij},
 \end{eqnarray}

\begin{eqnarray}
\label{eq:equationdensity}
\partial_{t}\rho+\frac{\gamma_{ad}\rho N\nabla_{i}v^{i}}{a[1+g(v,v)]^{1/2}}+\frac{(N-1)n\gamma_{ad}\rho\dot{a}}{a}\\\nonumber=\frac{1}{a}L_{X}\rho-\frac{\gamma_{ad}\rho g(\partial_{t}v,v)}{1+g(v,v)}-\frac{NL_{v}\rho}{a[1+g(v,v)]^{1/2}}+\frac{\gamma_{ad}\rho N}{1+g(v,v)}\left(\frac{1}{a}k^{tr}(v,v)-\frac{\dot{a}}{a}g(v,v)\right)\\\nonumber+\frac{\gamma_{ad}\rho v_{i}L_{X}v^{i}}{a[1+g(v,v)]}-\frac{\gamma_{ad}N\rho L_{v}N}{a[1+g(v,v)]^{1/2}},
 \end{eqnarray}
 \begin{eqnarray}
 \label{eq:eomf}
 \gamma_{ad}\rho u^{0}\partial_{t}v^{i}+\underbrace{\gamma_{ad}\rho u^{0}[2N-1-n(\gamma_{ad}-1)]\frac{\dot{a}}{a}}_{should~be~>0~to~generate~decay}v^{i}-\frac{2\gamma_{ad}\rho N u^{0}k^{tri}_{j}v^{j}}{a}\nonumber-\frac{\gamma_{ad}\rho u^{0}}{a}L_{X}v^{i}\\
+\frac{\gamma_{ad}\rho}{a}v^{j}\nabla_{j}v^{i}+\frac{\gamma_{ad}\rho[1+g(v,v)]}{a}\nabla^{i}N+\frac{(\gamma_{ad}-1)}{a}\left(\nabla^{i}\rho+v^{i}L_{v}\rho+au^{0}v^{i}\partial_{t}\rho\right.\\\nonumber
\left.-u^{0}v^{i}L_{X}\rho\right)=0,
 \end{eqnarray}
 \begin{eqnarray}
 \label{eq:hamilton}
 R(g)-|k^{tr}|^{2}-2a^{2-n\gamma_{ad}}\rho\{1+\gamma_{ad} g(v,v)\}+1+2C_{\rho}a^{2-n\gamma}=0,\\
 \label{eq:mc}
\nabla_{j}k^{trij}=-\gamma_{ad} a^{2-n\gamma_{ad}}\rho\sqrt{1+g(v,v)} v^{i},
\end{eqnarray}
 \begin{eqnarray}
 \label{eq:lapse1}
 \Delta_{g}N+\left(\underbrace{|k^{tr}|^{2}+a(t)^{2}(\frac{\tau^{2}}{n}-\frac{2\Lambda}{n-1})+a^{2-n\gamma_{ad}}[\frac{(n\gamma_{ad}-2)}{n-1}+\gamma_{ad}\nonumber g(v,v)]\rho}_{should~be~>0~for~ elliptic~regularity}\right)N\\
 =a^{2}\frac{\partial\tau}{\partial t},
 \end{eqnarray}
 \begin{eqnarray}
 \label{eq:shiftin}
\Delta_{g}X^{i}-R^{i}_{j}X^{j}=-n(1-\frac{2}{n})\dot{a} \nabla^{i}N-2\nabla^{j}Nk^{tri}_{j}-\underbrace{2N\nabla_{j}k^{trij}}_{use~momentum~constraint}\\\nonumber
+(2Nk^{trjk}-2\nabla^{j}X^{k})\left(\Gamma[g]^{i}_{jk}-\Gamma[\gamma]^{i}_{jk}\right)-ag^{ij}\partial_{t}\Gamma[\gamma]_{ij}^{k}.
\end{eqnarray}
In CMCSH gauge, the time function is obtained by setting $\tau=$monotonic function of $t$ alone. Now, we have $\tau$ evaluated for the background solution which is a monotonic function of $t$ alone since utilizing the Hamiltonian constraint and the lapse equation at the background solution one obtains 
\begin{eqnarray}
\frac{\partial \tau}{\partial t}&=&\frac{1}{n}(\tau^{2}-\frac{2n\Lambda}{n-1})+\frac{n\gamma_{ad}-2}{n-1}a^{-n\gamma_{ad}}C_{\rho}\\\nonumber 
&=&\frac{1}{a(t)^{2}(n-1)}\left(1+n\gamma_{ad} C_{\rho}a^{2-n\gamma_{ad}}\right)>0.
\end{eqnarray}
We set the time function to be the solution of this equation
\begin{eqnarray}
\frac{\partial \tau}{\partial t}=\frac{1}{a(t)^{2}(n-1)}\left(1+n\gamma_{ad} C_{\rho}a(t)^{2-n\gamma_{ad}}\right),
\end{eqnarray}
 i.e., $t=t(\tau)$ which settles the business of choosing a time coordinate. One subtlety we need to address here is that regarding the structure of the background solution space for dimensions strictly greater than $3+1$. We do so in the next section. For now, we will state a local well-posedness theorem which may be proved using the method developed by \cite{andersson2003elliptic} or \cite{majda2012compressible}. The proof will require an energy current for the fluid sector, which will be defined in the next section. However, since the proof of a local existence theorem follows in a standard way by using techniques developed by \cite{andersson2003elliptic}, we will not state the proof here. In addition, one needs to preserve the gauge conditions and constraints along the solution curve as well. The constraints and gauges read 
 \begin{eqnarray}
 A_{1}:=\tr_{g}k-\int\left(\frac{1}{a(t)^{2}(n-1)}\left(1+n\gamma_{ad} C_{\rho}a(t)^{2-n\gamma_{ad}}\right)\right)dt,\\
 A^{i}_{2}:g^{kl}(\Gamma[g]^{i}_{kl}-\Gamma[\gamma]^{i}_{kl}),\\
 A^{i}_{3}:=\nabla_{j}k^{trij}+\gamma_{ad} a^{2-n\gamma_{ad}}\rho\sqrt{1+g(v,v)} v^{i},\\
 A_{4}:=R(g)-|k^{tr}|^{2}-2a^{2-n\gamma_{ad}}\rho\{1+\gamma_{ad} g(v,v)\}+1+2C_{\rho}a^{2-n\gamma}.
 \end{eqnarray}
 Explicit computations utilizing the evolution equations leads to a hyperbolic system for $(A_{1},A^{i}_{2},A^{i}_{3},A_{4})$ in CMCSH gauge. Using an energy similar to the one presented in section $4$ of \cite{andersson2003elliptic}, one shows that if $(A_{1},A^{i}_{2},A^{i}_{3},A_{4})=\textbf{0}$ initially then $(A_{1},A^{i}_{2},A^{i}_{3},A_{4})\equiv\textbf{0}$ along the solution curve $t\mapsto (g(t),k^{tr}(t),\rho(t),v(t),N(t),X(t))$. The local well-posedness theorem is stated as follows  
 \\
 \textbf{Local Well-posedness Theorem:} \textit{Let $(g_{0},k^{tr}_{0},\rho_{0},v_{0})\in H^{s}\times H^{s-1}\times H^{s-1}\times H^{s-1},~s>\frac{n}{2}+2$ be the initial data for the Cauchy problem of the re-scaled Einstein-Euler-$\Lambda$ evolution equations (\ref{eq:metricevol}-\ref{eq:eomf}) in Constant Mean Extrinsic Curvature Spatial Harmonic (CMCSH) gauge satisfying the constraint equations (\ref{eq:hamilton}-\ref{eq:mc}). This CMCSH Cauchy problem is well posed in $C\left([0,t^{*}];H^{s}\times H^{s-1}\times H^{s-1}\times H^{s-1}\right)$. In particular, there exists a time $t^{*}>0$ dependent on $||g_{0}||_{H^{s}},||k^{tr}_{0}||_{H^{s-1}},||\rho_{0}||_{H^{s-1}}, ||v_{0}||_{H^{s-1}}$ such that the solution map $(g_{0},k^{tr}_{0},\rho_{0},v_{0}) \mapsto (g(t),k^{tr}(t),\rho(t),v(t),N(t),X(t))$ is continuous as a map 
\begin{eqnarray}
H^{s}\times H^{s-1}\times H^{s-1}\times H^{s-1}\\
\to H^{s}\times H^{s-1}\nonumber\times H^{s-1}\times H^{s-1}\times H^{s+1}\times H^{s+1}.  
\end{eqnarray}
Let $t^{*}$ be the maximal time of existence of a solution to the CMCSH Cauchy problem with data $(g_{0},k^{tr}_{0},\rho_{0},v_{0})$, then either $t^{*}=\infty$ or 
\begin{eqnarray}\lim_{t\to t^{*}}\sup \max
\left(C(\Omega_{t}),||\nabla[\gamma]g(t)||_{L^{\infty}},||k^{tr}||_{L^{\infty}},||\nabla\rho(t)||_{L^{\infty}},||\rho(t)||_{L^{\infty}},\right.\\\nonumber\left.
||\nabla[\gamma]v(t)||_{L^{\infty}},||v(t)||_{L^{\infty}}\right)=\infty,
\end{eqnarray}
where $C(\Omega_{t})$ is a constant dependent on the ellipticity constant of the metric $\hat{g}$ defined to be $\Omega_{t^{*}}:=\sup_{t\in[0,t^{*}]}\Omega[\hat{g}(t)]$, $\Omega[\hat{g}]:=\Omega[g]+||N||_{L^{\infty}}+||N^{-1}||_{L^{\infty}}+||X||_{L^{\infty}}$, and $\Omega^{-1}\gamma(\xi,\xi)\leq g(\xi,\xi)\leq \Omega\gamma(\xi,\xi),~\forall \xi\in sections\{T\Sigma\}$.\\
In addition the gauges and constraints are preserved along the solution curve $t\mapsto (g(t),k^{tr}(t),\rho(t),v(t),N(t),X(t))$ if they are satisfied initially.}
 
 \subsection{\textbf{Center Manifold of the dynamics}}
The following mathematics concerning the center manifold dynamics in the setting of the Einstein flow was first established by the work of Andersson-Moncrief \cite{andersson2004future}. We now present some of this work and adapt it to the Einstein-Euler-$\Lambda$ setting. A center manifold of a dynamical system is associated with fixed points of the dynamics. Essentially a center manifold of a fixed point corresponds to the nearby solutions (in phase space) that do not exhibit exponential growth or decay. This may be zero-dimensional or a subspace of the phase space and admits a manifold structure (so the name center `manifold'). For a more precise and mathematically rigorous definition, we refer the reader to \cite{marsden2012hopf, lanford1973bifurcation}. In a finite-dimensional setting, the existence of a center manifold roughly corresponds to the case when the linearization of the flow vector field has purely imaginary spectra. In an infinite-dimensional setting, moduli spaces play the role of center manifolds (in an appropriate sense of course). As we shall see, the center manifold in this particular occasion is played by the Einstein moduli space, which consists of non-isolated fixed points of the Einstein-Euler-$\Lambda$ flow and is characterized by the negative Einstein metrics modulo gauge transformations. 

Unlike the $n=3$ dimensional case where the negative Einstein spaces are hyperbolic (and the center manifold of the gravitational dynamics consists of a point), the higher dimensional case is more subtle. The matter degrees of freedom corresponding to the background solution (in its re-scaled version) in CMCSH gauge is essentially described by $v^{i}=0, \rho=\rho_{B.G}$. It becomes more interesting when we consider the gravitational degrees of freedom. In the case of $n=3$, a negative Einstein structure implies hyperbolic structure through the Mostow rigidity theorem and therefore the manifold describing the fixed point of the dynamics is essentially zero-dimensional. In other words, the particular solution $(\{\gamma|Ricci[\gamma]=-\frac{1}{3}\gamma, k^{tr}=0, \rho=C_{\rho}, v^{i}=0)$  serves as an isolated fixed point of the Einstein-Euler dynamics. However, in the higher dimensional cases ($n>3$), a new possibility arises of having non-hyperbolic negative Einstein spaces. When we linearize about any member of such a non-isolated family of negative Einstein metrics, the linearized equations will always admit a finite-dimensional space of neutral modes. Naturally, this represents (modulo the matter degrees of freedom) the tangent space to the background spacetimes (\ref{eq:model}). These smooth families of background spacetimes determined by the corresponding families of negative Einstein metrics and zero-dimensional matter degrees of freedom ($v^{i}=0, \rho=C_{\rho}$) form the `center manifold' for the dynamical system defined by the re-scaled Einstein-Euler-$\Lambda$ equations. The spatial metric satisfies 
\begin{eqnarray}
\label{eq:ein}
R_{ij}(\gamma)=-\frac{1}{n}\gamma_{ij}.
\end{eqnarray}
Let us denote the space of metrics satisfying equation (\ref{eq:ein}) by $\mathcal{E}in_{-\frac{1}{n}}$.

Let $\gamma^{*}\in \mathcal{E}in_{-\frac{1}{n}}$ and $\mathcal{V}$ be its connected component. Also consider $\mathcal{S}_{\gamma}$ to be the harmonic slice of the identity diffeomorphism i.e., the set of $\gamma\in\mathcal{E}in_{-\frac{1}{n}}$ for which the identity map $id: (\Sigma,\gamma)\to (\Sigma,\gamma^{*})$ is harmonic (since any metric $\gamma\in \mathcal{E}in_{-\frac{1}{n}}$ verifies the fixed point criteria $R[\gamma]_{ij}=-\frac{1}{n}\gamma_{ij}$, it should also satisfy the CMCSH gauge condition with respect to a background metric and in this case the background metric is simply chosen to be $\gamma^{*}$, another element of $\mathcal{E}in_{-\frac{1}{n}}$ and $||\gamma-\gamma^{*}||<\delta, \delta>0$ sufficiently small). This condition is equivalent to the vanishing of the tension field $-V^{k}$ that is
\begin{eqnarray}
\label{eq:harmonic}
-V^{k}=-\gamma^{ij}(\Gamma[\gamma]^{k}_{ij}-\Gamma[\gamma^{*}]^{k}_{ij})=0.
\end{eqnarray} 
For $\gamma\in \mathcal{E}in_{-\frac{1}{n}}$, $\mathcal{S}_{\gamma}$ is a submanifold of $\mathcal{M}_{\Sigma}$ for $\gamma$ sufficiently close to $\gamma^{*}$ (easily proven using a standard procedure and therefore we omit the proof, see \cite{andersson2011einstein} for details). The deformation space $\mathcal{N}$ of $\gamma^{*}\in \mathcal{E}in_{-\frac{1}{n}}$ is defined as the intersection of the $\gamma^{*}-$connected component $ \mathcal{V}\subset \mathcal{E}in_{-\frac{1}{n}}$ and the harmonic slice $\mathcal{S}_{\gamma}$ i.e.,
\begin{eqnarray}
\label{eq:ds}
\mathcal{N}:= \mathcal{V}\cap \mathcal{S}_{\gamma}.
\end{eqnarray}
$\mathcal{N}$ is assumed to be smooth (i.e., equipped with $C^{\infty}$ topology). 
In the case of $n=3$, following the Mostow rigidity theorem, the negative Einstein structure is rigid \cite{lebrun1994einstein}. This rigid Einstein structure in $n=3$ corresponds to the hyperbolic structure up to isometry. For higher genus Riemann surfaces $\Sigma_{genus}$ ($genus>1$), the space of distinct hyperbolic structures is the classical Teichm\"uller space diffeomorphic to $\mathbf{R}^{6genus-6}$. The tangent space to any point in the Teichm\"uller space corresponds to the `neutral modes' of $2+1$ gravity defined by transverse-traceless tensors. However, $2+1$ gravity is fundamentally different from the higher dimensional cases since the former is devoid of gravitational waves degrees of freedom. For $n>3$, the centre manifold $\mathcal{N}$ is a finite-dimensional submanifold of $\mathcal{M}_{\Sigma}$ (and in particular of $\mathcal{M}_{-1}$). 
Following the analysis of \cite{andersson2011einstein}, the formal tangent space $T_{\gamma}\mathcal{N}$ in local coordinates is expressible as 
\begin{eqnarray}
\label{eq:tangentspace}
\frac{\partial \gamma}{\partial q^{a}}&=&l^{TT||}_{a}+L_{W^{||}_{a}}\gamma,
\end{eqnarray}
where $l^{TT||}_{a}\in \ker(\mathcal{L}_{\gamma,\gamma})=C^{TT||}(S^{2}\Sigma)\subset C^{TT}(S^{2}\Sigma)$, $W^{||}\in \mathfrak{X}(\Sigma)$ satisfies
through the time derivative of the CMCSH gauge condition $-\gamma^{ij}(\Gamma[\gamma]^{k}_{ij}-\Gamma[\gamma^{*}]^{k}_{ij})=0$
\begin{eqnarray}
\label{eq:parallelvector}
-[\nabla[\gamma]^{m}\nabla[\gamma]_{m}W^{||i}+R[\gamma]^{i}_{m}W^{||m}]+(l^{TT||}\nonumber+L_{W^{||}}\gamma)^{mn}(\Gamma[\gamma]^{k}_{mn}-\Gamma[\gamma^{*}]^{k}_{mn})\\
=0,
\end{eqnarray} 
and, $\{q^{a}\}_{a=1}^{dim(\mathcal{N})}$ is a local chart on $\mathcal{N}$, $\mathfrak{X}(\Sigma)$ is the space of vector fields on $\Sigma$ (in a suitable function space setting), $C^{TT}(S^{2}\Sigma)$ is the space of $\gamma-$fransverse-traceless $2-$tensors on $\Sigma$. $\gamma^{*}$ is close to $\gamma$ i.e., $||\gamma-\gamma^{*}||<\delta,\delta>0$ sifficiently small (the norm is arbitrary since both $\gamma$ and $\gamma^{*}$ are elements of a finite dimensional vector space $\mathcal{N}$). An important thing to note is that all known examples of closed negative Einstein spaces have integrable deformation spaces and that the deformation spaces are stable i.e., $Spec\{\mathcal{L}_{\gamma,\gamma}\}\geq 0$. The spectrum of the operator $\mathcal{L}_{\gamma,\gamma}$ plays an important role determining the decay rates in the pure vacuum case considered by \cite{andersson2004future, andersson2011einstein}. We stress that the equation (\ref{eq:parallelvector}) plays a crucial role in the energy argument. 

\subsection{\textbf{Kinematics of the perturbations and shadow gauge}}
As discussed in the previous section, the Einstein moduli space is finite-dimensional for $n>3$ and serves as the center manifold of the dynamics (the deformation space $\mathcal{N}$ to be precise). On the other hand, for the $n=3$ case, following Mostow rigidity, negative Einstein structure implies hyperbolicity and therefore the moduli space reduces to a point. We will consider the general case when the moduli space is finite-dimensional since the zero-dimensional moduli space case does not require special treatment. Let $\mathcal{N}$ be the deformation space of $\gamma^{*}$. Now suppose, we perturb the metric about $\gamma^{*}$. The perturbation is defined to be $u:=g-\gamma^{*}$ such that $||u||_{H^{s}}<\delta$ for a sufficiently small $\delta>0$. Due to the finite dimensionality of $\mathcal{N}$, $g$ in fact could be an element of $\mathcal{N}$ lying within a $\delta-$ball of $\gamma^{*}$ i.e., while perturbing one may choose directions tangent to $\mathcal{N}$. But if $g$ lies on $\mathcal{N}$, then corresponding solution satisfies the fixed point condition $Ric(g)=-\frac{1}{n}g$. Therefore these `trivial' perturbations do not see the dynamics. The question is how to treat such perturbations which are tangent to $\mathcal{N}$. This was precisely handled by invoking a shadow gauge condition in the study of the fully nonlinear stability problem of the vacuum Einstein' equations by \cite{andersson2011einstein} and later used in \cite{mondal2019attractors} to handle the case with a positive cosmological constant. The shadow gauge reads
\begin{eqnarray}
(g-\gamma)\perp \mathcal{N}
\end{eqnarray}
for $\gamma\in\mathcal{N}$, where the orthogonality is in the $L^{2}$ sense with respect to the metric $\gamma$. This precisely takes care of the motion of the orthogonal perturbations by studying the time evolution of $h:=g-\gamma$, while the motion of the tangential perturbations is taken care of by evolving $\gamma\in\mathcal{N}$. This does not affect the characteristics of the background spacetimes as $\gamma$ does not leave the moduli space i.e., it is still a negative Einstein metric with Einstein constant $-\frac{1}{n}$. However, if we write the perturbation as $u_{ij}=g_{ij}-\gamma_{ij}$, then $\frac{\partial g_{ij}}{\partial t}=\frac{\partial u_{ij}}{\partial t}+\frac{\partial \gamma_{ij}}{\partial t}$ and therefore, $\frac{\partial \gamma_{ij}}{\partial t}$ enters into the evolution equation for the metric. Therefore this additional contribution arising as a result of finite dimensionality of the moduli space $\mathcal{N}$ needs to be estimated properly since this will inevitably show up in the time derivative of the energy functional as well.

\begin{center}
\begin{figure}
\begin{center}
\includegraphics[width=13cm,height=60cm,keepaspectratio,keepaspectratio]{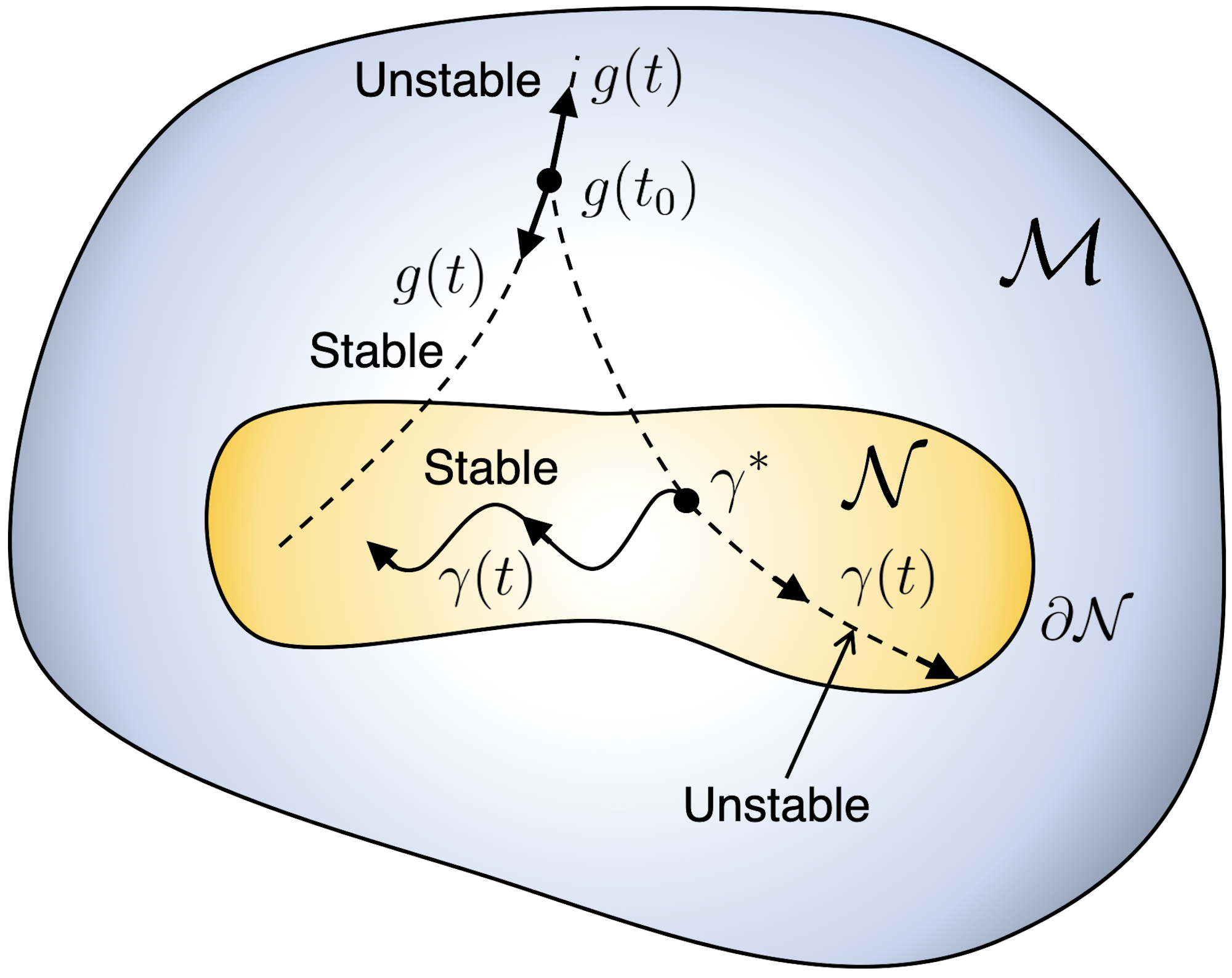}
\end{center}
\begin{center}
\caption{Kinematics of the perturbations for the Einstein-Euler system. Instability materializes in $\mathcal{N}$ iff $\gamma(t)$ leaves every compact set of $\mathcal{N}$ i.e., approaches $\partial\mathcal{N}$. Stability occurs when $\gamma(t)$ evolves from one point of $\mathcal{N}$ to simply another point of $\mathcal{N}$. The instability of $g(t)$ is naturally defined by the unbounded growth of $g(t)-\gamma(t)$ in a suitable function space.}
\label{fig:pdf}
\end{center}
\end{figure}
\end{center}

The shadow gauge condition is explicitly written as follows 
\begin{eqnarray}
\label{eq:shadow}
\int_{\Sigma} (g_{ij}-\gamma_{ij})\frac{\partial\gamma^{ij}}{\partial q^{a}}\mu_{\gamma}=0,
\end{eqnarray}
where $\frac{\partial \gamma_{ij}}{\partial q^{a}}$ is a local basis for the tangent space of $\mathcal{N}$ in local coordinates $\{q^{a}\}_{a=1}^{\dim(\mathcal{N})}$ and $\frac{\partial \gamma^{ij}}{\partial q^{a}}=-\gamma^{ik}\gamma^{jl}\frac{\partial\gamma_{kl}}{\partial q^{a}}$. Technically, this shadow gauge condition is equivalent to projecting $g$ onto $\gamma\in \mathcal{N}$. In another words, there is a projection map $\mathcal{P}:\mathcal{M}\to \mathcal{N}$ such that 
\begin{eqnarray}
\gamma=\mathcal{P}[g].
\end{eqnarray}
Since the deformation space $\mathcal{N}$ (finite dimensional) is assumed to have a $C^{\infty}$ structure, this projection map $\mathcal{P}$ can also be interpreted as a smoothing map. Through explicit computations under the assumption of $||g-\gamma||_{H^{s}}<\delta$, where $\delta>0$ is sufficiently small, one may explicitly show that the following estimates hold
\begin{eqnarray}
||\frac{\partial \gamma}{\partial t}||\leq C||\frac{\partial g}{\partial t}||_{H^{s-1}}, 
\end{eqnarray}
where $C$ is a fixed constant depending only on the dimension and the norm on the left hand side may be arbitrary since $\mathcal{N}$ is finite dimensional yielding an equivalence between Sobolev norms of all orders.

\section{Commuted equations for perturbations and Energy current}
In this section, we obtain evolution equations for the perturbations and construct the commuted equations for the fluid variables. We explicitly construct an energy current motivated by the energy current that was constructed by Christodoulou \cite{christodoulou2007formation} in the context of special relativistic perfect fluid dynamics. This current has a remarkable property that while contracted with a time-like vector field it yields positive definite energy. Since the Euler's equations are theoretically derivable from a Lagrangian (see \cite{choquet2008general} for a symmetric hyperbolic formulation of Euler's equations and \cite{moncrief1977hamiltonian, demaret1980hamiltonian, bao1985hamiltonin} for a Hamiltonian formulation of a relativistic perfect fluid coupled to gravity), one naturally expects a compatible current (see Christodoulou \cite{christodoulou2016action}) associated with Euler equations. Due to the reasons mentioned in the introduction, we will not use the energy current constructed by \cite{christodoulou2007formation} but rather define a current by hand which satisfies the necessary properties. Let us define the following perturbation entities 
\begin{eqnarray}
u:=g-\gamma,~k^{tr}:=k^{tr}-0,~\omega:=N-1,~X:=X-0,~\delta\rho:=\rho-C_{\rho},\\\nonumber 
~v:=v-0.
\end{eqnarray}
The evolution equations for the perturbations to the gravitational degrees of freedom are as follows 
\begin{eqnarray}
 \label{eq:fixed_new}
 \frac{\partial u_{ij}}{\partial t}&=&2\frac{\dot{a}}{a}\omega g_{ij}-\frac{2}{a}Nk^{tr}_{ij}+h^{TT||}_{ij}+\frac{1}{a}(L_{X}u)_{ij}+(L_{Y^{||}+\frac{1}{a}X}\gamma)_{ij},
 \end{eqnarray}
 \begin{eqnarray}
 \label{eq:momentumscaled}
\frac{\partial k^{tr}_{ij}}{\partial t}&=&-\frac{\dot{a}}{a}\left(\omega(n-2)+n-1\right)k^{tr}_{ij}-\frac{2}{a}Nk^{tr}_{ik}k^{trk}_{j} \\\nonumber 
 &&+\left(a^{1-n\gamma_{ad}}\frac{\omega\gamma_{ad}C_{\rho}+N(\gamma_{ad}-2)\delta\rho}{n-1}+\frac{\omega}{a(n-1)}\right)g_{ij}\\\nonumber &&-\frac{1}{a}\nabla_{i}\nabla_{j}\omega-a^{1-n\gamma_{ad}}N\gamma\rho v_{i}v_{j} +\frac{1}{a}(L_{X}k^{tr})_{ij}+\frac{1}{a}N\left(\frac{1}{2}\mathcal{L}_{g,\gamma}u_{ij}+\mathfrak{J}_{ij}\right),
 \end{eqnarray}
where we have used the fact that $\partial_{t}\gamma_{ij}=-h^{TT||}_{ij}-L_{Y^{||}}\gamma_{ij}$ (\ref{eq:tangentspace}) ($h^{TT||}$ is parallel to $\mathcal{N}$ i.e., belongs to the kernel of $\mathcal{L}_{\gamma,\gamma}$ and $Y^{||}$ satisfies the equation (\ref{eq:parallelvector}) with $l^{TT||}$ replaced by $h^{TT||}$) and $\mathfrak{J}$ contains first derivatives of $u$, is quasi-linear and satisfies $||\mathfrak{J}||_{H^{s-1}}\leq C||u||_{H^{s}},~s>\frac{n}{2}+2$ which directly follows from its expression 
\begin{eqnarray}
\mathfrak{J}_{ij}=\frac{1}{2}\left(u_{il}u^{mn}R[\gamma]^{l}_{mjn}+u_{jl}u^{mn}R[\gamma]^{l}_{min}\right)+\nonumber\frac{1}{2}g^{mn}\gamma^{ls}\left(\nabla[\gamma]_{j}u_{ns}\nabla[\gamma]_{l}u_{im}\right.\\\nonumber 
\left.+\nabla[\gamma]_{i}u_{lm}\nabla[\gamma]_{s}u_{jn}-\frac{1}{2}\nabla[\gamma]_{j}u_{ns}\nabla[\gamma]_{i}u_{lm}+\nabla[\gamma]_{m}u_{il}\nabla[\gamma]_{n}u_{js}\right.\\\nonumber 
\left.-\nabla[\gamma]_{m}u_{il}\nabla[\gamma]_{s}u_{jn}\right),
\end{eqnarray}
where $u^{ij}:=\gamma^{ik}\gamma^{jl}u_{kl}$ and SH gauge condition is implemented. From the structure of the above equations (\ref{eq:fixed_new}-\ref{eq:momentumscaled}), it is clear that these are hyperbolic evolution equations since the map $u\mapsto \mathcal{L}_{g,\gamma}u$ is elliptic. Since the scenario is slightly different for the case of Euler's equations, we will first define the commuted entities and obtain their evolution equations. 

Multiple applications of the operator $\mathcal{D}:=\langle \nabla[\gamma]\rangle=(1+\Delta_{\gamma})^{1/2}$ on the fields would produce the desired higher order energy since $W^{s,p}(\Sigma)=\mathcal{D}^{-s}L^{p}(\Sigma),~1<p<\infty,~s\in \mathbb{R}$, ($\mathcal{D}^{s}$ acts on the spectral side) with norm defined as 
\begin{eqnarray}
||v||_{W^{s,p}(\Sigma)}=||\mathcal{D}^{s}v||_{L^{p}(\Sigma)}.
\end{eqnarray}
Here $\Delta_{\gamma}$ is defined as $\Delta_{\gamma}:=-\gamma^{ij}\nabla[\gamma]_{i}\nabla[\gamma]_{j}$. If $s$ is a non-negative integer, then $W^{s,p},~1\leq p\leq \infty$ is the closure of $C^{\infty}(\Sigma)$ with respect to the equivalent norm $\sum_{|I|\leq s}||\nabla[\gamma]^{I}v||_{L^{p}(\Sigma)}$. Let us define the following entities corresponding to the fluid variables 
\begin{eqnarray}
\dot{\rho}:=\mathcal{D}^{s-1}(\rho-C_{\rho}),~\dot{v}^{i}:=\mathcal{D}^{s-1}v^{i},~~s>\frac{n}{2}+2.
\end{eqnarray}
The evolution equations for the commuted entities read 
\begin{eqnarray}
\label{eq:commuted1}
\partial_{t}\dot{\rho}+\frac{\gamma_{ad}\rho N\nabla_{i}\dot{v}^{i}}{a[1+g(v,v)]^{1/2}}=\frac{1}{a}L_{X}\dot{\rho}-\frac{\gamma_{ad}\rho g(\partial_{t}\dot{v},v)}{1+g(v,v)}-\frac{NL_{v}\dot{\rho}}{a[1+g(v,v)]^{1/2}}\\\nonumber+\frac{\gamma_{ad}\rho v_{i}L_{X}\dot{v}^{i}}{a[1+g(v,v)]}+\mathcal{T}_{1}
\end{eqnarray}

\begin{eqnarray}
\label{eq:commuted2}
\gamma_{ad}\rho u^{0}\partial_{t}\dot{v}^{i}-\frac{\gamma_{ad}\rho u^{0}}{a}L_{X}\dot{v}^{i}+\frac{\gamma_{ad}\rho}{a}v^{j}\nabla_{j}\dot{v}^{i}+\frac{(\gamma_{ad}-1)}{a}\left(\nabla^{i}\dot{\rho}\nonumber+v^{i}L_{v}\dot{\rho}+au^{0}v^{i}\partial_{t}\dot{\rho}\right.\\\nonumber
\left.-u^{0}v^{i}L_{X}\dot{\rho}\right)=\mathcal{T}^{i}_{2}
\end{eqnarray}
where $\mathcal{T}_{1}$ and $\mathcal{T}_{2}$ are defined as follows 
\begin{eqnarray}
\label{eq:T1}
\mathcal{T}_{1}:=-[\mathcal{D}^{s-1},\partial_{t}]\rho-[\mathcal{D}^{s-1},\frac{\gamma_{ad}\rho N\nabla_{i}}{a[1+g(v,v)]^{1/2}}]v^{i}+\frac{1}{a}[\mathcal{D}^{s-1},L_{X}]\rho\\\nonumber
-[\mathcal{D}^{s-1},\frac{\gamma_{ad}\rho g_{ij}v^{j}\partial_{t}}{1+g(v,v)}]v^{i}-\frac{1}{a}[\mathcal{D}^{s-1},\frac{NL_{v}}{[1+g(v,v)]^{1/2}}]\rho+\gamma_{ad}[\mathcal{D}^{s-1},\frac{\rho v_{i}L_{X}}{a[1+g(v,v)]}]v^{i}\\\nonumber 
+\mathcal{D}^{s-1}\left[\frac{\gamma_{ad}\rho N}{1+g(v,v)}\left(\frac{1}{a}k^{tr}(v,v)-\frac{\dot{a}}{a}g(v,v)\right)\right]-\mathcal{D}^{s-1}\left[\frac{\gamma_{ad}N\rho L_{v}N}{a[1+g(v,v)]^{1/2}}\right]\\\nonumber 
-\mathcal{D}^{s-1}\left[\frac{(N-1)n\gamma_{ad}\dot{a}}{a}\rho\right],
\end{eqnarray}
\begin{eqnarray}
\label{eq:T2}
\mathcal{T}^{i}_{2}:=-\underbrace{\gamma_{ad}\rho u^{0}[2N-1-n(\gamma_{ad}-1)]\frac{\dot{a}}{a}\dot{v}^{i}}_{one~of~the~most~important~term:~generates~decay}\\\nonumber
+\gamma_{ad}[\mathcal{D}^{s-1},\rho u^{0}\partial_{t}]v^{i}-\gamma_{ad}\left[\mathcal{D}^{s-1},\rho u^{0}[2N-1-n(\gamma_{ad}-1)]\right]v^{i}+\frac{2\gamma_{ad}}{a}\mathcal{D}^{s-1}[\rho N u^{0}k^{tri}_{j}v^{j}]\\\nonumber 
+\frac{\gamma_{ad}}{a}[\mathcal{D}^{s-1},\rho u^{0}L_{X}]v^{i}-\frac{\gamma_{ad}}{a}[\mathcal{D}^{s-1},\rho v^{j}\nabla_{j}]v^{i}+\frac{\gamma_{ad}}{a}\mathcal{D}^{s-1}\left[\frac{\rho[1+g(v,v)]}{N}\nabla^{i}N\right]\\\nonumber 
+\frac{(\gamma_{ad}-1)}{a}\left\{[\mathcal{D}^{s-1},\nabla^{i}]\rho+[\mathcal{D}^{s-1},v^{i}L_{v}]\rho+a[\mathcal{D}^{s-1},u^{0}v^{i}\partial_{t}]\rho-[\mathcal{D}^{s-1},u^{0}v^{i}L_{X}]\rho\right\}.
\end{eqnarray}
Notice that we have only retained the principal and decay generating terms in the evolution equations for commuted entities and collected remaining terms as $\mathcal{T}_{1}$ and $\mathcal{T}_{2}$. These terms need to be estimated carefully, which we will do in a later section. Now we define the energy current. The components of the energy current $\mathcal{C}$ associated with the commuted fluid variables are defined as follows 
\begin{eqnarray}
\mathcal{C}^{0}:=\frac{(\gamma_{ad}-1)}{\gamma_{ad}\rho N}[1+g(v,v)]^{1/2}\dot{\rho}^{2}\nonumber+2(\gamma_{ad}-1)\frac{g_{ij}v^{i}\dot{v}^{j}}{N[1+g(v,v)]^{1/2}}\dot{\rho}\\\nonumber 
+\frac{\gamma_{ad}\rho[1+g(v,v)]^{1/2}}{N}\left(-\frac{g(v,\dot{v})^{2}}{1+g(v,v)}+g_{ij}\dot{v}^{i}\dot{v}^{j}\right)
\end{eqnarray}
and 
\begin{eqnarray}
\label{eq:current}
\mathcal{C}^{i}:=\frac{(\gamma_{ad}-1)v^{i}\dot{\rho}^{2}}{\gamma_{ad}\rho}+2(\gamma_{ad}-1)\dot{v}^{i}\dot{\rho}+\gamma_{ad}\rho v^{i}(-\frac{g(v,\dot{v})^{2}}{1+g(v,v)}+g_{ij}\dot{v}^{i}\dot{v}^{j}).
\end{eqnarray}
The energy current vector field is $\mathcal{C}:=C^{0}\partial_{t}+\frac{1}{a}\mathcal{C}^{i}\partial_{i}$. Now we show how this energy current may be used to obtain an energy. The energy is naturally defined by contracting the current with the unit time-like Cauchy hypersurface orthogonal vector field $\mathbf{n}:=\frac{1}{N}\partial_{t}-\frac{1}{aN}X$
\begin{eqnarray}
E_{fluid}:=-\int_{\Sigma}\mathcal{C}^{\mu}\mathbf{n}_{\mu}\mu_{g},
\end{eqnarray}
positivity, coercivity, and convexity of which will be shown in lemma (2). We need to define an energy associated with the gravitational degrees of freedom as well. Motivated by the energy associated with the linear stability analysis \cite{mondal2021linear}, we define a natural wave equation type of energy (which can be read off from the evolution equations) for the gravitational degrees of freedom introduced by \cite{andersson2011einstein} 
\begin{eqnarray}
\label{eq:energygravity}
E_{Ein}=
\sum_{I=1}^{s}\left(\frac{1}{2}\langle k^{tr}|\mathcal{L}^{I-1}_{g,\gamma}k^{tr}\rangle_{L^{2}}+\frac{1}{8}\langle u|\mathcal{L}^{I}_{g,\gamma}u\rangle_{L^{2}}\right).
\end{eqnarray}
One may explicitly write down the lowest order gravitational energy as follows 
\begin{eqnarray}
E^{1}_{Ein}\\\
=\frac{1}{8}\int_{M}(\nabla[\gamma]_{k}u_{ij}\nabla[\gamma]_{l}u_{mn}g^{kl}\gamma^{im}\gamma^{jn}\nonumber+(R[\gamma]^{k}~_{ij}~^{m}+R[\gamma]^{k}~_{ji}~^{m})u_{km}u_{ln}\gamma^{il}\gamma^{jn})\mu_{g}\\\nonumber 
+\frac{1}{2}\int_{M}k^{tr}_{ij}k^{tr}_{kl}\gamma^{ik}\gamma^{jl}\mu_{g}.
\end{eqnarray}
We want to establish the coercive property of these energies. In other words we would like to show that these energies control the desired norm of the data $(u,k^{tr},\dot{\rho},\dot{v})$. Let us define the total energy as follows 
\begin{eqnarray}
\label{eq:totalenergy}
E_{total}:=E_{fluid}+E_{Ein}\\\nonumber 
=\int_{\Sigma}\left(\frac{(\gamma_{ad}-1)}{\gamma_{ad}\rho }[1+g(v,v)]^{1/2}\dot{\rho}^{2}\nonumber+2(\gamma_{ad}-1)\frac{g_{ij}v^{i}\dot{v}^{j}}{[1+g(v,v)]^{1/2}}\dot{\rho}\right.\\\nonumber\left. 
+\gamma_{ad}\rho[1+g(v,v)]^{1/2}\left(-\frac{g(v,\dot{v})^{2}}{1+g(v,v)}+g_{ij}\dot{v}^{i}\dot{v}^{j}\right)\right)\mu_{g}\\\nonumber
+\sum_{i=1}^{s}\left(\frac{1}{2}\langle k^{tr}|\mathcal{L}^{i-1}_{g,\gamma}k^{tr}\rangle_{L^{2}}+\frac{1}{8}\langle u|\mathcal{L}^{i}_{g,\gamma}u\rangle_{L^{2}}\right).
\end{eqnarray}
The following two lemmas establish the coercive property of the energy for small data. Let us write $u=g-\gamma$ as follows 
\begin{eqnarray}
u=g-\gamma=u^{||}+u^{\perp}+z,
\end{eqnarray}
where $u^{||}\in \ker(\mathcal{L}_{\gamma,\gamma})$, $u^{\perp}\perp \ker(\mathcal{L}_{\gamma,\gamma})$ and $||z||_{H^{s}}\lesssim ||u^{||}||^{2}_{H^{s}}+||u^{\perp}||^{2}_{H^{s}}$. In particular, $u^{||},u^{\perp}\in C^{TT}(S^{2}\Sigma)$.
\\
\textbf{Lemma 1:} \textit{ Let ($\gamma, g, k^{tr}$) be such that $(g,\gamma)$ satisfy the shadow gauge and $g-\gamma=u$. There exists $\delta>0$ sufficiently small, and a constant $C=C(\delta)>0$ such that if $(u,k^{tr})\in \mathcal{B}_{\delta}(\mathbf{0})\in H^{s}\times H^{s-1}$,$s>\frac{n}{2}+2$ the following estimate holds 
\begin{eqnarray}
||u^{||}||_{H^{s}}\leq C \left(||u^{\perp}||^{2}_{H^{s}}+||k^{tr}||^{2}_{H^{s-1}}\right).
\end{eqnarray} 
}
\textbf{Proof:} Following the shadow gauge ($u$ is $L^{2}$-orthogonal to $\mathcal{N}$) (\ref{eq:shadow}), we may write 
\begin{eqnarray}
\langle(g-\gamma),h^{TT||}+L_{Y^{||}}\gamma\rangle_{L^{2}}&=&0\\\nonumber
\langle u^{||}+u^{\perp}+z,h^{TT||}+L_{Y^{||}}\gamma\rangle_{L^{2}}&=&0\\\nonumber
\langle u^{||},h^{TT||}\rangle_{L^{2}}+\langle z,h^{TT||}+L_{Y^{||}}\gamma\rangle_{L^{2}}&=&0
\end{eqnarray}   
where we have used the facts that $\langle u^{\perp},h^{TT||}\rangle_{L^{2}}=0$ and $\langle u^{||}+u^{\perp},L_{Y^{||}}\gamma\rangle_{L^{2}}=0$. Using the relation obtained, we may say that $u^{||}$ is a smooth function of $z$ which satisfies 
\begin{eqnarray}
||z||_{H^{s}}\leq C(||u^{||}||^{2}_{H^{s}}+||u^{\perp}||^{2}_{H^{s}}),
\end{eqnarray} 
and therefore, $u^{||}$ satisfies the following estimate 
\begin{eqnarray}
||u^{||}||_{H^{s}}\leq C\left(||u^{\perp}||^{2}_{H^{s}}+||k^{tr}||^{2}_{H^{s-1}}\right).~~~~~~~~~~~~~~~~~~~~~~~~~~~~~~~~~~~~~~~~\Box
\end{eqnarray}
\textbf{Lemma 2:} \textit{Let $s>\frac{n}{2}+2$, $\gamma\in\mathcal{N}$, and $E_{total}$ be the the total energy defined in (\ref{eq:totalenergy}) and $\gamma_{ad}\in (1,\frac{n+1}{n})$. Then $\exists~\delta>0, C=C(\delta)>0$, such that $\forall (u,k^{tr},\delta\rho,v)\in \mathcal{B}_{\delta}(\mathbf{0})\in H^{s}\times H^{s-1}\times H^{s-1}\times H^{s-1}$, the following estimate holds
\begin{eqnarray}
||u^{\perp}||^{2}_{H^{s}}+||v||^{2}_{H^{s-1}}+||\delta\rho||^{2}_{H^{s-1}}+||v||^{2}_{H^{s-1}}\leq CE_{total}.
\end{eqnarray} 
}
\textbf{Proof:} Firstly notice that the following holds by definition
\begin{eqnarray}
||\delta\rho||^{2}_{H^{s-1}}=||\dot{\rho}||^{2}_{L^{2}},~||v||^{2}_{H^{s-1}}=||\dot{v}||^{2}_{L^{2}}.
\end{eqnarray}
First we show that the total energy is positive definite. Assuming $spec\{\mathcal{L}_{\gamma,\gamma}\}>0$, it follows that the gravitational energy is non-negative definite i.e., 
\begin{eqnarray}
E_{Ein}=
\sum_{i=1}^{s}\left(\frac{1}{2}\langle k^{tr}|\mathcal{L}^{i-1}_{g,\gamma}k^{tr}\rangle_{L^{2}}+\frac{1}{8}\langle u|\mathcal{L}^{i}_{g,\gamma}u\rangle_{L^{2}}\right)\geq 0
\end{eqnarray}
for $u\notin \ker(\mathcal{L}_{\gamma,\gamma})$ and under the small data assumption. The equality holds only for $u\equiv0,~k^{tr}\equiv 0$.
The non-negativity of the fluid energy follows due to the following calculations. Cauchy-Schwartz yields $g(v,\dot{v})^{2}\leq g(v,v)g(\dot{v},\dot{v})$ and therefore 
\begin{eqnarray}
-\frac{g(v,\dot{v})^{2}}{1+g(v,v)}+g_{ij}\dot{v}^{i}\dot{v}^{j}\geq g(\dot{v},\dot{v})-\frac{g(v,v)g(\dot{v},\dot{v})}{1+g(v,v)}=\frac{g(\dot{v},\dot{v})}{1+g(v,v)}
\end{eqnarray}
and 
\begin{eqnarray}
-\frac{g(v,\dot{v})^{2}}{1+g(v,v)}+g_{ij}\dot{v}^{i}\dot{v}^{j}\leq g(\dot{v},\dot{v})+\frac{g(v,v)g(\dot{v},\dot{v})}{1+g(v,v)}\nonumber=\frac{g(\dot{v},\dot{v})(1+2g(v,v))}{1+g(v,v)}.
\end{eqnarray}
Now consider the following 
\begin{eqnarray}
2\frac{g_{ij}v^{i}\dot{v}^{j}}{[1+g(v,v)]^{1/2}}\dot{\rho}\leq \frac{\dot{\rho}^{2}[1+g(v,v)]^{1/2}}{\alpha(n)\gamma_{ad}\rho}+\frac{\alpha(n)\gamma_{ad}\rho g(v,\dot{v})^{2}}{(1+g(v,v))^{3/2}}\\\nonumber 
\leq \frac{\dot{\rho}^{2}[1+g(v,v)]^{1/2}}{\alpha(n)\gamma_{ad}\rho}+\frac{\alpha(n)\gamma_{ad}\rho g(v,v)g(\dot{v},\dot{v})}{(1+g(v,v))^{3/2}}
\end{eqnarray}
which yields 
\begin{eqnarray}
E_{fluid}\geq \\\nonumber 
\int_{\Sigma}\left(\frac{\gamma_{ad}-1}{\gamma_{ad}\rho}(1-\frac{1}{\alpha(n)})[1+g(v,v)]^{1/2}\dot{\rho}^{2}+\frac{\gamma_{ad}\rho g(\dot{v},\dot{v})}{[1+g(v,v)]^{1/2}}\right.\\\nonumber
\left.-\frac{\alpha(n)\gamma_{ad}(\gamma_{ad}-1)\rho g(v,v)g(\dot{v},\dot{v})}{(1+g(v,v))^{3/2}}\right)\mu_{g}\\\nonumber 
=\int_{\Sigma}\left(\frac{\gamma_{ad}-1}{\gamma_{ad}\rho}(1-\frac{1}{\alpha(n)})[1+g(v,v)]^{1/2}\dot{\rho}^{2}\right.\\\nonumber 
\left.+\frac{\gamma_{ad}\rho g(\dot{v},\dot{v})}{(1+g(v,v))^{3/2}}\left(1+g(v,v)-\alpha(n)(\gamma_{ad}-1)g(v,v)\right)\right)\mu_{g}\geq 0
\end{eqnarray}
provided $\alpha(n)(\gamma_{ad}-1)\leq 1$ and $\alpha(n)>1$. The assumption $\gamma_{ad}<\frac{n+1}{n}$ yields a solution $\alpha(n)$ depending only on the dimension $n$.

We now compute the Hessian of $E_{total}$ at $(u,k^{tr},\dot{\rho},\dot{v})=(0,0,0,0)$ for variations $(p,q,r,y)$ and obtain
\begin{eqnarray}
D^{2}E_{total}|_{(0,0,0,0)}\cdot ((p,q,r,y),(p,q,r,y))= \sum_{i=1}^{s}\left(\frac{1}{4}\langle p,\mathcal{L}^{i}_{\gamma,\gamma}p\rangle_{L^{2}}\nonumber+\langle q,\mathcal{L}^{i-1}_{\gamma,\gamma}q\rangle_{L^{2}}\right)\\\nonumber 
+\int_{\Sigma}(\frac{\gamma_{ad}-1}{\gamma_{ad}\rho}r^{2}+\gamma_{ad}C_{\rho}\gamma(y,y))\mu_{\gamma}\geq 0
\end{eqnarray}
and equality holds if and only if $(q,r,y)=(0,0,0)$ and $p\in \ker(\mathcal{L}_{\gamma,\gamma})$. Therefore, $D^{2}E_{total}|_{(0,0,0,0)}: (H^{s}-\ker(\mathcal{L}_{\gamma,\gamma}))\times H^{s-1}\times L^{2}\times L^{2}\to Image(D^{2}E_{total}|_{(0,0,0,0)})$ is an isomorphism yielding 
\begin{eqnarray}
||u^{\perp}||^{2}_{H^{s}}+||k^{tr}||^{2}_{H^{s-1}}+||\dot{\rho}||^{2}_{L^{2}}+||\dot{v}||^{2}_{L^{2}}\\\nonumber 
\leq C D^{2}E_{total}|_{(0,0,0,0)}\cdot ((p,q,r,y),(p,q,r,y))
\end{eqnarray}
which upon utilizing the Hilbert space version of the Morse lemma (or Taylor's theorem) yields 
\begin{eqnarray}
||u^{\perp}||^{2}_{H^{s}}+||v||^{2}_{H^{s-1}}+||\delta\rho||^{2}_{H^{s-1}}+||v||^{2}_{H^{s-1}}\leq CE_{total}.~~~~~~~~~~~~~~~~~~~\Box
\end{eqnarray}

Now using the previous lemma (lemma 1), we obtain the following coercive property of the total energy
\begin{eqnarray}
||u||^{2}_{H^{s}}+||v||^{2}_{H^{s-1}}+||\delta\rho||^{2}_{H^{s-1}}+||v||^{2}_{H^{s-1}}\leq CE_{total}.
\end{eqnarray}
This is one of the most important properties that is necessary for the global existence proof. Now we will obtain a differential equation for the total energy. The time derivative of the total energy reads 
\begin{eqnarray}
\label{eq:energyevolution}
\frac{dE_{total}}{dt}=\frac{dE_{Ein}}{dt}+\frac{dE_{fluid}}{dt}.
\end{eqnarray}
Here the gravitational part $\frac{dE_{Ein}}{dt}$ can be evaluated the same way as in \cite{andersson2011einstein} and only the matter coupling terms need to be kept track of. The delicate part is the computation of the time derivative of the fluid energy. Therefore we focus on evaluating $\frac{dE_{fluid}}{dt}$ first. The following lemma yields a differential equation for the fluid energy.

\textbf{Lemma 3:} \textit{Let the adiabatic index $\gamma_{ad}$ lie in the interval $(1,\frac{n+1}{n})$ and $a(t)$, $t\in (0,\infty)$ be the scale factor.  Assume $(u,k^{tr},\dot{\rho},\dot{v})$ solve the Einstein's and commuted Euler's equations (\ref{eq:metricevol}-\ref{eq:eomf}). Then the fluid energy $E_{fluid}:=-\int_{\Sigma}\mathcal{C}^{\mu}\mathbf{n}_{\mu}\mu_{g}$ satisfies the following differential equation in time 
\begin{eqnarray}
\label{eq:energyidentity}
\frac{dE_{fluid}}{dt}=\frac{n\dot{a}}{a}\int_{\Sigma}N(N-1)\mathcal{C}^{0}\mu_{g}+\int_{\Sigma}\left(\partial_{t}\mathcal{C}^{0}+\frac{1}{a}\nabla_{i}\mathcal{C}^{i}\nonumber+(\frac{1}{a}\nabla_{i}X^{i}+\frac{1}{N}\partial_{t}N)\mathcal{C}^{0}\right.\\\nonumber 
\left.+\frac{1}{aN}\mathcal{C}^{i}\nabla_{i}N\right)N\mu_{g}. 
\end{eqnarray}
}

\textbf{Proof:} The proof is a consequence of an elementary divergence identity for the energy current $\mathcal{C}$. Integrating the divergence identity (on the scale-free geometry)
\begin{eqnarray}
\nabla_{\mu}\mathcal{C}^{\mu}=\frac{1}{N\mu_{g}}\partial_{\mu}(N\mu_{g}\mathcal{C}^{\mu})
\end{eqnarray}
over a Cauchy hypersurface $\Sigma$ yields 
\begin{eqnarray}
\int_{\Sigma}\nabla_{\mu}\mathcal{C}^{\mu}N\mu_{g}=\int_{\Sigma}\partial_{\mu}(N\mu_{g}\mathcal{C}^{\mu})=\int_{\Sigma}\partial_{t}(N\mu_{g}\mathcal{C}^{0})+\frac{1}{a}\int_{\Sigma}\partial_{i}(N\mu_{g}\mathcal{C}^{i})\\\nonumber 
=\partial_{t}\int_{\Sigma_{t}}N\mathcal{C}^{0}\mu_{g}=-\frac{d}{dt}\int_{\Sigma_{t}}\mathcal{C}^{\mu} \mathbf{n}_{\mu}\mu_{g}=\frac{dE_{fluid}}{dt},
\end{eqnarray}
where we have used Stokes's theorem on closed manifold to remove the total derivative term.
Explicit computation of $\nabla_{\mu}\mathcal{C}^{\mu}$ yields the following differential equation for the fluid energy $E_{fluid}$ 
\begin{eqnarray}
\label{eq:energyidentity}
\frac{dE_{fluid}}{dt}=\frac{n\dot{a}}{a}\int_{\Sigma}N(N-1)\mathcal{C}^{0}\mu_{g}+\int_{\Sigma}\left(\partial_{t}\mathcal{C}^{0}+\frac{1}{a}\nabla_{i}\mathcal{C}^{i}\nonumber+(\frac{1}{a}\nabla_{i}X^{i}+\frac{1}{N}\partial_{t}N)\mathcal{C}^{0}\right.\\\nonumber 
\left.+\frac{1}{aN}\mathcal{C}^{i}\nabla_{i}N\right)N\mu_{g}.~~~~~~~~~~~~~~~~~~~~~~~~~~~~~~~~~~~~~~~~~~~~~~~~~~~~~~~~~~~~~~~~~~~\Box 
\end{eqnarray}
Notice that this energy equation contains the top order term $\partial_{t}\mathcal{C}^{0}+\frac{1}{a}\nabla_{i}\mathcal{C}^{i}$ which from the definition of the energy current, will consist of terms that are derivatives of the commuted variables $\dot{\rho}$ and $\dot{v}^{i}$. This is dangerous since the energy does not contain any differentiated $\dot{\rho}$ and $\dot{v}^{i}$ terms. However, by using explicit computations, we will show that such dangerous terms cancel point-wise with their negative counter-parts or yield terms that are of total derivative forms and therefore vanish by integration. This cancellation property exhibits the hyperbolic nature of Euler's equations. 

\textbf{Claim:} \textit{The entity $\partial_{t}\mathcal{C}^{0}+\frac{1}{a}\nabla_{i}\mathcal{C}^{i}$ contains purely algebraic terms of $\dot{\rho}$ and $\dot{v}$ modulo terms involving Lie derivatives with respect to the shift vector field $X$.}

\textbf{Proof:} We compute the term $\partial_{t}\mathcal{C}^{0}+\frac{1}{a}\nabla_{i}\mathcal{C}^{i}$ explicitly. First we compute $\partial_{t}\mathcal{C}^{0}$ and $\nabla_{i}\mathcal{C}^{i}$ separately. We write the following 
\begin{eqnarray}
\partial_{t}\mathcal{C}^{0}=[\partial_{t}\mathcal{C}^{0}]_{principal}+[\partial_{t}\mathcal{C}^{0}]_{remainder},\\
\nabla_{i}\mathcal{C}^{i}=[\nabla_{i}\mathcal{C}^{i}]_{principal}+[\nabla_{i}\mathcal{C}^{i}]_{remainder},
\end{eqnarray}
where the subscript $principal$ denotes the terms that involve highest derivatives of $\dot{\rho}$ and $\dot{v}$ and $remainder$ denotes the terms that are purely algebraic in $(\dot{\rho},\dot{v})$. Explicit calculation yields 

\begin{eqnarray}
[\partial_{t}\mathcal{C}^{0}]_{principal}=\frac{2(\gamma_{ad}-1)}{\gamma_{ad}\rho N}[1+g(v,v)]^{1/2}\dot{\rho}\partial_{t}\dot{\rho}\nonumber+2(\gamma_{ad}-1)\left(\frac{\dot{\rho}g_{ij}v^{i}}{N[1+g(v,v)]^{1/2}}\partial_{t}\dot{v}^{j}\right.\\\nonumber
\left.+\frac{g_{ij}v^{i}\dot{v}^{j}}{N[1+g(v,v)]^{1/2}}\partial_{t}\dot{\rho}\right)
+\frac{\gamma_{ad}\rho[1+g(v,v)]^{1/2}}{N}\left[(2g_{ij}(\partial_{t}\dot{v}^{i})\dot{v}^{j}-\frac{2g(v,\dot{v})g_{ij}v^{i}\partial_{t}\dot{v}^{j}}{1+g(v,v)}\right],
\end{eqnarray}

\begin{eqnarray}
[\nabla_{i}\mathcal{C}^{i}]_{principal}=\frac{2(\gamma_{ad}-1)}{\gamma_{ad}\rho}\dot{\rho}v^{i}\nabla_{i}\dot{\rho}+2(\gamma_{ad}-1)\left[\dot{\rho}\nabla_{i}\dot{v}^{i}+\dot{v}^{i}\nabla_{i}\dot{\rho}\right]\\\nonumber 
+\gamma_{ad}\rho v^{i}\left[2\dot{v}_{k}\nabla_{i}\dot{v}^{k}-\frac{2g(v,\dot{v})g(v,\nabla_{i}\dot{v})}{1+g(v,v)}\right],
\end{eqnarray}

\begin{eqnarray}
[\partial_{t}\mathcal{C}^{0}]_{remainder}\\\nonumber
=\frac{\gamma_{ad}-1}{\gamma_{ad}}\left(\frac{[1+g(v,v)]^{-1/2}}{2\rho N}(\partial_{t}g(v,v)+2g(\partial_{t}v,v))-\frac{[1+g(v,v)]^{1/2}}{\rho^{2}N^{2}}[\rho\partial_{t}N+N\partial_{t}\rho]\right)\dot{\rho}^{2}\\\nonumber 
+2(\gamma_{ad}-1)\dot{\rho}\dot{v}^{j}\left(-\frac{\partial_{t}Ng_{ij}v^{i}}{N^{2}[1+g(v,v)]^{1/2}}-\frac{g_{ij}v^{i}}{N[1+g(v,v)]^{3/2}}[\partial_{t}g(v,v)+2g(\partial_{t}v,v)]\right.\\\nonumber 
\left.+\frac{1}{N[1+g(v,v)]^{1/2}}(\partial_{t}g_{ij}v^{i}+g_{ij}\partial_{t}v^{i})\right)+\gamma_{ad}\left[g(\dot{v},\dot{v})-\frac{g(v,\dot{v})^{2}}{1+g(v,v)}\right]\left(\frac{[1+g(v,v)]^{1/2}}{N}\partial_{t}\rho\right.\\\nonumber 
\left.+\frac{\rho[1+g(v,v)]^{-1/2}}{2N}(\partial_{t}g(v,v)+2g(\partial_{t}v,v))-\frac{\rho[1+g(v,v)]^{1/2}}{N^{2}}\partial_{t}N\right)\\\nonumber 
+\frac{\gamma_{ad}\rho[1+g(v,v)]^{1/2}}{N}\left[(\partial_{t}g_{ij})\dot{v}^{i}\dot{v}^{j}+\frac{g(v,\dot{v})^{2}}{[1+g(v,v)]^{2}}(\partial_{t}g(v,v)+2g(\partial_{t}v,v))\right.\\\nonumber \left. -\frac{g(v,\dot{v})}{1+g(v,v)}(\partial_{t}g_{ij}v^{i}\dot{v}^{j}+g_{ij}\dot{v}^{i}\partial_{t}v^{j})\right],
\end{eqnarray}
\begin{eqnarray}
[\nabla_{i}\mathcal{C}^{i}]_{remainder}\\\nonumber 
=-\frac{\gamma_{ad}-1}{\gamma_{ad}}\frac{\dot{\rho}^{2}}{\rho^{2}}v^{i}\nabla_{i}\rho+\frac{\gamma_{ad}-1}{\gamma_{ad}}\frac{\dot{\rho}^{2}}{\rho}\nabla_{i}v^{i}\nonumber+\gamma_{ad}[v^{i}\nabla_{i}\rho+\rho\nabla_{i}v^{i}]\\\nonumber 
\left[-\frac{g(v,\dot{v})^{2}}{1+g(v,v)}+g(\dot{v},\dot{v})\right]+\gamma_{ad}\rho v^{i}\left(\frac{2g(v,\nabla_{i}v)g(v,\dot{v})^{2}}{[1+g(v,v)]^{2}}-\frac{2g(v,\dot{v})g(\dot{v},\nabla_{i}v)}{1+g(v,v)}\right).
\end{eqnarray}

Now notice that there are several terms involving derivatives of the commuted fluid variables $(\dot{\rho},\dot{v})$ in the expressions of $[\partial_{t}\mathcal{C}^{0}]_{principal}$ and $[\nabla_{i}\mathcal{C}^{i}]_{principal}$ that appear in a complicated fashion. Therefore we need to carefully combine these terms to obtain cancellation of the dangerous terms. Using the commuted equation (\ref{eq:commuted1}), we expand the first term $\frac{2(\gamma_{ad}-1)}{\gamma_{ad}\rho N}[1+g(v,v)]^{1/2}\dot{\rho}\partial_{t}\dot{\rho}$ of $[\partial_{t}\mathcal{C}^{0}]_{principal}$ as follows 
\begin{eqnarray}
\frac{2(\gamma_{ad}-1)}{\gamma_{ad}\rho N}[1+g(v,v)]^{1/2}\dot{\rho}\partial_{t}\dot{\rho}\\\nonumber 
=\frac{2(\gamma_{ad}-1)}{\gamma_{ad}\rho N}[1+g(v,v)]^{1/2}\dot{\rho}\left(-\frac{\gamma_{ad}\rho N\nabla_{i}\dot{v}^{i}}{a[1+g(v,v)]^{1/2}}+\frac{1}{a}L_{X}\dot{\rho}-\frac{\gamma_{ad}\rho g(\partial_{t}\dot{v},v)}{1+g(v,v)}\right.\\\nonumber
\left.-\frac{NL_{v}\dot{\rho}}{a[1+g(v,v)]^{1/2}}+\frac{\gamma_{ad}\rho v_{i}L_{X}\dot{v}^{i}}{a[1+g(v,v)]}+\mathcal{T}_{1}\right).
\end{eqnarray}
Now collection of the principal terms together with the previous expression yields 
\begin{eqnarray}
\left(\partial_{t}\mathcal{C}^{0}+\frac{1}{a}\nabla_{i}\mathcal{C}^{i}\right)_{principal}\\
=\frac{2(\gamma_{ad}-1)}{\gamma_{ad}\rho N}[1+g(v,v)]^{1/2}\dot{\rho}\left(-\frac{\gamma_{ad}\rho N\nabla_{i}\dot{v}^{i}}{a[1+g(v,v)]^{1/2}}\nonumber+\frac{1}{a}L_{X}\dot{\rho}-\underbrace{\frac{\gamma_{ad}\rho g(\partial_{t}\dot{v},v)}{1+g(v,v)}}_{B}\right.\\\nonumber
\left.-\frac{NL_{v}\dot{\rho}}{a[1+g(v,v)]^{1/2}}+\frac{\gamma_{ad}\rho v_{i}L_{X}\dot{v}^{i}}{a[1+g(v,v)]}+\mathcal{T}_{1}\right)+\underbrace{\frac{2(\gamma_{ad}-1)\dot{\rho}g(v,\partial_{t}\dot{v})}{N[1+g(v,v)]^{1/2}}}_{-B}\\\nonumber 
+\underbrace{\frac{2(\gamma_{ad}-1)g(v,\dot{v})\partial_{t}\dot{\rho}}{N[1+g(v,v)]^{1/2}}+\frac{2\gamma_{ad}\rho[1+g(v,v)]^{1/2}}{N}g(\dot{v},\partial_{t}\dot{v})-\frac{2\gamma_{ad}\rho g(v,\dot{v})g(v,\partial_{t}\dot{v})}{N[1+g(v,v)]^{1/2}}}_{1A:~Combine~and~use~equations~of~variations}\\\nonumber 
+\frac{1}{a}\left(\frac{2(\gamma_{ad}-1)}{\gamma_{ad}\rho}\dot{\rho}v^{i}\nabla_{i}\dot{\rho}+2(\gamma_{ad}-1)[\dot{\rho}\nabla_{i}\dot{v}^{i}+\dot{v}^{i}\nabla_{i}\dot{\rho}]+\gamma_{ad}\rho v^{i}[2\dot{v}_{k}\nabla_{i}\dot{v}^{k}\right.\\\nonumber\left.-\frac{2g(v,\dot{v})g(v,\nabla_{i}\dot{v})}{1+g(v,v)}]\right).
\end{eqnarray}
We have numbered the terms here so that we can keep track of each dangerous term and how they cancel with their negative counterpart. First notice that the terms denoted by $B$ and $-B$ cancel each other (we denote them accordingly). Now we consider the combination of three terms denoted by $1A$
\begin{eqnarray}
1A=\frac{2(\gamma_{ad}-1)g(v,\dot{v})\partial_{t}\dot{\rho}}{N[1+g(v,v)]^{1/2}}+\frac{2\gamma_{ad}\rho[1+g(v,v)]^{1/2}}{N}g(\dot{v},\partial_{t}\dot{v})\nonumber-\frac{2\gamma_{ad}\rho g(v,\dot{v})g(v,\partial_{t}\dot{v})}{N[1+g(v,v)]^{1/2}}\\\nonumber 
=\frac{2g(v,\dot{v})}{N[1+g(v,v)]^{1/2}}\left((\gamma_{ad}-1)\partial_{t}\dot{\rho}-\gamma_{ad}\rho g(v,\partial_{t}\dot{v})\right)+\frac{2\gamma_{ad}\rho[1+g(v,v)]^{1/2}}{N}g(\dot{v},\partial_{t}\dot{v})\\\nonumber 
=\frac{2g(v,\dot{v})}{N[1+g(v,v)]^{1/2}}\left((\gamma_{ad}-1)\partial_{t}\dot{\rho}-g(v,\frac{\gamma_{ad}\rho}{a}L_{X}\dot{v}-\frac{\gamma_{ad}\rho}{au^{0}}v^{k}\nabla_{k}\dot{v}-\frac{\gamma_{ad}-1}{au^{0}}\nabla\dot{\rho}\right.\\\nonumber 
\left.-\frac{\gamma_{ad}-1}{au^{0}}vL_{v}\dot{\rho}-(\gamma_{ad}-1)v\partial_{t}\dot{\rho}+\frac{\gamma_{ad}-1}{a}vL_{X}\dot{\rho}+\frac{\mathcal{T}_{2}}{u^{0}})\right)\\\nonumber 
+\frac{2[1+g(v,v)]^{1/2}}{N}g\left(\dot{v},\frac{\gamma_{ad}\rho}{a}L_{X}\dot{v}-\frac{\gamma_{ad}\rho}{au^{0}}v^{k}\nabla_{k}\dot{v}-\frac{\gamma_{ad}-1}{au^{0}}\nabla\dot{\rho}-\frac{\gamma_{ad}-1}{au^{0}}vL_{v}\dot{\rho}\right.\\\nonumber 
\left.-(\gamma_{ad}-1)v\partial_{t}\dot{\rho}+\frac{\gamma_{ad}-1}{a}vL_{X}\dot{\rho}+\frac{\mathcal{T}_{2}}{u^{0}}\right)\\\nonumber 
=\underbrace{\frac{2g(v,\dot{v})}{N[1+g(v,v)]^{1/2}}\left((\gamma_{ad}-1)\partial_{t}\dot{\rho}(1+g(v,v))\right)-\frac{2[1+g(v,v)]^{1/2}}{N}(\gamma_{ad}-1)g(v,\dot{v})\partial_{t}\dot{\rho}}_{point-wise~cancellation}\\\nonumber 
+\frac{2g(v,\dot{v})}{N[1+g(v,v)]^{1/2}}\left(-\frac{\gamma_{ad}\rho}{a}g(v,L_{X}\dot{v})+\frac{\gamma_{ad}\rho}{au^{0}}g(v,v^{k}\nabla_{k}\dot{v})+\frac{\gamma_{ad}-1}{au^{0}}L_{v}\dot{\rho}\right.\\\nonumber
\left.+\frac{\gamma_{ad}-1}{au^{0}}g(v,v)L_{v}\dot{\rho}-\frac{\gamma_{ad}-1}{a}g(v,v)L_{X}\dot{\rho}-g(v,\frac{\mathcal{T}_{2}}{u^{0}})\right)\\\nonumber +\frac{2[1+g(v,v)]^{1/2}}{N}\left(\frac{\gamma_{ad}\rho}{a}g(\dot{v},L_{X}\dot{v})-\frac{\gamma_{ad}\rho}{au^{0}}g(\dot{v},v^{k}\nabla_{k}\dot{v})-\frac{\gamma_{ad}-1}{au^{0}}\dot{v}^{k}\nabla_{k}\dot{\rho}\right.\\\nonumber 
\left.-\frac{\gamma_{ad}-1}{au^{0}}g(v,\dot{v})L_{v}\dot{\rho}+\frac{\gamma_{ad}-1}{a}g(v,\dot{v})L_{X}\dot{\rho}+g(\dot{v},\frac{\mathcal{T}_{2}}{u^{0}})\right)+\frac{2(\gamma_{ad}-1)[1+g(v,v)]^{1/2}}{\gamma_{ad}\rho N}\dot{\rho}\mathcal{T}_{1},\\\nonumber 
=\frac{2g(v,\dot{v})}{N[1+g(v,v)]^{1/2}}\left(-\frac{\gamma_{ad}\rho}{a}g(v,L_{X}\dot{v})+\frac{\gamma_{ad}\rho}{au^{0}}g(v,v^{k}\nabla_{k}\dot{v})+\frac{\gamma_{ad}-1}{au^{0}}L_{v}\dot{\rho}\right.\\\nonumber
\left.+\frac{\gamma_{ad}-1}{au^{0}}g(v,v)L_{v}\dot{\rho}-\frac{\gamma_{ad}-1}{a}g(v,v)L_{X}\dot{\rho}-g(v,\frac{\mathcal{T}_{2}}{u^{0}})\right)\\\nonumber +\frac{2[1+g(v,v)]^{1/2}}{N}\left(\frac{\gamma_{ad}\rho}{a}g(\dot{v},L_{X}\dot{v})-\frac{\gamma_{ad}\rho}{au^{0}}g(\dot{v},v^{k}\nabla_{k}\dot{v})-\frac{\gamma_{ad}-1}{au^{0}}\dot{v}^{k}\nabla_{k}\dot{\rho}\right.\\\nonumber 
\left.-\frac{\gamma_{ad}-1}{au^{0}}g(v,\dot{v})L_{v}\dot{\rho}+\frac{\gamma_{ad}-1}{a}g(v,\dot{v})L_{X}\dot{\rho}+g(\dot{v},\frac{\mathcal{T}_{2}}{u^{0}})\right)+\frac{2(\gamma_{ad}-1)[1+g(v,v)]^{1/2}}{\gamma_{ad}\rho N}\dot{\rho}\mathcal{T}_{1}.
\end{eqnarray}

Substitution of $1A$ in the expression of $(\partial_{t}\mathcal{C}^{0}+\frac{1}{a}\nabla_{i}\mathcal{C}^{i})_{principal}$ (after carrying out the possible point-wise cancellations) yields 
\begin{eqnarray}
(\partial_{t}\mathcal{C}^{0}+\frac{1}{a}\nabla_{i}\mathcal{C}^{i})_{principal}=\underbrace{-\frac{2(\gamma_{ad}-1)}{a}\dot{\rho}\nabla_{i}\dot{v}^{i}}_{I}\nonumber+\frac{2(\gamma_{ad}-1)(1+g(v,v))^{1/2}}{a\gamma_{ad}\rho N}\dot{\rho}L_{X}\dot{\rho}\\\nonumber 
-\underbrace{\frac{2(\gamma_{ad}-1)}{a\gamma_{ad}\rho}\dot{\rho}L_{v}\dot{\rho}}_{II}+\underbrace{\frac{2(\gamma_{ad}-1)}{aN[1+g(v,v)]^{1/2}}\dot{\rho}g(v,L_{X}\dot{v})}_{III}\\\nonumber 
+\frac{1}{a}\left(\underbrace{\frac{2(\gamma_{ad}-1)}{\gamma_{ad}\rho}\dot{\rho}v^{i}\nabla_{i}\dot{\rho}}_{-II}+2(\gamma_{ad}-1)[\underbrace{\dot{\rho}\nabla_{i}\dot{v}^{i}}_{-I}+\underbrace{\dot{v}^{i}\nabla_{i}\dot{\rho}}_{IV}]+\gamma_{ad}\rho v^{i}[\underbrace{2\dot{v}_{k}\nabla_{i}\dot{v}^{k}}_{V}\right.\\\nonumber \left.-\underbrace{\frac{2g(v,\dot{v})g(v,\nabla_{i}\dot{v})}{1+g(v,v)}]}_{VI}\right)
+\frac{2g(v,\dot{v})}{N[1+g(v,v)]^{1/2}}\left(-\frac{\gamma_{ad}\rho}{a}g(v,L_{X}\dot{v})\right.\\\nonumber\left.
+\underbrace{\frac{\gamma_{ad}\rho}{au^{0}}g(v,v^{k}\nabla_{k}\dot{v})}_{-VI}+\underbrace{\frac{\gamma_{ad}-1}{au^{0}}L_{v}\dot{\rho}+\frac{\gamma_{ad}-1}{au^{0}}g(v,v)L_{v}\dot{\rho}}_{VII}\right.\\\nonumber 
\left.-\underbrace{\frac{\gamma_{ad}-1}{a}g(v,v)L_{X}\dot{\rho}}_{III}-g(v,\frac{\mathcal{T}_{2}}{u^{0}})\right)\\\nonumber +\frac{2[1+g(v,v)]^{1/2}}{N}\left(\frac{\gamma_{ad}\rho}{a}g(\dot{v},L_{X}\dot{v})-\underbrace{\frac{\gamma_{ad}\rho}{au^{0}}g(\dot{v},v^{k}\nabla_{k}\dot{v})}_{-V}-\underbrace{\frac{\gamma_{ad}-1}{au^{0}}\dot{v}^{k}\nabla_{k}\dot{\rho}}_{-IV}\right.\\\nonumber 
\left. -\underbrace{\frac{\gamma_{ad}-1}{au^{0}}g(v,\dot{v})L_{v}\dot{\rho}}_{-VII}+\underbrace{\frac{\gamma_{ad}-1}{a}g(v,\dot{v})L_{X}\dot{\rho}}_{III}+g(\dot{v},\frac{\mathcal{T}_{2}}{u^{0}})\right)\\\nonumber
+\frac{2(\gamma_{ad}-1)[1+g(v,v)]^{1/2}}{\gamma_{ad}\rho N}\dot{\rho}\mathcal{T}_{1}.
\end{eqnarray}
Here the terms that are denoted by the same number belong to the same category, that is, the combination of these terms yields lower-order innocuous terms. The remaining terms are denoted by numbers and their negative counterparts are denoted by the negative of their numbers. Note that there are three terms that fall under the category $III$ that requires the following few lines of additional calculations to generate lower-order innocuous terms
\begin{eqnarray}
\dot{\rho}g(v,L_{X}\dot{v})=\dot{\rho}g(v,X^{k}\nabla_{k}\dot{v}-\dot{v}^{k}\nabla_{k}X)\\\nonumber 
=\dot{\rho}X^{k}\nabla_{k}g(v,\dot{v})-\dot{\rho}g(X^{k}\nabla_{k}v,\dot{v})-\dot{\rho}g(v,\dot{v}^{k}\nabla_{k}X)\\\nonumber 
=X^{k}\nabla_{k}(\dot{\rho}g(v,\dot{v}))-g(v,\dot{v})L_{X}\dot{\rho}-\dot{\rho}g(X^{k}\nabla_{k}v,\dot{v})-\dot{\rho}g(v,\dot{v}^{k}\nabla_{k}X)\\\nonumber 
=-g(v,\dot{v})L_{X}\dot{\rho}-L_{X}(\dot{\rho}g(v,\dot{v}))-\dot{\rho}g(X^{k}\nabla_{k}v,\dot{v})-\dot{\rho}g(v,\dot{v}^{k}\nabla_{k}X).
\end{eqnarray}
After all the possible cancellations, we collect the remaining terms to obtain the following final expression for $\partial_{t}\mathcal{C}^{0}+\frac{1}{a}\nabla_{i}\mathcal{C}^{i}$
\begin{eqnarray}
\label{eq:reducedexpression}
\int_{\Sigma}(\partial_{t}\mathcal{C}^{0}+\frac{1}{a}\nabla_{i}\mathcal{C}^{i})N\mu_{g}\\\nonumber
=\int_{\Sigma}\left(\underbrace{\frac{2(\gamma_{ad}-1)(1+g(v,v))^{1/2}}{a\gamma_{ad}\rho N}\dot{\rho}L_{X}\dot{\rho}}_{I}-\underbrace{\frac{2\gamma_{ad}\rho g(v,\dot{v})}{aN[1+g(v,v)]^{1/2}}g(v,L_{X}\dot{v})}_{II}\right.\\\nonumber 
\left.-\frac{2g(v,\dot{v})}{N[1+g(v,v)]^{1/2}}g(v,\frac{\mathcal{T}_{2}}{u^{0}})+\underbrace{\frac{2\gamma_{ad}\rho[1+g(v,v)]^{1/2}}{aN}g(\dot{v},L_{X}\dot{v})}_{III}+\frac{2[1+g(v,v)]^{1/2}}{N}g(\dot{v},\frac{\mathcal{T}_{2}}{u^{0}})\right.\\\nonumber 
\left.-\frac{2(\gamma_{ad}-1)}{aN[1+g(v,v)]^{1/2}}\left(\underbrace{L_{X}(\dot{\rho}g(v,\dot{v}))}_{IV}+\dot{\rho}g(X^{k}\nabla_{k}v,\dot{v})+\dot{\rho}g(v,\dot{v}^{k}\nabla_{k}X)\right)\right.\\\nonumber 
\left.+\frac{2(\gamma_{ad}-1)[1+g(v,v)]^{1/2}}{\gamma_{ad}\rho N}\dot{\rho}\mathcal{T}_{1}\right.\\\nonumber 
\left.+\frac{\gamma_{ad}-1}{\gamma_{ad}}\left(\frac{[1+g(v,v)]^{-1/2}}{2\rho N}(\partial_{t}g(v,v)+2g(\partial_{t}v,v))-\frac{[1+g(v,v)]^{1/2}}{\rho^{2}N^{2}}[\rho\partial_{t}N+N\partial_{t}\rho]\right)\dot{\rho}^{2}\right.\\\nonumber 
\left.+2(\gamma_{ad}-1)\dot{\rho}\dot{v}^{j}\left(-\frac{\partial_{t}Ng_{ij}v^{i}}{N^{2}[1+g(v,v)]^{1/2}}-\frac{g_{ij}v^{i}}{N[1+g(v,v)]^{3/2}}[\partial_{t}g(v,v)+2g(\partial_{t}v,v)]\right.\right.\\\nonumber 
\left.\left.+\frac{1}{N[1+g(v,v)]^{1/2}}(\partial_{t}g_{ij}v^{i}+g_{ij}\partial_{t}v^{i})\right)+\gamma_{ad}\left[g(\dot{v},\dot{v})-\frac{g(v,\dot{v})^{2}}{1+g(v,v)}\right]\left(\frac{[1+g(v,v)]^{1/2}}{N}\partial_{t}\rho\right.\right.\\\nonumber 
\left.\left.+\frac{\rho[1+g(v,v)]^{-1/2}}{2N}(\partial_{t}g(v,v)+2g(\partial_{t}v,v))-\frac{\rho[1+g(v,v)]^{1/2}}{N^{2}}\partial_{t}N\right)\right.\\\nonumber 
\left.+\frac{\gamma_{ad}\rho[1+g(v,v)]^{1/2}}{N}\left[(\partial_{t}g_{ij})\dot{v}^{i}\dot{v}^{j}+\frac{g(v,\dot{v})^{2}}{[1+g(v,v)]^{2}}(\partial_{t}g(v,v)+2g(\partial_{t}v,v))\right.\right.\\\nonumber \left.\left. -\frac{g(v,\dot{v})}{1+g(v,v)}(\partial_{t}g_{ij}v^{i}\dot{v}^{j}+g_{ij}\dot{v}^{i}\partial_{t}v^{j})\right]+\frac{1}{a}\left(-\frac{\gamma_{ad}-1}{\gamma_{ad}}\frac{\dot{\rho}^{2}}{\rho^{2}}v^{i}\nabla_{i}\rho+\frac{\gamma_{ad}-1}{\gamma_{ad}}\frac{\dot{\rho}^{2}}{\rho}\nabla_{i}v^{i}\nonumber\right.\right.\\\nonumber
\left.\left.+\gamma_{ad}[v^{i}\nabla_{i}\rho+\rho\nabla_{i}v^{i}] 
\left[-\frac{g(v,\dot{v})^{2}}{1+g(v,v)}+g(\dot{v},\dot{v})\right]+\gamma_{ad}\rho v^{i}\left(\frac{2g(v,\nabla_{i}v)g(v,\dot{v})^{2}}{[1+g(v,v)]^{2}}\right.\right.\right.\\\nonumber\left.\left.\left.-\frac{2g(v,\dot{v})g(\dot{v},\nabla_{i}v)}{1+g(v,v)}\right)\right)\right) N\mu_{g}.
\end{eqnarray}
The terms that are numbered $I,II,III$, and $IV$ contain derivatives of $\dot{\rho}$ and $\dot{v}$. However, these are innocuous since these derivatives involve Lie derivatives with respect to the shift vector field $X$. We will reduce these terms to lower order terms through an integration by parts operation which will be discussed in the next section. The remaining terms are purely algebraic in nature. This completes the proof of the claim. Recall from equation (\ref{eq:energyevolution}) and lemma 3 that the time derivative of the total energy reads \begin{eqnarray}
\frac{dE_{total}}{dt}=\frac{dE_{Ein}}{dt}+\int_{\Sigma}\left(\partial_{t}\mathcal{C}^{0}+\frac{1}{a}\nabla_{i}\mathcal{C}^{i}\nonumber+(\frac{1}{a}\nabla_{i}X^{i}+\frac{1}{N}\partial_{t}N)\mathcal{C}^{0}\right.\\\nonumber 
\left.+\frac{1}{aN}\mathcal{C}^{i}\nabla_{i}N\right)N\mu_{g}+\frac{n\dot{a}}{a}\int_{\Sigma}N(N-1)\mathcal{C}^{0}\mu_{g}.
\end{eqnarray}
Using the reduced expression for $\partial_{t}\mathcal{C}^{0}+\frac{1}{a}\nabla_{i}\mathcal{C}^{i}$ (\ref{eq:reducedexpression}), we will estimate $\frac{dE_{total}}{dt}$.

\section{Energy estimates}
In this section, we will obtain a differential inequality for the total energy. For this purpose, we will need the following set of inequalities.
\subsection{Important inequalities}
\textbf{1.} \cite{globalcauchy} \textit{Let $f_{1}\in H^{t_{1}},~f_{2}\in H^{t_{2}}$ and $k\leq \min(t_{1},t_{2},t_{1}+t_{2}-\frac{n}{2}),~t_{i}\geq 0$ and some $t_{i}>0$, then the following estimate holds
\begin{eqnarray}
\label{eq:product1}
||f_{1}f_{2}||_{H^{k}}\leq C ||f_{1}||_{H^{t_{1}}}||f_{2}||_{H^{t_{2}}}
\end{eqnarray}}
\\ 
\textbf{2.} \cite{Taylor} \textit{For $\mathcal{P}\in \mathcal{OP}^{s},~s\in \mathbb{R}$, the following holds
\begin{eqnarray}
\label{eq:commutator}
||\mathcal{P}f||_{H^{r}}\leq C ||f||_{H^{s+r}},~~r\in \mathbb{R}.
\end{eqnarray}
Here $\mathcal{OP}^{s}$ denotes the space of  pseudo-differential operators with symbol in Hormander's class $S^{s}_{1,0}$ (see \cite{Taylor}). For example $\nabla[\gamma]\in \mathcal{OP}^{1}$.
}\\
\textbf{3.}\cite{Taylor} \textit{Assume $\mathcal{P}\in \mathcal{OP}^{s},~s>0$, then
\begin{eqnarray}
||[P,f_{1}]f_{2}||_{H^{\sigma}} \leq C \left(||\nabla[\gamma] f_{1}||_{L^{\infty}}||f_{2}||_{H^{s-1+\sigma}}+||f_{1}||_{H^{s+\sigma}}||f_{2}||_{L^{\infty}}\right)
\end{eqnarray}
}\\
\textbf{4.}\cite{Taylor} \textit{Let $1\leq s<\mu$ and $F\in C^{\mu}(\mathbb{R})$ with $F(0)=0$, where $C^{\mu}(\mathbb{R})$ denotes the H\"older space. Then for $f\in H^{s}\cap L^{\infty}$, the following inequality holds 
\begin{eqnarray}
\label{eq:composition}
||F(f)||_{H^{s}}\leq C||f||_{H^{s}}(1+||f||^{\mu-1}_{L^{\infty}})
\end{eqnarray}
}\\
\textbf{5.}\cite{kato1988commutator} \textit{If $s>0$, then $H^{s}\cap L^{\infty}$ is an algebra i.e., 
\begin{eqnarray}
\label{eq:product}
||f_{1}f_{2}||_{H^{s}}\leq C( ||f_{1}||_{L^{\infty}}||f_{2}||_{H^{s}}+||f_{1}||_{H^{s}}||f_{2}||_{L^{\infty}})
\end{eqnarray}
and for $s>\frac{n}{2}+1$ 
\begin{eqnarray}
||f_{1}f_{2}||_{H^{s}}\leq C ||f_{1}||_{H^{s}}||f_{2}||_{H^{s}}
\end{eqnarray}
due to Sobolev embedding.
}

In addition to these inequalities, we will also use H\"older's and Minkowski's inequality whenever necessary. First we obtain the necessary elliptic estimates 
\subsection{Elliptic estimates}
In this section, we will obtain estimates for the perturbation to the lapse function $\omega=N-1$, the shift vector field $X$, and the shifted shift vector field $Y^{||}+\frac{1}{a}X$. These estimates will be obtained by utilizing the elliptic equations that are obtained by means of gauge fixing. Before moving to the analysis of the elliptic equations, we need to obtain an estimate for $\partial_{t}\gamma$ describing the motion parallel to the center manifold $\mathcal{N}$. The following lemma provides an estimate for $\partial_{t}\gamma$

\textbf{Lemma 4:} \textit{Let $\gamma\in \mathcal{N}$ be the shadow of g such that $||g-\gamma||_{H^{s}}<\delta$ for a suitable $\delta>0$, $s>\frac{n}{2}+2$. The $\mathcal{N}-$tangential velocity $\partial_{t}\gamma$ verifies the following estimate
\begin{eqnarray}
||\partial_{t}\gamma||_{H^{s-1}}\\\nonumber 
\leq C\left(\frac{\dot{a}}{a}||\omega||_{H^{s+1}}(1+||u||_{H^{s}})+\frac{1}{a}(1+||\omega||_{H^{s+1}})||k^{tr}||_{H^{s-1}}\right.\\\nonumber 
\left.+\frac{1}{a}||X||_{H^{s+1}}||u||_{H^{s}}+\frac{1}{a}||X||_{H^{s+1}}\right),
\end{eqnarray}
for a constant $C$ that depends on a fixed background geometry.
}\\
\textbf{Proof:} In the light of the shadow gauge condition (\ref{eq:shadow}), the $\mathcal{N}-$tangential velocity $\partial_{t}\gamma$ satisfies 
\begin{eqnarray}
||\partial_{t}\gamma||\leq C||\partial_{t}g||_{H^{s-1}},
\end{eqnarray}
where the norm in the left hand side refers to a Sobolev norm of arbitrary order (since $\partial_{t}\gamma$ is an element of a finite dimensional vector space). Utilizing the evolution equation (\ref{eq:metricevol}) and Minkowski's inequality together with the fact that $H^{s_{1}}\hookrightarrow H^{s_{2}}$ for $s_{1}>s_{2}$, we obtain 
\begin{eqnarray}
||\partial_{t}\gamma||\\\nonumber 
\leq C\left(\frac{\dot{a}}{a}||\omega||_{H^{s+1}}(1+||u||_{H^{s}})+\frac{1}{a}(1+||\omega||_{H^{s+1}})||k^{tr}||_{H^{s-1}}\right.\\\nonumber 
\left.+\frac{1}{a}||X||_{H^{s+1}}||u||_{H^{s}}+\frac{1}{a}||X||_{H^{s+1}}\right),
\end{eqnarray}
and in particular 
\begin{eqnarray}
||\partial_{t}\gamma||_{H^{s-1}}\\\nonumber 
\leq C\left(\frac{\dot{a}}{a}||\omega||_{H^{s+1}}(1+||u||_{H^{s}})+\frac{1}{a}(1+||\omega||_{H^{s+1}})||k^{tr}||_{H^{s-1}}\right.\\\nonumber 
\left.+\frac{1}{a}||X||_{H^{s+1}}||u||_{H^{s}}+\frac{1}{a}||X||_{H^{s+1}}\right)
\end{eqnarray}
with a possible modification of the constant $C$.$~~~~~~~~~~~~~~~~~~~~~~~~~~~~~~~~~~~~~~~~~~~~~~~~\Box$

Now we will obtain the estimates for $\omega$, $\partial_{t}N$ and $X$ utilizing the elliptic equations obtained through gauge fixing. The following lemma provides an estimate for the perturbation to the elapse function.

\textbf{Lemma 5:} \textit{Let $B_{\delta}(\mathbf{0})\subset H^{s}\times H^{s-1}\times H^{s-1}\times H^{s-1}$ be a ball for sufficiently small $\delta>0$ and $(u,k^{tr},\delta\rho,v)\in B_{\delta}(\mathbf{0})$ solve the re-scaled Einstein-Euler-$\Lambda$ system in CMCSH gauge. Then the following estimate holds for the perturbation to the lapse function ($\omega=N-1$) 
\begin{eqnarray}
\label{eq:lapsepert}
||\omega||_{H^{s+1}}\leq\\\nonumber 
C(||k^{tr}||^{2}_{H^{s-1}}+a^{2-n\gamma_{ad}}||\delta\rho||_{H^{s-1}}+a^{2-n\gamma_{ad}}||v||^{2}_{H^{s-1}}(1+||\delta\rho||_{H^{s-1}})).
\end{eqnarray}
}
\textbf{Proof:} The elliptic equation for the perturbation $\omega=N-1$ to the lapse function $N$ reads 
\begin{eqnarray}
\Delta_{g}\omega+\underbrace{\left(|k^{tr}|^{2}+\frac{2C_{\rho}a^{2-n\gamma_{ad}}}{n-1}+\frac{1}{n-1}+a^{2-n\gamma_{ad}}[\frac{(n\gamma_{ad}-2)}{n-1}+\gamma_{ad}\nonumber g(v,v)](C_{\rho}+\delta\rho)\right)}_{>0}\omega\\\nonumber 
=-|k^{tr}|^{2}_{g}-\frac{n\gamma_{ad}-2}{n-1}a^{2-n\gamma_{ad}}\delta\rho-\gamma_{ad}C_{\rho}a^{2-n\gamma_{ad}}g(v,v)-\gamma_{ad}a^{2-n\gamma_{ad}}g(v,v)\delta\rho.
\end{eqnarray}
Now the elliptic operator 
$\Delta_{g}+\left(|k^{tr}|^{2}+\frac{2C_{\rho}a^{2-n\gamma_{ad}}}{n-1}+\frac{1}{n-1}+a^{2-n\gamma_{ad}}[\frac{(n\gamma_{ad}-2)}{n-1}+\gamma_{ad} g(v,v)](C_{\rho}+\delta\rho)\right)id: H^{s+1}\to H^{s-1}$ is an isomorphism yielding the following estimate
\begin{eqnarray}
||\omega||_{H^{s+1}}\leq\\\nonumber 
C(||k^{tr}||^{2}_{H^{s-1}}+a^{2-n\gamma_{ad}}||\delta\rho||_{H^{s-1}}+a^{2-n\gamma_{ad}}||v||^{2}_{H^{s-1}}(1+||\delta\rho||_{H^{s-1}})).
\end{eqnarray}
A short proof by contradiction may be as follows. Let us denote the elliptic operator in question by $\mathcal{Q}$. Assume that an estimate of the type $||\omega||_{H^{s+1}}\lesssim ||\mathcal{Q}\omega||_{H^{s-1}}$ does not hold. Then there exists a sequence $\{\omega_{i}\}_{i=1}^{\infty}$ with $||\omega_{i}||_{H^{s+1}}=1$ and $||\mathcal{Q}\omega_{i}||_{H^{s-1}}\to 0$ as $i\to \infty$. Now $\Sigma$ is compact and therefore $H^{s+1}$ is compactly embedded into $H^{s-1}$ (notice $s>\frac{n}{2}+2$). This yields a sub-sequence $\{\omega_{i_{j}}\}$ converging to $\omega^{*}\in H^{s+1}$ strongly in $H^{s-1}$, which by construction satisfies $\mathcal{Q}\omega^{*}=0$. This contradicts the fact that the operator $\Omega$ is injective. Therefore an estimate of the type $||\omega||_{H^{s+1}}\lesssim ||A\omega||_{H^{s-1}}$ holds $\forall s\geq 1$ (which is satisfied here since $s>\frac{n}{2}+2$). 

Estimation of the shift vector field is slightly more involved since the CMCSH gauge is known to exhibit the so called \textit{Gribov} degeneracy \cite{moncrief1979gribov, fischer1996quantum}. The shift equation reads 
\begin{eqnarray}
\label{eq:shiftin}
\Delta_{g}X^{i}-R^{i}_{j}X^{j}=-n(1-\frac{2}{n})\dot{a} \nabla^{i}N-2\nabla^{j}Nk^{tri}_{j}-2N\nabla_{j}k^{trij}\\\nonumber
+(2Nk^{trjk}-2\nabla^{j}X^{k})\left(\Gamma[g]^{i}_{jk}-\Gamma[\gamma]^{i}_{jk}\right)-ag^{ij}\partial_{t}\Gamma[\gamma]_{ij}^{k}.
\end{eqnarray}
In order to estimate $X$ in terms of the remaining variables, we want to show that the map $X\mapsto \Delta_{g}X^{i}-R^{i}_{j}X^{j}+2\nabla^{j}X^{k}\left(\Gamma[g]^{i}_{jk}-\Gamma[\gamma]^{i}_{jk}\right)$ is an isomorphism. However, apart from the operator $\Delta_{g}$, the remaining two terms do not have a definite sign. Therefore, it is not obvious that such an isomorphism property holds. However, here we will explicitly prove the isomorphism property of this operator using the small data assumption ($||g-\gamma||_{H^{s}}<\delta$). The following lemma provides the necessary estimate for the shift vector field.

\textbf{Lemma 6:} \textit{Let $B_{\delta}(\mathbf{0})\subset H^{s}\times H^{s-1}\times H^{s-1}\times H^{s-1}$ be a ball for sufficiently small $\delta>0$ and $(u,k^{tr},\delta\rho,v)\in B_{\delta}(\mathbf{0})$ solve the re-scaled Einstein-Euler-$\Lambda$ system in CMCSH gauge. Then the following estimate holds for the shift vector field 
\begin{eqnarray}
\label{eq:shiftpert}
||X||_{H^{s+1}}\leq 
C\left(\dot{a}||\omega||_{H^{s+1}}+||\omega||_{H^{s+1}}||k^{tr}||_{H^{s-1}}+a^{2-n\gamma_{ad}}\right.\\\nonumber \left.\{||v||^{3}_{H^{s-1}}(1+||u||_{H^{s}})+||v||_{H^{s-1}}\}+||k^{tr}||_{H^{s-1}}||u||_{H^{s}}\right)\\\nonumber 
+C\left(\dot{a}||\omega||_{H^{s+1}}(1+||u||_{H^{s}})+||k^{tr}||_{H^{s-1}}\right)
\end{eqnarray}
}
\textbf{Proof:} The first step in obtaining the estimate for $X$ is to prove that the map $X\mapsto \Delta_{g}X^{i}-R^{i}_{j}X^{j}+2\nabla^{j}X^{k}\left(\Gamma[g]^{i}_{jk}-\Gamma[\gamma]^{i}_{jk}\right)$ is an isomorphism between suitable function spaces. Let us denote the flow of $X$ on a constant time hypersurface by $\varphi_{s}$ ($s$ being the flow parameter). Let us pull back $V^{\flat}$ (where $V^{i}:=g^{kl}(\Gamma[g]^{i}_{kl}-\Gamma[\gamma]^{i}_{kl})$ is the negative of the tension vector field) by the flow of $X$ to obtain the following
\begin{eqnarray}
[(\varphi^{*}_{s}V^{\flat})^{\sharp}]^{i}=(\varphi^{*}_{s}g)^{jk}(\varphi^{*}_{s}(\Gamma[g]-\Gamma[\gamma]))^{i}_{kl}\nonumber=(\varphi^{*}_{s}g)^{jk}((\Gamma[\varphi^{*}_{s}g]-\Gamma[\varphi^{*}_{s}\gamma]))^{i}_{kl}
\end{eqnarray}
which holds because of the fact that the difference of connections transforms as a section of a suitable vector bundle. Differentiating both sides with respect to $s$ and setting $s=0$ yields 
\begin{eqnarray}
L_{X}V^{i}=(L_{X}g)^{jk}(\Gamma[g]^{i}_{jk}-\Gamma[\gamma]^{i}_{jk})\nonumber+g^{jk}\frac{d}{ds}\left((\Gamma[\varphi^{*}_{s}g]-\Gamma[\varphi^{*}_{s}\gamma])^{i}_{jk}\right)|_{s=0}\\\nonumber
=(L_{X}g)^{jk}(\Gamma[g]^{i}_{jk}-\Gamma[\gamma]^{i}_{jk})+\frac{1}{2}g^{jk}g^{il}\left[\nabla_{j}(L_{X}g_{lk})+\nabla_{k}(L_{X}g_{jl})\right.\\\nonumber \left.-\nabla_{l}(L_{X}g_{jk})\right]
-\frac{1}{2}g^{jk}\gamma^{il}\left[\nabla[\gamma]_{j}(L_{X}\gamma_{lk})+\nabla[\gamma]_{k}(L_{X}\gamma_{jl})-\nabla[\gamma]_{l}(L_{X}\gamma_{jk})\right]
\end{eqnarray}
Explicit computation in CMCSH gauge ($V^{i}=0$) yields  
\begin{eqnarray}
g^{kl}\nabla_{k}\nabla_{l}X^{i}+R^{i}_{j}X^{j}+2\nabla^{k}X^{l}(\Gamma[g]^{i}_{kl}-\Gamma[\gamma]^{i}_{kl})\\\nonumber 
=g^{kl}\nabla[\gamma]_{k}\nabla[\gamma]_{l}X^{i}_{\gamma}+g^{kl}R[\gamma]_{mk}~^{i}~_{l}X^{m}_{\gamma},
\end{eqnarray}
where $X^{i}_{\gamma}:=X_{j}\gamma^{ij}$. Now the map $P: X\mapsto -g^{kl}\nabla[\gamma]_{k}\nabla[\gamma]_{l}X^{i}_{\gamma}-g^{kl}R[\gamma]_{mk}~^{i}~_{l}X^{m}_{\gamma}$ is an isomorphism which may be shown as follows. Let's consider that $\ker(P)$ is not empty and $0\neq Z\in \ker(P)$ i.e., 
\begin{eqnarray}
-g^{kl}\nabla[\gamma]_{k}\nabla[\gamma]_{l}Z^{i}-g^{kl}R[\gamma]_{mk}~^{i}~_{l}Z^{m}=0.
\end{eqnarray}
Multiplying both sides of the above equation by $Z$ and integrating over $\Sigma$ yields
\begin{eqnarray}
\int_{\Sigma}\left(-g^{kl}Z_{i}\nabla[\gamma]_{k}\nabla[\gamma]_{l}Z^{i}-g^{kl}Z_{i}R[\gamma]_{mk}~^{i}~_{l}Z^{m}\right)\mu_{\gamma}=0\\\nonumber 
\int_{\Sigma}\left(\gamma^{kl}\nabla[\gamma]_{k}Z_{i}\nabla[\gamma]_{l}Z^{i}-\gamma^{kl}R[\gamma]_{mkil}Z^{i}Z^{m}-(g^{kl}-\gamma^{kl})Z_{i}\nabla[\gamma]_{k}\nabla[\gamma]_{l}Z^{i}\right.\\\nonumber 
\left.-(g^{kl}-\gamma^{kl})R[\gamma]_{mkil}Z^{i}Z^{m}\right)\mu_{\gamma}=0
\end{eqnarray}
Now for $||g-\gamma||_{H^{s}}<\delta$ sufficiently small, we have 
\begin{eqnarray}
\left(\gamma^{kl}\nabla[\gamma]_{k}Z_{i}\nabla[\gamma]_{l}Z^{i}-\gamma^{kl}R[\gamma]_{mkil}Z^{i}Z^{m}-(g^{kl}-\gamma^{kl})Z_{i}\nabla[\gamma]_{k}\nabla[\gamma]_{l}Z^{i}\right.\\\nonumber 
\left.-(g^{kl}-\gamma^{kl})R[\gamma]_{mkil}Z^{i}Z^{m}\right)>0
\end{eqnarray}
since $Ricci[\gamma](Z,Z)<0$ yielding 
\begin{eqnarray}
Z=0
\end{eqnarray}
i.e., $\ker(P)=\{0\}$. Since $\Sigma$ is compact and $P$ is second order elliptic, showing injectivity is enough to ensure its isomorphism property. Therefore in CMCSH gauge, we establish the isomorphism of the map $X\mapsto g^{kl}\nabla_{k}\nabla_{l}X^{i}+R^{i}_{j}X^{j}+2\nabla^{k}X^{l}(\Gamma[g]^{i}_{kl}-\Gamma[\gamma]^{i}_{kl})$. Using the equation for the shift (\ref{eq:shiftin}), this isomorphism yields the following estimate for the shift vector field
\begin{eqnarray}
\label{eq:shift1}
||X||_{H^{s+1}}\leq C||\left((1-\frac{2}{n})a\tau \nabla^{i}N\nonumber-2\nabla^{j}Nk^{tri}_{j}\right.\\
\left.-2N\left(\gamma_{ad} a^{2-n\gamma_{ad}}\rho(Nu^{0}-1)v^{i}+\gamma_{ad} a^{2-n\gamma_{ad}}\rho v^{i}\right)\nonumber+2Nk^{trjk}(\Gamma[g]^{i}_{jk}-\Gamma[\gamma]^{i}_{jk})\right.\\r\left.
-ag^{ij}\partial_{t}\Gamma[\gamma]_{ij}^{k}\right)||_{H^{s-1}}.
\end{eqnarray}
The source term that is present in the right-hand side of the previous estimate (\ref{eq:shift1}) contains several terms that are multiplied by $\dot{a}$ which may be potentially dangerous. We need to keep track of these terms carefully. The most problematic term $ag^{ij}\partial_{t}\Gamma[\gamma]_{ij}^{k}$ needs to be estimated (notice that it is multiplied by $a$). Since $\partial_{t}\gamma$ is tangent to $\mathcal{N}$, it may be written as follows 
\begin{eqnarray}
\partial_{t}\gamma&=&-h^{TT||}-L_{Y^{||}}\gamma,
\end{eqnarray}
with $Y^{||}$ satisfying (\ref{eq:parallelvector})
\begin{eqnarray}
[\nabla[\gamma]^{m}\nabla[\gamma]_{m}Y^{||i}+R[\gamma]^{||i}_{m}Y^{||m}]-(h^{TT||}+L_{Y^{||}}\gamma)^{mn}(\Gamma[\gamma]^{i}_{mn}\\\nonumber 
-\Gamma[\gamma^{*}]^{i}_{mn})=0
\end{eqnarray}
and $h^{TT||}\in \ker(\mathcal{L}_{\gamma,\gamma})$.
Now  $\gamma^{mn}\partial_{t}\Gamma^{i}_{mn}[\gamma]=\gamma^{mn}D\Gamma^{i}_{mn}[\gamma]\cdot \partial_{t}\gamma$ for $\partial_{t}\gamma\in T_{\gamma}\mathcal{N}$ may be written as 
\begin{eqnarray}
\label{eq:dgamma}
 \gamma^{mn}D\Gamma[\gamma]^{i}_{mn} \cdot \partial_{t}\gamma&=&\gamma^{mn}\gamma^{ik}\left(\nabla[\gamma]_{m}\partial_{t}\gamma_{kn}+\nabla[\gamma]_{n}\partial_{t}\gamma_{mk}\nonumber-\nabla[\gamma]_{k}\partial_{t}\gamma_{mn}\right)\\\nonumber
 &=&2\gamma^{ik}\nabla[\gamma]^{n}\partial_{t}\gamma_{kn}-\nabla[\gamma]^{i}(tr_{\gamma}\partial_{t}\gamma),
\end{eqnarray}
from which $\gamma^{mn}D\Gamma^{i}_{mn}[\gamma]h^{TT||}=0$ follows immediately since $\nabla[\gamma]^{i}h^{TT||}_{ij}=0,~\tr_{\gamma}h^{TT||}=0$. The remaining terms lead to the following
\begin{eqnarray}
 -\gamma^{mn}D\Gamma[\gamma]^{i}_{mn} \cdot \partial_{t}\gamma=2\gamma^{ik}\nabla[\gamma]^{n}\left(\nabla[\gamma]_{k}Y^{||}_{n}+\nabla[\gamma\nonumber]_{n}Y^{||}_{k}\right)-2\nabla[\gamma]^{i}(\nabla[\gamma]_{m}Y^{||m})\\\nonumber
 =2\gamma^{ik}(\nabla[\gamma]_{k}\nabla[\gamma]_{n}Y^{n}+R[\gamma]_{mk}Y^{||m})+2\nabla[\gamma]^{n}\nabla[\gamma]_{n}Y^{||i}-2\nabla[\gamma]^{i}(\nabla[\gamma]_{m}Y^{||m})\\\nonumber
 =2(\nabla[\gamma]^{n}\nabla[\gamma]_{n}Y^{||i}+R[\gamma]^{i}_{m}Y^{||m})\\\nonumber
 =2(L_{Y^{||}}\gamma+h^{TT})^{mn}(\Gamma[\gamma]^{i}_{mn}-\Gamma[\gamma^{*}]^{i}_{mn})\\\nonumber
 =2(\partial_{t}\gamma)^{mn}(\Gamma[\gamma]^{i}_{mn}-\Gamma[\gamma^{*}]^{i}_{mn}).
\end{eqnarray} 
This yields 
\begin{eqnarray}
 ||\gamma^{mn}D\Gamma[\gamma]^{i}_{mn} \partial_{t}\gamma||_{H^{s-1}}\leq C||\gamma-\gamma^{*}||_{H^{s}}||\partial_{t}\gamma||_{H^{s-1}}
\end{eqnarray}
Therefore, using the triangle inequality, we have the following 
\begin{eqnarray}
 ||g^{mn}D\Gamma[\gamma]^{i}_{mn} \partial_{t}\gamma||_{H^{s-1}}\leq C(||g-\gamma||_{H^{s}}+||\gamma-\gamma^{*}||_{H^{s}})||\partial_{t}\gamma||_{H^{s-1}}.
\end{eqnarray}
Utilizing the estimate of $\partial_{t}\gamma$ from lemma 4, we obtain the following estimate for the shift vector field $X$
\begin{eqnarray}
||X||_{H^{s+1}}\left(1-C(||u||_{H^{s}}+||\gamma-\gamma^{*}||_{H^{s}})(1+||u||_{H^{s}})\right)\\\nonumber 
\leq C\left(\dot{a}||\omega||_{H^{s+1}}+||\omega||_{H^{s+1}}||k^{tr}||_{H^{s-1}}+a^{2-n\gamma_{ad}}(1+||\delta\rho||_{H^{s-1}})\right.\\\nonumber \left.(||v||^{2}_{H^{s-1}}(1+||u||_{H^{s}})||v||_{H^{s-1}}+||v||_{H^{s-1}})+||k^{tr}||_{H^{s-1}}||u||_{H^{s}}\right)\\\nonumber 
+C\left(\dot{a}||\omega||_{H^{s+1}}(1+||u||_{H^{s}})+(1+||\omega||_{H^{s+1}})||k^{tr}||_{H^{s-1}}\right)
\end{eqnarray}
which for sufficiently small $||u||_{H^{s}}+||\gamma-\gamma^{*}||_{H^{s}}$ yields 
\begin{eqnarray}
||X||_{H^{s+1}}\leq C\left(\dot{a}||\omega||_{H^{s+1}}\nonumber+||\omega||_{H^{s+1}}||k^{tr}||_{H^{s-1}}+a^{2-n\gamma_{ad}}(1+||\delta\rho||_{H^{s-1}})\right.\\\nonumber \left.(||v||^{2}_{H^{s-1}}(1+||u||_{H^{s}})||v||_{H^{s-1}}+||v||_{H^{s-1}})+||k^{tr}||_{H^{s-1}}||u||_{H^{s}}\right)\\\nonumber 
+C\left(\dot{a}||\omega||_{H^{s+1}}(1+||u||_{H^{s}})+||k^{tr}||_{H^{s-1}}\right).~~~~~~~~~~~~~~~~~~~~~~~~~~~~~~~~~~~~~~\Box
\end{eqnarray}
Here several terms contain the factor $\dot{a}$ which may be potentially dangerous. However, the structure of Einstein's equations is such that these potentially problematic terms are always multiplied by $\frac{1}{a}$ in the energy estimate thereby turning these into innocuous terms. We will discover this fact in a later section where we study the energy estimates. 

It is worth mentioning that the smallness of $u:=g-\gamma$ is extremely important in the proof of the isomorphism property of $P$. The size of the Gribov domain is essentially defined by the largest value of $||g-\gamma||_{H^{s}}$ for which the isomorphism property of $P$ holds. If $g$ leaves the Gribov horizon, then the CMCSH gauge orbit develops a direction (or directions) of tangency to the gauge slice introducing degeneracy \cite{moncrief1979gribov, fischer1996quantum}. However, in the current context of stability considerations, such a problem does not arise due to the smallness of the data.     

In addition to estimating the shift vector field $X$, we note that a new vector field $Y^{||}+\frac{1}{a}X$ enters into the field equations (\ref{eq:fixed_new}). In order to estimate this new vector field, we will have to perform a series of calculations. The following lemma provides an estimate for $Y^{||}+\frac{1}{a}X$

\textbf{Lemma 7:} \textit{Let $B_{\delta}(\mathbf{0})\subset H^{s}\times H^{s-1}\times H^{s-1}\times H^{s-1},~s>\frac{n}{2}+2$ be a ball for sufficiently small $\delta>0$ and $(u,k^{tr},\delta\rho,v)\in B_{\delta}(\mathbf{0})$ solve the re-scaled Einstein-Euler-$\Lambda$ system in CMCSH gauge. Then the following estimate holds for the vector field $Y^{||}+\frac{1}{a}X$
\begin{eqnarray}
||Y^{||}+\frac{X}{a}||_{H^{s+1}}\leq C\left(\frac{\dot{a}}{a}||\omega||_{H^{s+1}}\nonumber+\frac{1}{a}||\omega||_{H^{s+1}}||k^{tr}||_{H^{s-1}}\right.\\\nonumber 
\left.+a^{1-n\gamma_{ad}}(1+||\omega||_{H^{s+1}})(1+||\delta\rho||_{H^{s-1 }})(||v||_{H^{s-1}}+||v||^{3}_{H^{s-1}}(1+||u||_{H^{s}}))\right.\\\nonumber 
\left.+\frac{1}{a}(1+||\omega||_{H^{s+1}})||k^{tr}||_{H^{s-1}}||u||_{H^{s}}+\frac{1}{a}||u||_{H^{s}}||X||_{H^{s+1}}+||u||_{H^{s}}||\partial_{t}\gamma||_{H^{s-1}}\right).
\end{eqnarray}
}\\
\textbf{Proof:} Let us recall that the $\mathcal{N}-$tangential velocity $\partial_{t}\gamma$ verifies $\partial_{t}\gamma=-h^{TT||}-L_{Y^{||}}\gamma$ and an explicit calculation shows that the following holds 
\begin{eqnarray}
\label{eq:parallelmotion}
\gamma^{mn}\partial_{t}\Gamma[\gamma]^{i}_{mn}=\gamma^{mn}D\Gamma[\gamma]^{i}_{mn}\cdot\partial_{t}\gamma=-\gamma^{mn}D\Gamma[\gamma]^{i}_{mn}\cdot (h^{TT||}+L_{Y^{||}}\gamma)\\\nonumber 
=-\gamma^{mn}D\Gamma[\gamma]^{i}_{mn}\cdot L_{Y^{||}}\gamma=-\gamma^{mn}\nabla[\gamma]_{m}\nabla[\gamma]_{n}Y^{||i}-R[\gamma]^{i}_{j}Y^{||j}.
\end{eqnarray}
The calculations from the previous lemma yield 
\begin{eqnarray}
g^{kl}\nabla_{k}\nabla_{l}X^{i}+R^{i}_{j}X^{j}+2\nabla^{k}X^{l}(\Gamma[g]^{i}_{kl}-\Gamma[\gamma]^{i}_{kl})\\\nonumber 
=g^{kl}\nabla[\gamma]_{k}\nabla[\gamma]_{l}X^{i}_{\gamma}+g^{kl}R[\gamma]_{mk}~^{i}~_{l}X^{m}_{\gamma}
\end{eqnarray}
which by the virtue of the shift equation (\ref{eq:shiftin}) yields 
\begin{eqnarray}
-g^{kl}\nabla[\gamma]_{k}\nabla[\gamma]_{l}\frac{X^{i}_{\gamma}}{a}-g^{kl}R[\gamma]_{mk}~^{i}~_{l}\frac{X^{m}_{\gamma}}{a}=-(1-\frac{2}{n})n\frac{\dot{a}}{a} \nabla^{i}N\nonumber\\\nonumber-\frac{2}{a}\nabla^{j}Nk^{tri}_{j}-\frac{2}{a}N\nabla_{j}k^{trij}
+\frac{2}{a}Nk^{trjk}\left(\Gamma[g]^{i}_{jk}-\Gamma[\gamma]^{i}_{jk}\right)-g^{ij}\partial_{t}\Gamma[\gamma]_{ij}^{k}
\end{eqnarray}
or 
\begin{eqnarray}
-\gamma^{kl}\nabla[\gamma]_{k}\nabla[\gamma]_{l}\frac{X^{i}_{\gamma}}{a}-\gamma^{kl}R[\gamma]_{mk}~^{i}~_{l}\frac{X^{m}_{\gamma}}{a}=-(1-\frac{2}{n})n\frac{\dot{a}}{a} \nabla^{i}N\nonumber-\frac{2}{a}\nabla^{j}Nk^{tri}_{j}\\\nonumber-\frac{2}{a}N\nabla_{j}k^{trij}
+\frac{2}{a}Nk^{trjk}\left(\Gamma[g]^{i}_{jk}-\Gamma[\gamma]^{i}_{jk}\right)-g^{ij}\partial_{t}\Gamma[\gamma]_{ij}^{k}+(g^{kl}-\gamma^{kl})\nabla[\gamma]_{k}\nabla[\gamma]_{l}\frac{X^{i}_{\gamma}}{a}\\\nonumber 
+(g^{kl}-\gamma^{kl})R[\gamma]_{mk}~^{i}~_{l}\frac{X^{m}_{\gamma}}{a}
\end{eqnarray}
which yields after adding $-\gamma^{kl}\nabla[\gamma]_{k}\nabla[\gamma]_{l}Y^{||i}_{\gamma}-\gamma^{kl}R[\gamma]_{mk}~^{i}~_{l}Y^{||m}_{\gamma}$ and using equation (\ref{eq:parallelmotion})
\begin{eqnarray}
-\gamma^{kl}\nabla[\gamma]_{k}\nabla[\gamma]_{l}(\frac{X^{i}_{\gamma}}{a}+Y^{||i})\nonumber-\gamma^{kl}R[\gamma]_{mk}~^{i}~_{l}(\frac{X^{m}_{\gamma}}{a}+Y^{||m})=-(1-\frac{2}{n})n\frac{\dot{a}}{a} \nabla^{i}N\\\nonumber-\frac{2}{a}\nabla^{j}Nk^{tri}_{j}-\frac{2}{a}N\nabla_{j}k^{trij}
+\frac{2}{a}Nk^{trjk}\left(\Gamma[g]^{i}_{jk}-\Gamma[\gamma]^{i}_{jk}\right)-g^{mn}\partial_{t}\Gamma[\gamma]_{mn}^{i}\\\nonumber +(g^{kl}\nonumber-\gamma^{kl})\nabla[\gamma]_{k}\nabla[\gamma]_{l}\frac{X^{i}_{\gamma}}{a} 
+(g^{kl}-\gamma^{kl})R[\gamma]_{mk}~^{i}~_{l}\frac{X^{m}_{\gamma}}{a}\\\nonumber 
+\gamma^{mn}\partial_{t}\Gamma[\gamma]^{i}_{mn}
\end{eqnarray}
which yields using momentum constraint (\ref{eq:mc}) and $H^{s_{1}}\hookrightarrow H^{s_{2}},~s_{1}>s_{2}$
\begin{eqnarray}
||Y^{||}+\frac{X}{a}||_{H^{s+1}}\leq C\left(\frac{\dot{a}}{a}||\omega||_{H^{s+1}}\nonumber+\frac{1}{a}||\omega||_{H^{s+1}}||k^{tr}||_{H^{s-1}}\right.\\\nonumber 
\left.+a^{1-n\gamma_{ad}}(1+||\omega||_{H^{s+1}})(1+||\delta\rho||_{H^{s-1 }})(||v||_{H^{s-1}}+||v||^{3}_{H^{s-1}}(1+||u||_{H^{s}}))\right.\\\nonumber 
\left.+\frac{1}{a}(1+||\omega||_{H^{s+1}})||k^{tr}||_{H^{s-1}}||u||_{H^{s}}+\frac{1}{a}||u||_{H^{s}}||X||_{H^{s+1}}+||u||_{H^{s}}||\partial_{t}\gamma||_{H^{s-1}}\right).
\end{eqnarray}

In addition to the above estimates, we need to estimate $\partial_{t}N$ as well since it appears in the time derivative of the fluid energy current (\ref{eq:energyidentity}). We commute the vector field $\partial_{t}$ with the elliptic equation for the lapse function (\ref{eq:lapse1}) to obtain
\begin{eqnarray}
\label{eq:timederivativelapse}
\Delta_{g}(\partial_{t}N)+\underbrace{\left(|k^{tr}|^{2}+\frac{2C_{\rho}a^{2-n\gamma_{ad}}}{n-1}+\frac{1}{n-1}+a^{2-n\gamma_{ad}}[\frac{(n\gamma_{ad}-2)}{n-1}+\gamma_{ad}\nonumber g(v,v)]\rho\right)}_{>0}\\\nonumber\partial_{t}N\\
=-N\partial_{t}|k^{tr}|^{2}_{g}-\frac{\gamma_{ad}N(n\gamma_{ad}-2)}{n-1}\partial_{t}(a^{2-n\gamma_{ad}}\delta\rho)-\gamma_{ad}N\partial_{t}(a^{2-n\gamma_{ad}}g(v,v))\\\nonumber 
-\partial_{t}\left(\frac{2C_{\rho}a^{2-n\gamma_{ad}}}{n-1}+\frac{1}{n-1}+C_{\rho}a^{2-n\gamma_{ad}}\frac{(n\gamma_{ad}-2)}{n-1}\right)(N-1)+\partial_{t}g^{ij}\nabla[g]_{i}\nabla_{j}N\\\nonumber 
-g^{ij}\partial_{t}\Gamma[g]^{k}_{ij}\nabla_{k}N.
\end{eqnarray}
The following lemma provides the necessary estimate for $\partial_{t}N$.

\textbf{Lemma 8:} \textit{Let $B_{\delta}(\mathbf{0})\subset H^{s}\times H^{s-1}\times H^{s-1}\times H^{s-1}$ be a ball for sufficiently small $\delta>0$ and $(u,k^{tr},\delta\rho,v)\in B_{\delta}(\mathbf{0})$ solve the re-scaled Einstein-Euler-$\Lambda$ system in CMCSH gauge. Then the following estimate holds for the time derivative of the lapse function 
\begin{eqnarray}
||\partial_{t}N||_{H^{s}}\leq C\left(||\partial_{t}|k^{tr}|^{2}_{g}||_{H^{s-2}}+\dot{a}a^{1-n\gamma_{ad}}||\delta\rho||_{H^{s-1}}\nonumber+a^{2-n\gamma_{ad}}||\partial_{t}\delta\rho||_{H^{s-2}}\right.\\\nonumber 
\left.+\dot{a}a^{1-n\gamma_{ad}}||v||^{2}_{H^{s-1}}+a^{2-n\gamma_{ad}}(||\partial_{t}g(v,v)+2g(\partial_{t}v,v)||_{H^{s-2}})+\dot{a}a^{1-n\gamma_{ad}}||\omega||_{H^{s+1}}\right.\\\nonumber\left.+||\partial_{t}g||_{H^{s-2}}(||\omega||_{H^{s+1}}+||u||_{H^{s}}||\omega||_{H^{s+1}})+ C||\omega||_{H^{s+1}}(1+||u||_{H^{s}})\right.\\\nonumber 
\left.\left(\frac{\dot{a}}{a}||\omega||_{H^{s+1}}(1+||u||_{H^{s}})+\frac{1}{a}(1+||\omega||_{H^{s+1}})||k^{tr}||_{H^{s-1}}+\frac{1}{a}(||X||_{H^{s+1}}||u||_{H^{s}}\right.\right.\\\nonumber
\left.\left.+(1+||u||_{H^{s}})||X||_{H^{s+1}})\right)\right)
\end{eqnarray}
}
\textbf{Proof:} Elliptic regularity theory when applied to the equation for the time derivative of the lapse function (\ref{eq:timederivativelapse}) yields
\begin{eqnarray}
||\partial_{t}N||_{H^{s}}\\\nonumber
\leq C ||-N\partial_{t}|k^{tr}|^{2}_{g}-\frac{\gamma_{ad}N(n\gamma_{ad}-2)}{n-1}\partial_{t}(a^{2-n\gamma_{ad}}\delta\rho)-\gamma_{ad}N\partial_{t}(a^{2-n\gamma_{ad}}g(v,v))\\\nonumber 
-\partial_{t}\left(\frac{2C_{\rho}a^{2-n\gamma_{ad}}}{n-1}+\frac{1}{n-1}+C_{\rho}a^{2-n\gamma_{ad}}\frac{(n\gamma_{ad}-2)}{n-1}\right)(N-1)+\partial_{t}g^{ij}\nabla[g]_{i}\nabla_{j}N\\\nonumber 
-g^{ij}\partial_{t}\Gamma[g]^{k}_{ij}\nabla_{k}N||_{H^{s-2}}\\\nonumber 
\leq C\left(||\partial_{t}|k^{tr}|^{2}_{g}||_{H^{s-2}}+\dot{a}a^{1-n\gamma_{ad}}||\delta\rho||_{H^{s-2}}+a^{2-n\gamma_{ad}}||\partial_{t}\delta\rho||_{H^{s-2}}+\dot{a}a^{1-n\gamma_{ad}}||v||^{2}_{H^{s-1}}\right.\\\nonumber 
\left.+a^{2-n\gamma_{ad}}(||\partial_{t}g(v,v)+2g(\partial_{t}v,v)||_{H^{s-2}})+\dot{a}a^{1-n\gamma_{ad}}||\omega||_{H^{s-2}}\right.\\\nonumber\left.+||\partial_{t}g||_{H^{s-2}}(||\omega||_{H^{s}}+||u||_{H^{s-1}}||\omega||_{H^{s-1}})+||g^{ij}\partial_{t}\Gamma[g]^{k}_{ij}\nabla_{k}N||_{H^{s-2}}\right).
\end{eqnarray}
Now we need to estimate the last term i.e., $||g^{ij}\partial_{t}\Gamma[g]^{k}_{ij}\nabla_{k}N||_{H^{s-2}}$. We do so in the following way. Explicit calculations yield
\begin{eqnarray}
||g^{ij}\partial_{t}\Gamma[g]^{k}_{ij}\nabla_{k}N||_{H^{s-2}}\\\nonumber 
=||g^{ij}g^{kl}(\nabla_{i}\partial_{t}g_{lj}+\nabla_{j}\partial_{t}g_{il}-\nabla_{l}\partial_{t}g_{ij})\nabla_{k}N||_{H^{s-2}}
\end{eqnarray}
Now before applying the algebra property of the Sobolev spaces, we do the following 
\begin{eqnarray}
\nabla_{i}\partial_{t}g_{lj}=\nabla[\gamma]_{i}\partial_{t}g_{lj}-(\Gamma[g]^{m}_{il}-\Gamma[\gamma]^{m}_{il})\partial_{t}g_{mj}-(\Gamma[g]^{m}_{ij}-\Gamma[\gamma]^{m}_{ij})\partial_{t}g_{ml}
\end{eqnarray}
and write $(L_{X}g)_{ij}=X^{k}\nabla[\gamma]_{k}u_{ij}+g_{ik}\nabla[\gamma]_{j}X^{k}+g_{kj}\nabla[\gamma]_{i}X^{k}$. Combining everything, we obtain the following estimate
\begin{eqnarray}
||g^{ij}\partial_{t}\Gamma[g]^{k}_{ij}\nabla_{k}N||_{H^{s-2}}\\\nonumber
\leq C||N-1||_{H^{s-1}}(1+||u||_{H^{s-2}})\left(\frac{\dot{a}}{a}||N-1||_{H^{s-2}}(1+||u||_{H^{s-2}})\right.\\\nonumber 
\left.+\frac{1}{a}(1+||N-1||_{H^{s-1}})||k^{tr}||_{H^{s-2}}+\frac{1}{a}(||X||_{H^{s-2}}||u||_{H^{s-1}}\right.\\\nonumber 
\left.+(1+||u||_{H^{s-2}})||X||_{H^{s-1}})\right).
\end{eqnarray}
Collecting all the terms and using $H^{s_{1}}\hookrightarrow H^{s_{2}},~s_{1}>s_{2}$, we obtain the result.~~~~~~~~~~~~~~~~~~~~~~~~~~~~~~~~~~~~~~~~~~~~~~~~~~~~~$\Box$

Notice that there are terms such as $||\partial_{t}|k^{tr}|^{2}_{g}||_{H^{s-2}}$, $||\partial_{t}\delta\rho||_{H^{s-2}}$, and $||g(\partial_{t}v,v)||_{H^{s-2}}$ estimation of which require the evolution equations. The following two lemmas provide the necessary estimates.

\textbf{Lemma 9:} \textit{Let $s>\frac{n}{2}+2$ and $B_{\delta}(\mathbf{0})\subset H^{s}\times H^{s-1}\times H^{s-1}\times H^{s-1}$ be a ball for sufficiently small $\delta>0$. Also let $(u,k^{tr},\delta\rho,v)\in B_{\delta}(\mathbf{0})$ solve the re-scaled Einstein-Euler-$\Lambda$ evolution equations. Then $\partial_{t}k^{tr}$ and $\partial_{t}|k^{tr}|^{2}_{g}$  verify the following estimates
\begin{eqnarray}
||\partial_{t}k^{tr}||_{H^{s-2}}\leq C\left(\frac{\dot{a}}{a}(1+||\omega||_{H^{s-2}})||k^{tr}||_{H^{s-2}}\nonumber+\frac{1}{a}(1+||\omega||_{H^{s-2}})||k^{tr}||^{2}_{H^{s-2}}\right.\\\nonumber 
\left.+a^{1-n\gamma_{ad}}(||\omega||_{H^{s-2}}+||\delta\rho||_{H^{s-2}}(1+||\omega||_{H^{s-2}}))\right.\\\nonumber 
\left.+\frac{1}{a}||\omega||_{H^{s}}+a^{1-n\gamma_{ad}}||v||^{2}_{H^{s-2}}(1+||\omega||_{H^{s-2}})(1+||\delta\rho||_{H^{s-2}})\right.\\\nonumber 
\left.+\frac{1}{a}(||X||_{H^{s-2}}||k^{tr}||_{H^{s-1}}+||X||_{H^{s-1}}||k^{tr}||_{H^{s-2}})\right.\\\nonumber 
\left.+\frac{1}{a}(||u||_{H^{s}}+||u||^{2}_{H^{s}})(1+||\omega||_{H^{s-2}})\right),
\end{eqnarray}
\begin{eqnarray}
||\partial_{t}|k^{tr}|^{2}_{g}||_{H^{s-2}}\leq C\left((\frac{\dot{a}}{a}||\omega||_{H^{s-2}}\nonumber+\frac{1}{a}(1+||\omega||_{H^{s-2}})||k^{tr}||_{H^{s-2}}\right.\\\nonumber 
\left.+\frac{1}{a}||X||_{H^{s-2}}||u||_{H^{s-1}}+||X||_{H^{s-1}}(1+||u||_{H^{s-2}}))||k^{tr}||^{2}_{H^{s-2}}\right.\\\nonumber 
\left.+(1+||u||_{H^{s-2}})||\partial_{t}k^{tr}||_{H^{s-2}}||k^{tr}||_{H^{s-2}}\right).
\end{eqnarray}
}
\textbf{Proof:} Let us write down the evolution equation for $k^{tr}_{ij}$ in the following way 
\begin{eqnarray}
\frac{\partial k^{tr}_{ij}}{\partial t}=-\frac{\dot{a}}{a}(n-1)k^{tr}_{ij}\nonumber-\frac{1}{a}\nabla[\gamma]_{i}\nabla_{j}\omega+\frac{1}{a}X^{k}\nabla[\gamma]_{k}k^{tr}_{ij}+\frac{1}{2a}N\mathcal{L}_{g,\gamma}u_{ij}+\mathcal{F}_{2ij},
\end{eqnarray}
where $\mathcal{F}_{2}$ is defined as follows \begin{eqnarray}
\mathcal{F}_{2ij}=-\frac{2}{a}(1+\omega)k^{tr}_{ik}k^{trk}_{j}\nonumber+\left(a^{1-n\gamma_{ad}}\frac{\omega\gamma_{ad}C_{\rho}+N(\gamma_{ad}-2)\delta\rho}{n-1}+\frac{\omega}{a(n-1)}\right)g_{ij}\\
+\frac{1}{2a}g^{lk}(\nabla[\gamma]_{i}u_{jk}+\nabla[\gamma]_{j}u_{ik}-\nabla[\gamma]_{k}u_{ij})\nabla_{l}\omega-\frac{\dot{a}}{a}(n-2)\omega k^{tr}_{ij}\\\nonumber 
-\gamma_{ad}a^{1-n\gamma_{ad}}(1+\omega)(C_{\rho}+\delta\rho)v_{i}v_{j}+\frac{1}{a}k^{tr}_{ik}\nabla[\gamma]_{j}X^{k}+\frac{1}{a}k^{tr}_{kj}\nabla[\gamma]_{i}X^{k}\\\nonumber
+\frac{1}{a}(1+\omega)\mathfrak{J}_{ij}.
\end{eqnarray}
Using product estimate (\ref{eq:product}), other elementary inequalities, and $||\mathfrak{J}||_{H^{s-1}}\leq C||u||^{2}_{H^{s}},$ the estimate for $\partial_{t}k^{tr}$ follows. For $\partial_{t}|k^{tr}|^{2}_{g}$, we write the following using symmetry of $k^{tr}$ and $g$
\begin{eqnarray}
\partial_{t}|k^{tr}|^{2}_{g}=\partial_{t}(g^{ik}g^{jl}k^{tr}_{ij}k^{tr}_{kl})=2\partial_{t}g^{ik}g^{jl}k^{tr}_{ij}k^{tr}_{kl}+2g^{ik}g^{jl}\partial_{t}k^{tr}_{ij}k^{tr}_{kl},
\end{eqnarray}
and use the product estimate (\ref{eq:product}), equation of motion for $g$ (\ref{eq:fixed_new}), and elementary calculus inequalities to obtain the desired estimate.

\textbf{Lemma 10:}  \textit{Let $s>\frac{n}{2}+2$ and $B_{\delta}(\mathbf{0})\subset H^{s}\times H^{s-1}\times H^{s-1}\times H^{s-1}$ be a ball for sufficiently small $\delta>0$. Also let $(u,k^{tr},\delta\rho,v)\in B_{\delta}(\mathbf{0})$ solve the re-scaled Einstein-Euler-$\Lambda$ evolution equations. Then $\partial_{t}\delta\rho$ and $\partial_{t}v$ verify the following estimates 
\begin{eqnarray}
\label{eq:rhoestimate}
||\partial_{t}\delta\rho||_{H^{s-2}}\leq C\left(\frac{\dot{a}}{a}||v||^{2}_{H^{s-1}}(1+||\omega||_{H^{s+1}})+\frac{1}{a}||k^{tr}||_{H^{s-1}}||v||^{2}_{H^{s-1}}\right.\\\nonumber 
\left.+\frac{1}{a}||X||_{H^{s+1}}||v||^{2}_{H^{s-1}}+\frac{1}{a}||v||^{3}_{H^{s-1}}+\frac{1}{a}||v||^{3}_{H^{s-1}}||u||_{H^{s}}+\frac{1}{a}||v||_{H^{s-1}}||\omega||_{H^{s+1}}\right.\\\nonumber 
\left.+\frac{1}{a}||\delta\rho||_{H^{s-1}}||v||_{H^{s-1}}(1+||v||_{H^{s-1}}+||X||_{H^{s+1}}||v||_{H^{s-1}})+\frac{1}{a}||v||_{H^{s-1}}+\frac{\dot{a}}{a}||\omega||_{H^{s+1}}\right.\\\nonumber \left.+\frac{1}{a}||X||_{H^{s+1}}||\delta\rho||_{H^{s-1}}+\frac{1}{a}||v||_{H^{s-1}}||\delta\rho||_{H^{s-1}}+\frac{1}{a}||k^{tr}||_{H^{s-1}}||v||^{2}_{H^{s-1}}\right.\\\nonumber 
\left.+\frac{\dot{a}}{a}||v||^{2}_{H^{s-1}}+\frac{1}{a}||v||^{2}_{H^{s-1}}||X||_{H^{s+1}}+\frac{1}{a}||\omega||_{H^{s+1}}||v||_{H^{s-1}}\right),\\
||\partial_{t}v||_{H^{s-2}}\leq C\left(\frac{\dot{a}}{a}||v||_{H^{s-1}}(1+||\omega||_{H^{s+1}})+\frac{1}{a}||k^{tr}||_{H^{s-1}}||v||_{H^{s-1}}\right.\\\nonumber 
\left.+\frac{1}{a}||X||_{H^{s+1}}||v||_{H^{s-1}}+\frac{1}{a}||v||^{2}_{H^{s-1}}+\frac{1}{a}||v||^{2}_{H^{s-1}}||u||_{H^{s}}+\frac{1}{a}||\omega||_{H^{s+1}}\right.\\\nonumber 
\left.+\frac{1}{a}||\delta\rho||_{H^{s-1}}(1+||v||_{H^{s-1}}+||X||_{H^{s+1}}||v||_{H^{s-1}})\right)(1+||v||^{2}_{H^{s-1}})\\\nonumber 
+C||v||_{H^{s-1}}\left(\frac{1}{a}||v||_{H^{s-1}}(1+||u||_{H^{s}})+\frac{\dot{a}}{a}||\omega||_{H^{s+1}}\nonumber+\frac{1}{a}||X||_{H^{s+1}}||\delta\rho||_{H^{s-1}}\right.\\\nonumber \left.+\frac{1}{a}||v||_{H^{s-1}}||\delta\rho||_{H^{s-1}}+\frac{1}{a}||k^{tr}||_{H^{s-1}}||v||^{2}_{H^{s-1}}+\frac{\dot{a}}{a}||v||^{2}_{H^{s-1}}+\frac{1}{a}||v||^{2}_{H^{s-1}}||X||_{H^{s+1}}\right.\\\nonumber 
\left.+\frac{1}{a}||\omega||_{H^{s+1}}||v||_{H^{s-1}}\right)
\end{eqnarray}
}\\
\textbf{Proof:} In order to estimate $\partial_{t}\delta\rho$ and $\partial_{t}v$, we need to obtain their explicit expressions. We do so by using the original evolution equations (\ref{eq:equationdensity}-\ref{eq:eomf}) and elementary linear algebra  
\begin{eqnarray}
\label{eq:rhoexplicit}
\partial_{t}\rho=-\frac{\gamma_{ad}\rho}{1+2g(v,v)-\gamma_{ad}g(v,v)}g(\mathcal{Z}_{2},v)+\frac{1+g(v,v)}{1+2g(v,v)-\gamma_{ad}g(v,v)}\mathcal{Z}_{1},\\
\label{eq:vexplicit}
\partial_{t}v^{i}=\left(\frac{(\gamma_{ad}-1)g(\mathcal{Z}_{2},v)}{1+2g(v,v)-\gamma_{ad}g(v,v)}-\frac{(\gamma_{ad}-1)(1+g(v,v))\mathcal{Z}_{1}}{\gamma_{ad}\rho(1+2g(v,v)-\gamma_{ad}g(v,v))}\right)v^{i}\\\nonumber 
+\mathcal{Z}^{i}_{2}.
\end{eqnarray}
Here $\mathcal{Z}_{1}$ and $\mathcal{Z}^{i}_{2}$ are defined as follows 
\begin{eqnarray}
\mathcal{Z}_{1}:=-\frac{\gamma_{ad}\rho N\nabla_{i}v^{i}}{a[1+g(v,v)]^{1/2}}-\frac{(N-1)n\gamma_{ad}\rho\dot{a}}{a}+\frac{1}{a}L_{X}\rho\\\nonumber 
-\frac{NL_{v}\rho}{a[1+g(v,v)]^{1/2}}+\frac{\gamma_{ad}\rho N}{1+g(v,v)}\left(\frac{1}{a}k^{tr}(v,v)-\frac{\dot{a}}{a}g(v,v)\right)+\frac{\gamma_{ad}\rho v_{i}L_{X}v^{i}}{a[1+g(v,v)]}\\\nonumber-\frac{\gamma_{ad}N\rho L_{v}N}{a[1+g(v,v)]^{1/2}},\\\nonumber 
\mathcal{Z}^{i}_{2}:=-[2N-1-n(\gamma_{ad}-1)]\frac{\dot{a}}{a}v^{i}\nonumber+\frac{2N k^{tri}_{j}v^{j}}{a}+\frac{1}{a}L_{X}v^{i}\\\nonumber 
-\frac{v^{j}\nabla_{j}v^{i}}{au^{0}}-\frac{1+g(v,v)}{aNu^{0}}\nabla^{i}N-\frac{\gamma_{ad}-1}{a\gamma_{ad}\rho u^{0}}\nabla^{i}\rho-\frac{\gamma_{ad}-1}{a\gamma_{ad}\rho u^{0}}v^{i}L_{v}\rho+\frac{\gamma_{ad}-1}{a\gamma_{ad}\rho }v^{i}L_{X}\rho.
\end{eqnarray}
Now since $\rho=\delta\rho+C_{\rho}$ ($C_{\rho}$ is a constant), action of derivatives on $\rho$ and $\delta\rho$ are the same and therefore $\partial_{t}\rho$ and $\partial_{t}\delta\rho$ will have the same meaning. Using the normalization condition $Nu^{0}=\sqrt{1+g(v,v)}$, the composition estimate (\ref{eq:composition}), the product estimate (\ref{eq:product}), and Sobolev embedding, we obtain the following for $\mathcal{Z}_{1}$ and $\mathcal{Z}_{2}$
\begin{eqnarray}
||\mathcal{Z}_{1}||_{H^{s-2}}\leq C\left(\frac{1}{a}||v||_{H^{s-1}}(1+||u||_{H^{s}})+\frac{\dot{a}}{a}||\omega||_{H^{s+1}}\nonumber+\frac{1}{a}||X||_{H^{s+1}}||\delta\rho||_{H^{s-1}}\right.\\\nonumber \left.+\frac{1}{a}||v||_{H^{s-1}}||\delta\rho||_{H^{s-1}}+\frac{1}{a}||k^{tr}||_{H^{s-1}}||v||^{2}_{H^{s-1}}+\frac{\dot{a}}{a}||v||^{2}_{H^{s-1}}+\frac{1}{a}||v||^{2}_{H^{s-1}}||X||_{H^{s+1}}\right.\\\nonumber 
\left.+\frac{1}{a}||\omega||_{H^{s+1}}||v||_{H^{s-1}}\right),
\end{eqnarray}
\begin{eqnarray}
||\mathcal{Z}_{2}||_{H^{s-2}}\leq C\left(\frac{\dot{a}}{a}||v||_{H^{s-1}}(1+||\omega||_{H^{s+1}})+\frac{1}{a}||k^{tr}||_{H^{s-1}}||v||_{H^{s-1}}\right.\\\nonumber 
\left.+\frac{1}{a}||X||_{H^{s+1}}||v||_{H^{s-1}}+\frac{1}{a}||v||^{2}_{H^{s-1}}+\frac{1}{a}||v||^{2}_{H^{s-1}}||u||_{H^{s}}+\frac{1}{a}||\omega||_{H^{s+1}}\right.\\\nonumber 
\left.+\frac{1}{a}||\delta\rho||_{H^{s-1}}(1+||v||_{H^{s-1}}+||X||_{H^{s+1}}||v||_{H^{s-1}})\right),
\end{eqnarray}
where we have replaced lower order Sobolev norms by the maximum available ones.
Using the equations (\ref{eq:rhoexplicit}-\ref{eq:vexplicit}), the estimates for $\partial_{t}\delta\rho$ and $\partial_{t}v$ follow. ~~~~~~~~~~~~~~~~~~$\Box$

\subsection{Time evolution of total energy}
Utilizing the elliptic estimates obtained in the previous section and the calculations associated with the fluid energy current in section (3), we will explicitly estimate the time derivative of the total energy.  
We need to estimate the time evolution of the total energy
$E_{total}$ 
\begin{eqnarray}
\frac{dE_{total}}{dt}=\frac{dE_{Ein}}{dt}+\frac{dE_{fluid}}{dt}.
\end{eqnarray}
Since we have already taken care of the principal terms that arise in the expression $\frac{dE_{fluid}}{dt}$, we will focus on on the gravity part instead. Explicit calculations yield
\begin{eqnarray}
\frac{dE_{Ein}}{dt}\\\nonumber
=\frac{1}{8}\sum_{I=1}^{s}\frac{d}{dt}\int_{\Sigma}\mathcal{L}^{I}_{g,\gamma}u_{ij}u_{kl}\gamma^{ik}\gamma^{jl}\mu_{g}+\frac{1}{2}\sum_{I=1}^{s}\frac{d}{dt}\int_{\Sigma}\mathcal{L}^{I-1}_{g,\gamma}k^{tr}_{ij}k^{tr}_{kl}\gamma^{ik}\gamma^{jl}\mu_{g}\\\nonumber 
=\sum_{I=1}^{s}\left(\frac{1}{8}\int_{\Sigma}(\mathcal{L}^{I}_{g,\gamma}\partial_{t}u_{ij})u_{kl}\gamma^{ik}\gamma^{jl}\mu_{g}+\frac{1}{8}\int_{\Sigma}\mathcal{L}^{I}_{g,\gamma}u_{ij}\partial_{t}u_{kl}\gamma^{ik}\gamma^{jl}\mu_{g}\right.\\\nonumber 
\left.+\frac{1}{8}\int_{\Sigma}[\partial_{t},\mathcal{L}^{I}_{g,\gamma}]u_{ij}u_{kl}\gamma^{ik}\gamma^{jl}\mu_{g} +\frac{1}{2}\int_{\Sigma}(\mathcal{L}^{I-1}_{g,\gamma}\partial_{t}k^{tr}_{ij})k^{tr}_{kl}\gamma^{ik}\gamma^{jl}\mu_{g}\right.\\\nonumber 
\left.+\frac{1}{2}\int_{\Sigma}\mathcal{L}^{I-1}_{g,\gamma}k^{tr}_{ij}\partial_{t}k^{tr}_{kl}\gamma^{ik}\gamma^{jl}\mu_{g}+\frac{1}{2}\int_{\Sigma}[\partial_{t},\mathcal{L}^{I-1}_{g,\gamma}]k^{tr}_{ij}k^{tr}_{kl}\gamma^{ik}\gamma^{jl}\mu_{g} 
\right.\\\nonumber 
\left.+\frac{1}{8}\int_{\Sigma}\mathcal{L}^{I}_{g,\gamma}u_{ij}u_{kl}\partial_{t}(\gamma^{ik}\gamma^{jl})\mu_{g}+\frac{1}{2}\int_{\Sigma}\mathcal{L}^{I-1}_{g,\gamma}k^{tr}_{ij}k^{tr}_{kl}\partial_{t}(\gamma^{ik}\gamma^{jl})\mu_{g}\right.\\\nonumber 
\left.+\frac{1}{16}\int_{\Sigma}\mathcal{L}^{I}_{g,\gamma}u_{ij}u_{kl}\gamma^{ik}\gamma^{jl}\mu_{g}\tr_{g}(\partial_{t}g)+\frac{1}{4}\int_{\Sigma}\mathcal{L}^{I-1}_{g,\gamma}k^{tr}_{ij}k^{tr}_{kl}\gamma^{ik}\gamma^{jl}\mu_{g}\tr_{g}(\partial_{t}g)\right).
\end{eqnarray}
Now using the self-adjoint property of $\mathcal{L}_{g,\gamma}$, we simplify the previous expression and obtain the following 
\begin{eqnarray}
\frac{dE_{Ein}}{dt}=\sum_{I=1}^{s}\left(\int_{\Sigma}\left(\frac{1}{4}\partial_{t}u_{ij}\mathcal{L}^{I}_{g,\gamma}u_{kl}\gamma^{ik}\gamma^{jl}+k^{tr}_{ij}\mathcal{L}^{I-1}_{g,\gamma}(\partial_{t}k^{tr}_{kl})\gamma^{ik}\gamma^{jl}\right.\right.\\\nonumber 
\left.\left.+\frac{1}{8}u_{ij}[\partial_{t},\mathcal{L}^{I}_{g,\gamma}]u_{kl}\gamma^{ik}\gamma^{jl}+\frac{1}{2} k^{tr}_{ij}[\partial_{t},\mathcal{L}^{I-1}_{g,\gamma}]k^{tr}_{kl}\gamma^{ik}\gamma^{jl}\right)\mu_{g}\right.\\\nonumber 
\left.+\int_{\Sigma}\left(\frac{1}{8}u_{ij}\mathcal{L}^{I}_{g,\gamma}u_{kl}(\partial_{t}\gamma^{ik}\gamma^{jl}+\gamma^{ik}\partial_{t}\gamma^{jl})+\frac{1}{2}k^{tr}_{ij}\mathcal{L}^{I-1}_{g,\gamma}k^{tr}_{kl}(\partial_{t}\gamma^{ik}\gamma^{jl}+\gamma^{ik}\partial_{t}\gamma^{jl})\right)\mu_{g}\right.\\\nonumber
\left.+\int_{\Sigma}\left(\frac{1}{8}u_{ij}\mathcal{L}^{I}_{g,\gamma}u_{kl}\gamma^{ik}\gamma^{jl}+\frac{1}{2}k^{tr}_{ij}\mathcal{L}^{I-1}_{g,\gamma}k^{tr}_{kl}\gamma^{ik}\gamma^{jl}\right)\frac{1}{2}\tr_{g}\partial_{t}g\mu_{g}\right).
\end{eqnarray}
We will handle the $I=1$ case first and then use the inequalities listed in section $4.1$ to control the higher order terms. Let us denote the energy corresponding to $I=1$ by $E^{1}_{Ein}$. First let us write down the evolution equations for $(u,k^{tr})$ as follows
\begin{eqnarray}
\label{eq:fixednew}
\frac{\partial u_{ij}}{\partial t}&=&-\frac{2}{a}Nk^{tr}_{ij}+h^{TT||}_{ij}+\frac{1}{a}X^{k}\nabla[\gamma]_{k}u_{ij}+\mathcal{F}_{1ij},
\end{eqnarray}
\begin{eqnarray}
\frac{\partial k^{tr}_{ij}}{\partial t}=-\frac{\dot{a}}{a}(n-1)k^{tr}_{ij}\nonumber-\frac{1}{a}\nabla[\gamma]_{i}\nabla_{j}\omega+\frac{1}{a}X^{k}\nabla[\gamma]_{k}k^{tr}_{ij}+\frac{1}{2a}N\mathcal{L}_{g,\gamma}u_{ij}+\mathcal{F}_{2ij},
 \end{eqnarray}
where $\mathcal{F}_{1}$ and $\mathcal{F}_{2}$ are non-linear terms which will be estimated separately
\begin{eqnarray}
\label{eq:nonlinear1}
\mathcal{F}_{1ij}=2\frac{\dot{a}}{a}\omega g_{ij}+(L_{Y^{||}+\frac{1}{a}X}\gamma)_{ij}+\frac{1}{a}u_{ik}\nabla[\gamma]_{j}X^{k}+\frac{1}{a}u_{kj}\nabla[\gamma]_{i}X^{k},\\
\label{eq:nonlinear2}
\mathcal{F}_{2ij}=-\frac{2}{a}(1+\omega)k^{tr}_{ik}k^{trk}_{j}\nonumber+\left(a^{1-n\gamma_{ad}}\frac{\omega\gamma_{ad}C_{\rho}+N(\gamma_{ad}-2)\delta\rho}{n-1}+\frac{\omega}{a(n-1)}\right)g_{ij}\\
+\frac{1}{2a}g^{lk}(\nabla[\gamma]_{i}u_{jk}+\nabla[\gamma]_{j}u_{ik}-\nabla[\gamma]_{k}u_{ij})\nabla_{l}\omega-\frac{\dot{a}}{a}(n-2)\omega k^{tr}_{ij}\\\nonumber 
-\gamma_{ad}a^{1-n\gamma_{ad}}(1+\omega)(C_{\rho}+\delta\rho)v_{i}v_{j}+\frac{1}{a}k^{tr}_{ik}\nabla[\gamma]_{j}X^{k}+\frac{1}{a}k^{tr}_{kj}\nabla[\gamma]_{i}X^{k}\\\nonumber
+\frac{1}{a}(1+\omega)\mathfrak{J}_{ij}.
\end{eqnarray}
The following lemma yields the estimate for the first order total energy coming from the gravity sector.

\textbf{Lemma 11:} \textit{Let $s>\frac{n}{2}+2$ and $B_{\delta}(\mathbf{0})\subset H^{s}\times H^{s-1}\times H^{s-1}\times H^{s-1}$ be a ball for sufficiently small $\delta>0$. Also let $(u,k^{tr},\delta\rho,v)\in B_{\delta}(\mathbf{0})$ solve the re-scaled Einstein-Euler-$\Lambda$ evolution equations and $E^{1}_{Ein}$ be the first order energy of the gravitational sector defined by equation (\ref{eq:energygravity}) for $I=1$. Then $\frac{dE^{1}_{Ein}}{dt}$ satisfies the equation
\begin{eqnarray}
\frac{dE^{1}_{Ein}}{dt}=-\frac{(n-1)\dot{a}}{a}\langle k^{tr}|k^{tr}\rangle_{L^{2}}+\mathcal{ER}_{1},
\end{eqnarray}
where $\mathcal{ER}_{1}$ satisfies 
\begin{eqnarray}
\mathcal{ER}_{1}\leq C\left(\frac{1}{a}||u||^{2}_{H^{s}}||k^{tr}||_{H^{s-1}}\nonumber+\frac{1}{a}||X||_{H^{s+1}}(1+||u||_{H^{s}})||u||^{2}_{H^{s}}\right.\\\nonumber 
\left.+\frac{1}{a}||\omega||_{H^{s+1}}||k^{tr}||_{H^{s-1}}+\frac{1}{a}||X||_{H^{s+1}}||k^{tr}||^{2}_{H^{s-1}}+\frac{\dot{a}}{a}||\omega||_{H^{s+1}}||u||_{H^{s}}\right.\\\nonumber 
\left.+||Y^{||}+\frac{1}{a}X||_{H^{s+1}}||u||_{H^{s}}+\frac{1}{a}||X||_{H^{s+1}}||u||^{2}_{H^{s}}+\frac{1}{a}(1+||\omega||_{H^{s+1}})||k^{tr}||^{3}_{H^{s-1}}\right.\\\nonumber \left.+a^{1-n\gamma_{ad}}||\omega||_{H^{s+1}}||k^{tr}||_{H^{s-1}}+a^{1-n\gamma_{ad}}||\delta\rho||_{H^{s-1}}||k^{tr}||_{H^{s-1}}\right.\\\nonumber 
\left.+\frac{1}{a}||u||_{H^{s}}||\omega||_{H^{s+1}}||k^{tr}||_{H^{s-1}}+\frac{\dot{a}}{a}||\omega||_{H^{s+1}}||k^{tr}||_{H^{s-1}}+\frac{\dot{a}}{a}||\omega||_{H^{s+1}}||u||^{2}_{H^{s}}\right.\\\nonumber 
\left.+a^{1-n\gamma_{ad}}(1+||\omega||_{H^{s+1}})(1+||\delta\rho||_{H^{s-1}})||v||^{2}_{H^{s-1}}||k^{tr}||_{H^{s-1}}+\frac{1}{a}||X||_{H^{s+1}}\right.\\\nonumber 
\left.||k^{tr}||^{2}_{H^{s-1}}+\frac{1}{a}(1+||\omega||_{H^{s+1}})||u||^{2}_{H^{s}}||k^{tr}||_{H^{s-1}}+\left[\frac{\dot{a}}{a}||\omega||_{H^{s+1}}(1+||u||_{H^{s}})\right.\right.\\\nonumber 
\left.\left.+\frac{1}{a}(1+||\omega||_{H^{s+1}})||k^{tr}||_{H^{s-1}}+\frac{1}{a}||X||_{H^{s+1}}||u||_{H^{s}}(1+||u||_{H^{s}})\right]\right.\\\nonumber \left.(||u||^{2}_{H^{s}}+||k^{tr}||^{2}_{H^{s-1}})\right).
\end{eqnarray}
}\\
\textbf{Proof:}

$\frac{dE^{1}_{Ein}}{dt}$ reads 
\begin{eqnarray}
\frac{dE^{1}_{Ein}}{dt}=\int_{\Sigma}(\underbrace{\frac{1}{4}\partial_{t}u_{ij}\mathcal{L}_{g,\gamma}u_{kl}\gamma^{ik}\gamma^{jl}\nonumber+k^{tr}_{ij}(\partial_{t}k^{tr}_{kl})\gamma^{ik}\gamma^{jl}}_{I}+\underbrace{\frac{1}{8}u_{ij}[\partial_{t},\mathcal{L}_{g,\gamma}]u_{kl}\gamma^{ik}\gamma^{jl})}_{II}\mu_{g}\\\nonumber 
\left.+\int_{\Sigma}\underbrace{\left(\frac{1}{8}u_{ij}\mathcal{L}_{g,\gamma}u_{kl}(\partial_{t}\gamma^{ik}\gamma^{jl}+\gamma^{ik}\partial_{t}\gamma^{jl})+\frac{1}{2}k^{tr}_{ij}k^{tr}_{kl}(\partial_{t}\gamma^{ik}\gamma^{jl}+\gamma^{ik}\partial_{t}\gamma^{jl})\right)}_{III}\mu_{g}\right.\\\nonumber
+\int_{\Sigma}\underbrace{\left(\frac{1}{8}u_{ij}\mathcal{L}_{g,\gamma}u_{kl}\gamma^{ik}\gamma^{jl}+\frac{1}{2}k^{tr}_{ij}k^{tr}_{kl}\gamma^{ik}\gamma^{jl}\right)\frac{1}{2}\tr_{g}\partial_{t}g}_{IV}\mu_{g}.
\end{eqnarray}
Once again, we consider collection of terms that would yield necessary cancellations to remove the dangerous terms.\\
\textbf{Estimating terms of type $I$:} We first control the terms of type $I$ as follows
\begin{eqnarray}
I=\int_{\Sigma}\left(\frac{1}{4}\partial_{t}u_{ij}\mathcal{L}_{g,\gamma}u_{kl}\gamma^{ik}\gamma^{jl}\nonumber+k^{tr}_{ij}(\partial_{t}k^{tr}_{kl})\gamma^{ik}\gamma^{jl}\right)\mu_{g}\\
=\int_{\Sigma}\left(\frac{1}{4}\left(-\frac{2N}{a}k^{tr}_{ij}+\frac{1}{a}X^{m}\nabla[\gamma]_{m}u_{ij}-h^{TT||}_{ij}+\mathcal{F}_{1ij}\right)\mathcal{L}_{g,\gamma}u_{kl}\gamma^{ik}\gamma^{jl}\right.\\\nonumber 
\left.+k^{tr}_{ij}\left(-\frac{\dot{a}}{a}(n-1)k^{tr}_{kl}-\frac{1}{a}\nabla[\gamma]_{k}\nabla_{l}\omega+\frac{1}{a}X^{m}\nabla[\gamma]_{m}k^{tr}_{kl}+\frac{N}{2a}\mathcal{L}_{g,\gamma}u_{kl}\right.\right.\\\nonumber
\left.\left.+\mathcal{F}_{2kl}\right)\gamma^{ik}\gamma^{jl}\right)\mu_{g}.
\end{eqnarray}
Now notice that the principal term $k^{tr}_{ij}\mathcal{L}_{g,\gamma}u_{kl}\gamma^{ik}\gamma^{jl}$ cancels point-wise with its negative counterpart. This is essential since controlling this term requires one order of extra regularity (since in the case of $I=1$, we can not have terms that involve the derivative of $k^{tr}$). We still do have derivatives of $k^{tr}$ (and of $u$) in the direction of the shift vector field. However, the derivatives in the direction of the shift vector field $X$ are good and can be controlled through integration by parts as follows 
\begin{eqnarray}
\int_{\Sigma}X^{m}\nabla[\gamma]_{m}u_{ij}\mathcal{L}_{g,\gamma}u_{kl}\gamma^{ik}\gamma^{jl}\mu_{g}\\\nonumber 
=\int_{\Sigma}X^{m}\nabla[\gamma]_{m}u_{ij}\left(-g^{rs}\nabla[\gamma]_{r}\nabla[\gamma]_{s}u_{kl}-V^{n}\nabla[\gamma]_{n}u_{kl}-2R[\gamma]_{kplq}u^{pq}\right)\gamma^{ik}\gamma^{jl}\mu_{g},\\\nonumber 
=\int_{\Sigma}X^{m}\nabla[\gamma]_{m}u_{ij}\left(-g^{rs}\nabla[\gamma]_{r}\nabla[\gamma]_{s}u_{kl}-2R[\gamma]_{kplq}u^{pq}\right)\gamma^{ik}\gamma^{jl}\mu_{g},\\\nonumber 
=\int_{\Sigma}g^{rs}\nabla[\gamma]_{r}(X^{m}\nabla[\gamma]_{m}u_{ij})\nabla[\gamma]_{s}u_{kl}\gamma^{ik}\gamma^{jl}-2\int_{\Sigma}X^{m}\nabla[\gamma]_{m}u_{ij}R[\gamma]_{kplq}u^{pq}\gamma^{ik}\gamma^{jl}\mu_{g},\\\nonumber 
=\int_{\Sigma}\left(g^{rs}\nabla[\gamma]_{r}X^{m}\nabla[\gamma]_{m}u_{ij}\nabla[\gamma]_{s}u_{kl}\gamma^{ik}\gamma^{jl}+g^{rs}X^{m}\nabla[\gamma]_{m}\nabla[\gamma]_{r}u_{ij}\nabla[\gamma]_{s}u_{kl}\gamma^{ik}\gamma^{jl}\right)\mu_{g}\\\nonumber 
-\int_{\Sigma}g^{rs}X^{m}\left(R[\gamma]^{p}~_{irm}u_{pj}+R[\gamma]^{p}`_{jrm}u_{ip}\right)\nabla[\gamma]_{s}u_{kl}\gamma^{ik}\gamma^{jl}\mu_{g}\\\nonumber 
-2\int_{\Sigma}X^{m}\nabla[\gamma]_{m}u_{ij}R[\gamma]_{kplq}u^{pq}\gamma^{ik}\gamma^{jl}\mu_{g}\\\nonumber 
=\frac{1}{2}\int_{\Sigma}g^{rs}X^{m}\nabla[\gamma]_{m}(\nabla[\gamma]_{r}u_{ij}\nabla[\gamma]_{s}u_{kl}\gamma^{ik}\gamma^{jl})\mu_{g}\\\nonumber +\int_{\Sigma}g^{rs}\nabla[\gamma]_{r}X^{m}\nabla[\gamma]_{m}u_{ij}\nabla[\gamma]_{s}u_{kl}\gamma^{ik}\gamma^{jl})\mu_{g}\\\nonumber 
-\int_{\Sigma}g^{rs}X^{m}\left(R[\gamma]^{p}~_{irm}u_{pj}+R[\gamma]^{p}`_{jrm}u_{ip}\right)\nabla[\gamma]_{s}u_{kl}\gamma^{ik}\gamma^{jl}\mu_{g}\\\nonumber 
-2\int_{\Sigma}X^{m}\nabla[\gamma]_{m}u_{ij}R[\gamma]_{kplq}u^{pq}\gamma^{ik}\gamma^{jl}\mu_{g}\\\nonumber 
=-\frac{1}{2}\int_{\Sigma}(X^{m}\nabla[\gamma]_{m}g^{rs}+g^{rs}\nabla[\gamma]_{m}X^{m}+\frac{1}{2}g^{pn}(\nabla[\gamma]_{m}u_{pn}+\nabla[\gamma]_{p}u_{mn}-\nabla[\gamma]_{n}u_{ap})\\\nonumber 
\nabla[\gamma]_{r}u_{ij}\nabla[\gamma]_{s}u_{kl}\gamma^{ik}\gamma^{jl}\mu_{g}+\int_{\Sigma}g^{rs}\nabla[\gamma]_{r}X^{m}\nabla[\gamma]_{m}u_{ij}\nabla[\gamma]_{s}u_{kl}\gamma^{ik}\gamma^{jl})\mu_{g}\\\nonumber 
-\int_{\Sigma}g^{rs}X^{m}\left(R[\gamma]^{p}~_{irm}u_{pj}+R[\gamma]^{p}`_{jrm}u_{ip}\right)\nabla[\gamma]_{s}u_{kl}\gamma^{ik}\gamma^{jl}\mu_{g}\\\nonumber 
-2\int_{\Sigma}X^{m}\nabla[\gamma]_{m}u_{ij}R[\gamma]_{kplq}u^{pq}\gamma^{ik}\gamma^{jl}\mu_{g},
\end{eqnarray}
where we have used the SH gauge condition (i.e., $V=0$) and Stokes's theorem on the closed manifold $\Sigma$. 
Notice that all the terms can be estimated in terms of $E^{1}_{Ein}$ since the terms of the type $X^{m}\nabla[\gamma]_{m}u_{ij}\mathcal{L}_{g,\gamma}u_{kl}$ have disappeared by the virture of integration by parts and $H^{s}\hookrightarrow W^{1,\infty}$ for $s>\frac{n}{2}+1$ (we assume $s>\frac{n}{2}+2$ throughout). Here we have used the definition of $\mathcal{L}_{g,\gamma}$ and $V^{i}=g^{kl}(\Gamma[g]^{i}_{kl}-\Gamma[\gamma]^{i}_{kl})=0$ in CMCSH gauge. Therefore the term $\int_{\Sigma}X^{m}\nabla[\gamma]_{m}u_{ij}\mathcal{L}_{g,\gamma}u_{kl}\gamma^{ik}\gamma^{jl}\mu_{g}$ is estimated as follows 
\begin{eqnarray}
|\int_{\Sigma}X^{m}\nabla[\gamma]_{m}u_{ij}\mathcal{L}_{g,\gamma}u_{kl}\gamma^{ik}\gamma^{jl}\mu_{g}|\\\nonumber 
\leq C(||X||_{L^{\infty}}||\nabla[\gamma]u||_{L^{\infty}}+||\nabla[\gamma]X||_{L^{\infty}}+||X||_{L^{\infty}})||u||^{2}_{H^{s}}
\end{eqnarray}
and following Sobolev embedding (since $s>\frac{n}{2}+2$), we obtain 
\begin{eqnarray}
|\int_{\Sigma}X^{m}\nabla[\gamma]_{m}u_{ij}\mathcal{L}_{g,\gamma}u_{kl}\gamma^{ik}\gamma^{jl}\mu_{g}|\\\nonumber 
\leq C||X||_{H^{s+1}}(1+||u||_{H^{s}})||u||^{2}_{H^{s}}.
\end{eqnarray}

The term containing the directional derivative of $k^{tr}$ may be reduced to lower-order terms in a similar fashion. Another potentially dangerous term that appears in $I$ is the following term 
\begin{eqnarray}
\int_{\Sigma}h^{TT||}_{ij}\mathcal{L}_{g,\gamma}u_{kl}\gamma^{ik}\gamma^{jl}\mu_{g}.
\end{eqnarray}
Upon utilizing the self-adjoint property of $\mathcal{L}_{g,\gamma}$ (with respect the inner product (\ref{eq:twistedinner}) used in this context), we obtain 
\begin{eqnarray}
\int_{\Sigma}h^{TT||}_{ij}\mathcal{L}_{g,\gamma}u_{kl}\gamma^{ik}\gamma^{jl}\mu_{g}=\int_{\Sigma}\mathcal{L}_{g,\gamma}h^{TT||}_{ij}u_{kl}\gamma^{ik}\gamma^{jl}\mu_{g}.
\end{eqnarray}
Now utilizing the definition of $\mathcal{L}_{g,\gamma}$ and the property that $\mathcal{L}_{\gamma,\gamma}h^{TT||}=0$, we will reduce this term to an innocuous term. We have 
\begin{eqnarray}
\mathcal{L}_{g,\gamma}h^{TT||}_{ij}&=&\Delta^{\gamma}_{g}h^{TT||}_{ij}-2R[\gamma]_{ikjl}h^{TT||kl},\\\nonumber
&=&-g^{mn}(\nabla[\gamma]_{m}\nabla[\gamma]_{n}h^{TT||}_{ij})-V^{m}\nabla[\gamma]_{m}h^{TT||}_{ij}-2R[\gamma]_{ikjl}h^{TT||kl},\\\nonumber
&=&-(g^{mn}-\gamma^{mn})(\nabla[\gamma]_{m}\nabla[\gamma]_{n}h^{TT||}_{ij})-\gamma^{mn}\nabla[\gamma]_{m}\nabla[\gamma]_{n}h^{TT||}_{ij}\\\nonumber &&-2R[\gamma]_{ikjl}h^{TT||kl},\\\nonumber
&=&-(g^{mn}-\gamma^{mn})(\nabla[\gamma]_{m}\nabla[\gamma]_{n}h^{TT||}_{ij})+\mathcal{L}_{\gamma,\gamma}h^{TT||}_{ij},
\end{eqnarray}
where we have used the identity $\nabla[\gamma]_{m}(\mu_{g}g^{-1})^{mn}=-V^{n}\mu_{g}$, and set $V^{m}=0$ since we are working in CMCSH gauge. Using $\mathcal{L}_{\gamma,\gamma}h^{TT||}=0$ (by the definition of $h^{TT||}$) 
\begin{eqnarray}
\label{eq:1}
\mathcal{L}_{g,\gamma}h^{TT||}_{ij}=-(g^{mn}-\gamma^{mn})(\nabla[\gamma]_{m}\nabla[\gamma]_{n}h^{TT||}_{ij}).
\end{eqnarray}
Notice that we have already gained a factor of $g-\gamma$ in the expression of $\mathcal{L}_{g,\gamma}h^{TT||}$. 
Now we will exploit the shadow gauge condition to obtain an expression for $h^{TT||}$. The shadow gauge together with the expression (\ref{eq:tangentspace}) of an $\mathcal{N}-$tangent vector $\partial_{q^{\alpha}}\gamma$ reads
\begin{eqnarray}
\langle g-\gamma|\frac{\partial \gamma}{\partial q^{\alpha}}\rangle_{L^{2}}&=&0,\\
\langle u|h^{TT||\alpha}+L_{Y^{||\alpha}}\gamma\rangle_{L^{2}}&=&0
\end{eqnarray}
which upon time differentiation becomes 
\begin{eqnarray}\nonumber
\langle\partial_{t}u|h^{TT||\alpha}+L_{Y^{||\alpha}}\gamma\rangle_{L^{2}}+second~order~terms =0,\\\nonumber
\langle-\frac{2}{a}(1+\omega)k^{tr}+h^{TT||}+\frac{1}{a}X^{k}\nabla[\gamma]_{k}u+\mathcal{F}_{1}|h^{TT||\alpha}+L_{Y^{||\alpha}}\rangle_{L^{2}}\\\nonumber +second~order~terms =0.
\end{eqnarray}
Now, in the light of elliptic estimates, $\omega$ satisfies a second order estimate. Since the remaining terms are clearly of second order, using the identity $\langle A^{TT}|L_{Z}\gamma\rangle_{L^{2}}=0$ for any transverse-traceless tensor $A^{TT}$ and vector field $Z\in \mathfrak{X}(M)$, we immediately obtain
\begin{eqnarray}
\langle\frac{2}{a}k^{tr}-h^{TT||}|h^{TT||\alpha}\rangle_{L^{2}}\nonumber+\langle-\frac{2}{a}\omega k^{tr}+\frac{1}{a}X^{k}\nabla[\gamma]_{k}u+\mathcal{F}_{1}|h^{TT||\alpha}+L_{Y^{||\alpha}}\rangle\\\nonumber 
-\langle\frac{2}{a}k^{tr}|L_{Y^{||\alpha}}\rangle=0
\end{eqnarray}
which leads to
\begin{eqnarray}
h^{TT||}=\frac{2}{a}k^{tr||}+\frac{\dot{a}}{a}(terms~linear~in~\omega)+\frac{1}{a}(second~order~terms),
\end{eqnarray}
where $k^{tr||}$ is the projection of $k^{tr}$ onto the subspace of TT tensors belonging to the kernel of $\mathcal{L}_{\gamma,\gamma}$. Now since $h^{TT||}$ is an element of a finite dimensional vector space, its Sobolev norm of any order may be estimated by the $H^{s-1}$ norm of $\frac{2}{a}k^{tr}$ and $H^{s+1}$ norm of $\frac{\dot{a}}{a}\omega$ i.e., 
\begin{eqnarray}
||h^{TT||}||\leq C\left( \frac{1}{a}||k^{tr}||_{H^{s-1}}+\frac{\dot{a}}{a}||\omega||_{H^{s+1}}\right),
\end{eqnarray}
where the norm on the right hand side denotes Sobolev norm of arbitrary order. Even though $\omega$ is harmless, we kept the term due to the presence of $\frac{\dot{a}}{a}$ factor which may be potentially dangerous. Therefore the term $\int_{\Sigma}h^{TT||}_{ij}\mathcal{L}_{g,\gamma}u_{kl}\gamma^{ik}\gamma^{jl}\mu_{g}$ is estimated as 
\begin{eqnarray}
|\int_{\Sigma}h^{TT||}_{ij}\mathcal{L}_{g,\gamma}u_{kl}\gamma^{ik}\gamma^{jl}\mu_{g}|\leq C\left( \frac{1}{a}||u||^{2}_{H^{s}}||k^{tr}||_{H^{s-1}}\nonumber+\frac{\dot{a}}{a}||u||^{2}_{H^{s}}||\omega||_{H^{s+1}}\right).
\end{eqnarray}
The term $\int_{\Sigma}\mathcal{F}_{1ij}\mathcal{L}_{g,\gamma}u_{kl}\gamma^{ik}\gamma^{jl}\mu_{g}$
may be controlled as follows. Using the expression of $\mathcal{F}_{1ij}$ (\ref{eq:nonlinear1}), integration by parts, and $H^{s_{1}}\hookrightarrow H^{s_{2}},~s_{1}>s_{2}$, we obtain
\begin{eqnarray}
|\int_{\Sigma}\mathcal{F}_{1ij}\mathcal{L}_{g,\gamma}u_{kl}\gamma^{ik}\gamma^{jl}\mu_{g}|\\\nonumber 
=|\int_{\Sigma}\left(2\frac{\dot{a}}{a}\omega g_{ij}+(L_{Y^{||}+\frac{1}{a}X}\gamma)_{ij}+\frac{1}{a}u_{im}\nabla[\gamma]_{j}X^{m}+\frac{1}{a}u_{mj}\nabla[\gamma]_{i}X^{m}\right)\\\nonumber \mathcal{L}_{g,\gamma}u_{kl}\gamma^{ik}\gamma^{jl}\mu_{g}|\\\nonumber 
\leq C\left(\frac{\dot{a}}{a}||\omega||_{H^{s+1}}||u||_{H^{s}}+||Y^{||}+\frac{1}{a}X||_{H^{s+1}}||u||_{H^{s}}+\frac{1}{a}||X||_{H^{s+1}}||u||^{2}_{H^{s}}\right).
\end{eqnarray}
Similarly we estimate the term $\int_{\Sigma}k^{tr}_{ij}\mathcal{F}_{2kl}\gamma^{ik}\gamma^{jl}\mu_{g}$  
\begin{eqnarray}
|\int_{\Sigma}k^{tr}_{ij}\nonumber\mathcal{F}_{2kl}\gamma^{ik}\gamma^{jl}\mu_{g}|\\\nonumber 
=|\int_{\Sigma}k^{tr}_{kl}\left(-\frac{2}{a}(1+\omega)k^{tr}_{im}k^{trm}_{j}\nonumber+\left(a^{1-n\gamma_{ad}}\frac{\omega\gamma_{ad}C_{\rho}\nonumber+N(\gamma_{ad}-2)\delta\rho}{n-1}+\frac{\omega}{a(n-1)}\right)g_{ij}\right.\\
\left.+\nonumber\frac{1}{2a}g^{mn}(\nabla[\gamma]_{i}u_{jm}+\nabla[\gamma]_{j}u_{im}-\nabla[\gamma]_{m}u_{ij})\nabla_{n}\omega-\frac{\dot{a}}{a}(n-2)\omega k^{tr}_{ij}\right.\\\nonumber 
\left.-\gamma_{ad}a^{1-n\gamma_{ad}}(1+\omega)(C_{\rho}+\delta\rho)v_{i}v_{j}+\frac{1}{a}k^{tr}_{im}\nabla[\gamma]_{j}X^{m}+\frac{1}{a}k^{tr}_{mj}\nabla[\gamma]_{i}X^{m}\right.\\\nonumber 
\left.+\frac{1}{a}(1+\omega)\mathfrak{J}_{ij}\right)\mu_{g}|\\\nonumber 
\leq C\left(\frac{1}{a}(1+||\omega||_{H^{s+1}})||k^{tr}||^{3}_{H^{s-1}}+a^{1-n\gamma_{ad}}||\omega||_{H^{s+1}}||k^{tr}||_{H^{s-1}}+a^{1-n\gamma_{ad}}||\delta\rho||_{H^{s-1}}\right.\\\nonumber 
\left.||k^{tr}||_{H^{s-1}}+\frac{1}{a}||\omega||_{H^{s+1}}||k^{tr}||_{H^{s-1}}+\frac{1}{a}||u||_{H^{s}}||\omega||_{H^{s+1}}||k^{tr}||_{H^{s-1}}+\frac{\dot{a}}{a}||\omega||_{H^{s+1}}||k^{tr}||_{H^{s-1}}\right.\\\nonumber 
\left.+a^{1-n\gamma_{ad}}(1+||\omega||_{H^{s+1}})(1+||\delta\rho||_{H^{s-1}})||v||^{2}_{H^{s-1}}||k^{tr}||_{H^{s-1}}+\frac{1}{a}||X||_{H^{s+1}}||k^{tr}||^{2}_{H^{s-1}}\right.\\\nonumber 
\left.+\frac{1}{a}(1+||\omega||_{H^{s+1}})||u||^{2}_{H^{s}}||k^{tr}||_{H^{s-1}}\right).
\end{eqnarray}

Collecting every individual terms, $I$ computed as follows
\begin{eqnarray}
I=-\frac{(n-1)\dot{a}}{a}\langle k^{tr}|k^{tr}\rangle_{L^{2}}+I_{er},
\end{eqnarray}
where $I_{er}$ is estimated as (we keep track of the factor $1/a$)
\begin{eqnarray}
|I_{er}|\leq C\left(\frac{1}{a}||u||^{2}_{H^{s}}||k^{tr}||_{H^{s-1}}\nonumber+\frac{1}{a}||X||_{H^{s+1}}(1+||u||_{H^{s}})||u||^{2}_{H^{s}}\right.\\\nonumber 
\left.+\frac{1}{a}||\omega||_{H^{s+1}}||k^{tr}||_{H^{s-1}}+\frac{1}{a}||X||_{H^{s+1}}||k^{tr}||^{2}_{H^{s-1}}+\frac{\dot{a}}{a}||\omega||_{H^{s+1}}||u||_{H^{s}}\right.\\\nonumber 
\left.+||Y^{||}+\frac{1}{a}X||_{H^{s+1}}||u||_{H^{s}}+\frac{1}{a}||X||_{H^{s+1}}||u||^{2}_{H^{s}}+\frac{1}{a}(1+||\omega||_{H^{s+1}})||k^{tr}||^{3}_{H^{s-1}}\right.\\\nonumber \left.+a^{1-n\gamma_{ad}}||\omega||_{H^{s+1}}||k^{tr}||_{H^{s-1}}+a^{1-n\gamma_{ad}}||\delta\rho||_{H^{s-1}}||k^{tr}||_{H^{s-1}}\right.\\\nonumber 
\left.+\frac{1}{a}||u||_{H^{s}}||\omega||_{H^{s+1}}||k^{tr}||_{H^{s-1}}+\frac{\dot{a}}{a}||\omega||_{H^{s+1}}||k^{tr}||_{H^{s-1}}+\frac{\dot{a}}{a}||\omega||_{H^{s+1}}||u||^{2}_{H^{s-1}}\right.\\\nonumber 
\left.+a^{1-n\gamma_{ad}}(1+||\omega||_{H^{s+1}})(1+||\delta\rho||_{H^{s-1}})||v||^{2}_{H^{s-1}}||k^{tr}||_{H^{s-1}}+\frac{1}{a}||X||_{H^{s+1}}||k^{tr}||^{2}_{H^{s-1}}\right.\\\nonumber 
\left.+\frac{1}{a}(1+||\omega||_{H^{s+1}})||u||^{2}_{H^{s}}||k^{tr}||_{H^{s-1}}\right).
\end{eqnarray}\\
\textbf{Estimating terms of type II,III, and IV:} The term $II$ reads 
\begin{eqnarray}
II=\frac{1}{8}\int_{\Sigma}u_{ij}[\partial_{t},\mathcal{L}_{g,\gamma}]u_{kl}\gamma^{ik}\gamma^{jl}\mu_{g}
\end{eqnarray}
We will control this term as follows 
\begin{eqnarray}
II=\frac{1}{8}\int_{\Sigma}u_{ij}[\partial_{t},\mathcal{L}_{g,\gamma}]u_{kl}\gamma^{ik}\gamma^{jl}\mu_{g}\\\nonumber 
=\frac{1}{8}\int_{\Sigma}u_{ij}\partial_{t}(\mathcal{L}_{g,\gamma}u_{kl})\gamma^{ik}\gamma^{jl}\mu_{g}-\frac{1}{8}\int_{\Sigma}u_{ij}\mathcal{L}_{g,\gamma}\partial_{t}u_{kl} \gamma^{ik}\gamma^{jl}\mu_{g}\\\nonumber 
=\frac{1}{8}\frac{d}{dt}\int_{\Sigma}u_{ij}\mathcal{L}_{g,\gamma}u_{kl}\gamma^{ik}\gamma^{jl}\mu_{g}-\frac{1}{8}\int_{\Sigma}(\partial_{t}u_{ij})\mathcal{L}_{g,\gamma}u_{kl}\gamma^{ik}\gamma^{jl}\mu_{g}\\\nonumber 
-\frac{1}{8}\int_{\Sigma}u_{ij}\mathcal{L}_{g,\gamma}u_{kl}\partial_{t}(\gamma^{ik}\gamma^{jl}\mu_{g})-\frac{1}{8}\int_{\Sigma}u_{ij}\mathcal{L}_{g,\gamma}\partial_{t}u_{kl} \gamma^{ik}\gamma^{jl}\mu_{g}\\\nonumber 
=\frac{1}{8}\frac{d}{dt}\int_{\Sigma}u_{ij}\mathcal{L}_{g,\gamma}u_{kl}\gamma^{ik}\gamma^{jl}\mu_{g}-\frac{1}{8}\int_{\Sigma}u_{ij}\mathcal{L}_{g,\gamma}u_{kl}\partial_{t}(\gamma^{ik}\gamma^{jl}\mu_{g})\\\nonumber 
-\frac{1}{4}\int_{\Sigma}u_{ij}\mathcal{L}_{g,\gamma}\partial_{t}u_{kl} \gamma^{ik}\gamma^{jl}\mu_{g},
\end{eqnarray}
where we have used the self-adjoint property of $\mathcal{L}_{g,\gamma}$. Now we explicitly write down $\mathcal{L}_{g,\gamma}$ and obtain 
\begin{eqnarray}
II=\frac{1}{8}\frac{d}{dt}\int_{\Sigma}u_{ij}(\Delta^{\gamma}_{g}u_{kl}-2R[\gamma]_{kplq}u^{pq})\gamma^{ik}\gamma^{jl}\mu_{g}-\frac{1}{8}\int_{\Sigma}u_{ij}\mathcal{L}_{g,\gamma}u_{kl}\partial_{t}(\gamma^{ik}\gamma^{jl}\mu_{g})\\\nonumber 
-\frac{1}{4}\int_{\Sigma}u_{ij}\mathcal{L}_{g,\gamma}\partial_{t}u_{kl} \gamma^{ik}\gamma^{jl}\mu_{g}\\\nonumber 
=\frac{1}{8}\frac{d}{dt}\int_{\Sigma}g^{mn}\nabla[\gamma]_{m}u_{ij}\nabla[\gamma]_{n}u_{kl}\gamma^{ik}\gamma^{jl}\mu_{g}+\frac{1}{4}\int_{\Sigma}\partial_{t}(u_{ij}R[\gamma]_{kplq}u^{pq}\gamma^{ik}\gamma^{jl}\mu_{g})\\\nonumber 
-\frac{1}{8}\int_{\Sigma}u_{ij}\mathcal{L}_{g,\gamma}u_{kl}\partial_{t}(\gamma^{ik}\gamma^{jl}\mu_{g})-\frac{1}{4}\int_{\Sigma}u_{ij}\mathcal{L}_{g,\gamma}\partial_{t}u_{kl} \gamma^{ik}\gamma^{jl}\mu_{g}\\\nonumber 
=\frac{1}{8}\int_{\Sigma}(\partial_{t}g^{mn})\nabla[\gamma]_{m}u_{ij}\nabla[\gamma]_{n}u_{kl}\gamma^{ik}\gamma^{jl}\mu_{g}+\frac{1}{4}\int_{\Sigma}g^{mn}\nabla[\gamma]_{m}\partial_{t}u_{ij}\nabla[\gamma]_{n}u_{kl}\gamma^{ik}\gamma^{jl}\mu_{g}\\\nonumber+\frac{1}{4}\int_{\Sigma}g^{mn}[\partial_{t},\nabla[\gamma]_{m}]u_{ij}\nabla[\gamma]_{n}u_{kl}\gamma^{ik}\gamma^{jl}\mu_{g}+\frac{1}{2}\int_{\Sigma}u_{ij}R[\gamma]_{kplq}\partial_{t}u^{pq}\gamma^{ik}\gamma^{jl}\mu_{g}\\\nonumber 
+\frac{1}{4}\int_{\Sigma}u_{ij}\partial_{t}R[\gamma]_{kplq}u^{pq}\gamma^{ik}\gamma^{jl}\mu_{g}+\frac{1}{4}\int_{\Sigma}u_{ij}R[\gamma]_{kplq}u^{pq}\partial_{t}(\gamma^{ik}\gamma^{jl}\mu_{g})\\\nonumber -\frac{1}{8}\int_{\Sigma}u_{ij}\mathcal{L}_{g,\gamma}u_{kl}\partial_{t}(\gamma^{ik}\gamma^{jl}\mu_{g})-\frac{1}{4}\int_{\Sigma}u_{ij}\mathcal{L}_{g,\gamma}\partial_{t}u_{kl}\\\nonumber 
=\frac{1}{8}\int_{\Sigma}(\partial_{t}g^{mn})\nabla[\gamma]_{m}u_{ij}\nabla[\gamma]_{n}u_{kl}\gamma^{ik}\gamma^{jl}\mu_{g}+\frac{1}{4}\int_{\Sigma}g^{mn}[\partial_{t},\nabla[\gamma]_{m}]u_{ij}\nabla[\gamma]_{n}u_{kl}\gamma^{ik}\gamma^{jl}\mu_{g}\\\nonumber 
+\frac{1}{4}\int_{\Sigma}u_{ij}\partial_{t}R[\gamma]_{kplq}u^{pq}\gamma^{ik}\gamma^{jl}\mu_{g}+\frac{1}{4}\int_{\Sigma}u_{ij}R[\gamma]_{kplq}u^{pq}\partial_{t}(\gamma^{ik}\gamma^{jl}\mu_{g})\\\nonumber -\frac{1}{8}\int_{\Sigma}u_{ij}\mathcal{L}_{g,\gamma}u_{kl}\partial_{t}(\gamma^{ik}\gamma^{jl}\mu_{g})
\end{eqnarray}
where we have used the fact that $\int_{\Sigma}u_{ij}\Delta^{\gamma}_{g}u_{kl}\gamma^{ik}\gamma^{jl}\mu_{g}=\int_{\Sigma}g^{mn}\nabla[\gamma]_{m}u_{ij}\nabla[\gamma]_{n}u_{kl}\gamma^{ik}\gamma^{jl}\mu_{g}$. Now we combine terms $II$, $III$, and $IV$ to yield 
\begin{eqnarray}
II+III+IV=\frac{1}{8}\int_{\Sigma}(\partial_{t}g^{mn})\nabla[\gamma]_{m}u_{ij}\nabla[\gamma]_{n}u_{kl}\gamma^{ik}\gamma^{jl}\mu_{g}\\\nonumber 
+\frac{1}{4}\int_{\Sigma}g^{mn}[\partial_{t},\nabla[\gamma]_{m}]u_{ij}\nabla[\gamma]_{n}u_{kl}\gamma^{ik}\gamma^{jl}\mu_{g}+\frac{1}{4}\int_{\Sigma}u_{ij}\partial_{t}R[\gamma]_{kplq}u^{pq}\gamma^{ik}\gamma^{jl}\mu_{g}\\\nonumber 
+\int_{\Sigma}(\frac{1}{2}k^{tr}_{ij}k^{tr}_{kl}(\partial_{t}\gamma^{ik}\gamma^{jl}+\gamma^{ik}\partial_{t}\gamma^{jl})\mu_{g}+\int_{\Sigma}\frac{1}{4}k^{tr}_{ij}k^{tr}_{kl}\gamma^{ik}\gamma^{jl}\tr_{g}\partial_{t}g\mu_{g}
\end{eqnarray}
which using the evolution equations together with $H^{s_{1}}\hookrightarrow H^{s_{2}},~s_{1}>s_{2}$ leads to the following estimate 
\begin{eqnarray}
|II+III+IV|\leq\\\nonumber C\left(\left[\frac{\dot{a}}{a}||\omega||_{H^{s+1}}(1+||u||_{H^{s}})+\frac{1}{a}(1+||\omega||_{H^{s+1}})||k^{tr}||_{H^{s-1}}+\frac{1}{a}||X||_{H^{s+1}}||u||_{H^{s}}\right.\right.\\\nonumber 
\left.\left.(1+||u||_{H^{s}})\right](||u||^{2}_{H^{s}}+||k^{tr}||^{2}_{H^{s-1}})+(||u||^{2}_{H^{s}}+||k^{tr}||^{2}_{H^{s-1}})||\partial_{t}\gamma||_{H^{s-1}}\right)\\\nonumber
\leq C\left(\left[\frac{\dot{a}}{a}||\omega||_{H^{s+1}}(1+||u||_{H^{s}})+\frac{1}{a}(1+||\omega||_{H^{s+1}})||k^{tr}||_{H^{s-1}}+\frac{1}{a}||X||_{H^{s+1}}||u||_{H^{s}}\right.\right.\\\nonumber 
\left.\left.(1+||u||_{H^{s}})\right](||u||^{2}_{H^{s}}+||k^{tr}||^{2}_{H^{s-1}})\right)
\end{eqnarray}
Collecting all the terms concludes the proof of the lemma.~~~~~~~~~~~~~~~~~~~~~~~~~~~~~~~~~$\Box$

Now we will obtain estimates for the higher order energies in a similar way as for the first order energy. Only, in this case, there will be extra innocuous commutator terms which may be estimated using the inequalities listed in section $4.1$. The following lemma provides the estimate for the total energy associated with the gravitational part. 

\textbf{Lemma 12:} \textit{Let $s>\frac{n}{2}+2$ and $B_{\delta}(\mathbf{0})\subset H^{s}\times H^{s-1}\times H^{s-1}\times H^{s-1}$ be a ball for sufficiently small $\delta>0$. Also let $(u,k^{tr},\delta\rho,v)\in B_{\delta}(\mathbf{0})$ solve the re-scaled Einstein-Euler-$\Lambda$ evolution equations and $E_{Ein}$ be the total energy of the gravitational sector defined by equation (\ref{eq:energygravity}). Then $\frac{dE_{Ein}}{dt}$ satisfies the equation
\begin{eqnarray}
\frac{dE_{Ein}}{dt}=-\frac{(n-1)\dot{a}}{a}\sum_{I=1}^{s}\langle k^{tr}|\mathcal{L}^{I-1}_{g,\gamma}k^{tr}\rangle_{L^{2}}+\mathcal{ER}_{s},
\end{eqnarray}
where $\mathcal{ER}_{s}$ satisfies 
\begin{eqnarray}
\mathcal{ER}_{s}\leq C\left(\frac{1}{a}||u||^{2}_{H^{s}}||k^{tr}||_{H^{s-1}}\nonumber+\frac{1}{a}||X||_{H^{s+1}}(1+||u||_{H^{s}})||u||^{2}_{H^{s}}\right.\\\nonumber 
\left.+\frac{1}{a}||\omega||_{H^{s+1}}||k^{tr}||_{H^{s-1}}+\frac{1}{a}||X||_{H^{s+1}}||k^{tr}||^{2}_{H^{s-1}}+\frac{\dot{a}}{a}||\omega||_{H^{s+1}}||u||_{H^{s}}\right.\\\nonumber 
\left.+||Y^{||}+\frac{1}{a}X||_{H^{s+1}}||u||_{H^{s}}+\frac{1}{a}||X||_{H^{s+1}}||u||^{2}_{H^{s}}+\frac{1}{a}(1+||\omega||_{H^{s+1}})||k^{tr}||^{3}_{H^{s-1}}\right.\\\nonumber \left.+a^{1-n\gamma_{ad}}||\omega||_{H^{s+1}}||k^{tr}||_{H^{s-1}}+a^{1-n\gamma_{ad}}||\delta\rho||_{H^{s-1}}||k^{tr}||_{H^{s-1}}\right.\\\nonumber 
\left.+\frac{1}{a}||u||_{H^{s}}||\omega||_{H^{s+1}}||k^{tr}||_{H^{s-1}}+\frac{\dot{a}}{a}||\omega||_{H^{s+1}}||k^{tr}||_{H^{s-1}}+\frac{\dot{a}}{a}||\omega||_{H^{s+1}}||u||^{2}_{H^{s-1}}\right.\\\nonumber 
\left.+a^{1-n\gamma_{ad}}(1+||\omega||_{H^{s+1}})(1+||\delta\rho||_{H^{s-1}})||v||^{2}_{H^{s-1}}||k^{tr}||_{H^{s-1}}+\frac{1}{a}||X||_{H^{s+1}}||k^{tr}||^{2}_{H^{s-1}}\right.\\\nonumber 
\left.+\frac{1}{a}(1+||\omega||_{H^{s+1}})||u||^{2}_{H^{s}}||k^{tr}||_{H^{s-1}}+\left[\frac{\dot{a}}{a}||\omega||_{H^{s+1}}(1+||u||_{H^{s}})\right.\right.\\\nonumber 
\left.\left.+\frac{1}{a}(1+||\omega||_{H^{s+1}})||k^{tr}||_{H^{s-1}}+\frac{1}{a}||X||_{H^{s+1}}||u||_{H^{s}}(1+||u||_{H^{s}})\right]\right.\\\nonumber \left.(||u||^{2}_{H^{s}}+||k^{tr}||^{2}_{H^{s-1}})\right)
\end{eqnarray}
}\\
\textbf{Proof:} The proof follows in an exact similar way as that of the previous lemma (lemma 11). We use the commutator estimates noted in section $4.1$ to handle the terms that arise due to commuting $u$ with $\mathcal{L}^{I-1}_{g,\gamma},~2\leq I\leq s$ (and $k^{tr}$ by $\mathcal{L}^{I-1}_{g,\gamma}$).~~~~~~~~~~~~~~~~~~~~~~~~~~~~~~~~$\Box$ 

Now we turn our attention to estimating the fluid energy. This is the most important part and therefore we will include every detail of the calculations. Recall the following expression (\ref{eq:energyidentity}) from lemma 3
\begin{eqnarray}
\frac{dE_{fluid}}{dt}=\frac{n\dot{a}}{a}\int_{\Sigma}N(N-1)\mathcal{C}^{0}\mu_{g}+\int_{\Sigma}\left(\partial_{t}\mathcal{C}^{0}+\frac{1}{a}\nabla_{i}\mathcal{C}^{i}\nonumber+(\frac{1}{a}\nabla_{i}X^{i}+\frac{1}{N}\partial_{t}N)\mathcal{C}^{0}\right.\\\nonumber 
\left.+\frac{1}{aN}\mathcal{C}^{i}\nabla_{i}N\right)N\mu_{g}\\\nonumber 
=\int_{\Sigma}\left(\partial_{t}\mathcal{C}^{0}+\frac{1}{a}\nabla_{i}\mathcal{C}^{i}\right)N\mu_{g}+\mathcal{ER}_{fluid1},
\end{eqnarray}
where $\mathcal{ER}_{fluid1}$ reads
\begin{eqnarray}
\mathcal{ER}_{fluid1}\\
=\frac{n\dot{a}}{a}\int_{\Sigma}N(N-1)\mathcal{C}^{0}\mu_{g}+\int_{\Sigma}\left(\nonumber(\frac{1}{a}\nabla_{i}X^{i}+\frac{1}{N}\partial_{t}N)\mathcal{C}^{0}+\frac{1}{aN}\mathcal{C}^{i}\nabla_{i}N\right)N\mu_{g}.\nonumber
\end{eqnarray}
We first estimate this error term $\mathcal{ER}_{fluid1}$. The following lemma provides the necessary estimate.

\textbf{Lemma 13:} \textit{Let $s>\frac{n}{2}+2$ and $B_{\delta}(\mathbf{0})\subset H^{s}\times H^{s-1}\times H^{s-1}\times H^{s-1}$ be a ball for sufficiently small $\delta>0$. Also let $(u,k^{tr},\delta\rho,v)\in B_{\delta}(\mathbf{0})$ solve the re-scaled Einstein-Euler-$\Lambda$ evolution equations. Then the following estimate holds for the fluid energy error term $\mathcal{ER}_{fluid1}$
\begin{eqnarray}
|\mathcal{ER}_{fluid1}|\leq C\left([\frac{\dot{a}}{a}||\omega||_{H^{s+1}}+\frac{1}{a}||X||_{H^{s+1}}(1+||u||_{H^{s}})\right.\\\nonumber 
\left.+(1+||\omega||_{H^{s+1}})||\partial_{t}N||_{H^{s}}]E_{fluid}+\frac{1}{a}||\omega||_{H^{s+1}}\left\{(1+||\delta\rho||_{H^{s-1}})||v||_{H^{s-1}}\right.\right.\\\nonumber
\left.\left.||\delta\rho||^{2}_{H^{s-1}}+||v||_{H^{s-1}}||\delta\rho||_{H^{s-1}}+||v||^{3}_{H^{s-1}}(1+||u||_{H^{s}})(1+||\delta\rho||_{H^{s-1}})\right\}\right)
\end{eqnarray}
}
\textbf{Proof:} Noting $N-1=\omega$ and $-\int_{\Sigma}\mathcal{C}^{\mu}n_{\mu}\mu_{g}=\int_{\Sigma}N\mathcal{C}^{0}\mu_{g}=E_{fluid}$, we have 
\begin{eqnarray}
|\mathcal{ER}_{fluid1}|\leq \left(\frac{n\dot{a}}{a}||\omega||_{L^{\infty}}+\frac{1}{a}||\nabla_{i}X^{i}||_{L^{\infty}}+||\frac{1}{N}\partial_{t}N||_{L^{\infty}}\right)E_{fluid}\\\nonumber 
+\frac{1}{a}||\nabla\omega||_{L^{\infty}}\int_{\Sigma}|\mathcal{C}|\mu_{g}.
\end{eqnarray}
Now writing $\nabla_{i}X^{i}=\nabla[\gamma]_{i}X^{i}+(\Gamma[g]^{k}_{kl}-\Gamma[\gamma]^{k}_{kl})X^{l}$, using the expression  for $\mathcal{C}^{i}$ (\ref{eq:current}) 
\begin{eqnarray}
\mathcal{C}^{i}:=\frac{(\gamma_{ad}-1)v^{i}\dot{\rho}^{2}}{\gamma_{ad}\rho}+2(\gamma_{ad}-1)\dot{v}^{i}\dot{\rho}+\gamma_{ad}\rho v^{i}(-\frac{g(v,\dot{v})^{2}}{1+g(v,v)}+g_{ij}\dot{v}^{i}\dot{v}^{j}),
\end{eqnarray}
and using Sobolev embedding, we obtain the desired estimate.~~~~~~~~~~~~~~~~~~~~~~~$\Box$

Now we will estimate $\int_{\Sigma}\left(\partial_{t}\mathcal{C}^{0}+\frac{1}{a}\nabla_{i}\mathcal{C}^{i}\right)N\mu_{g}$, the most important term in the fluid energy. The following lemma provides the necessary estimates for this term. The proof of this lemma is the longest one of this article. We will carefully keep track of each and every term and show the explicit calculations behind the estimates obtained. The most important point to note here is that there are terms that are multiplied by $\frac{\dot{a}}{a}$ which may be potentially dangerous. We will keep track of these terms very carefully. 

\textbf{Lemma 14:} \textit{Let $s>\frac{n}{2}+2$ and $B_{\delta}(\mathbf{0})\subset H^{s}\times H^{s-1}\times H^{s-1}\times H^{s-1}$ be a ball for sufficiently small $\delta>0$. Also let $(u,k^{tr},\delta\rho,v)\in B_{\delta}(\mathbf{0})$ solve the re-scaled Einstein-Euler-$\Lambda$ evolution equations. The principal terms $\int_{\Sigma}\left(\partial_{t}\mathcal{C}^{0}+\frac{1}{a}\nabla_{i}\mathcal{C}^{i}\right)N\mu_{g}$ verifies the following equation
\begin{eqnarray}
\int_{\Sigma}\left(\partial_{t}\mathcal{C}^{0}+\frac{1}{a}\nabla_{i}\mathcal{C}^{i}\right)N\mu_{g}=\nonumber-2\int_{\Sigma}\gamma_{ad}\rho Nu^{0}[2N-1-n(\gamma_{ad}-1)]\frac{\dot{a}}{a}g(\dot{v},\dot{v})\mu_{g}\\\nonumber 
+\mathcal{ER}_{fluid2},
\end{eqnarray}
where $\mathcal{ER}_{fluid2}$ satisfies the estimate 
\begin{eqnarray}
\label{eq:estimateerror1}
|\mathcal{ER}_{fluid2}|\leq C\left( \frac{1}{a}||X||_{H^{s+1}}\left[1+||u||_{H^{s}}+||v||_{H^{s-1}}\nonumber+||v||^{2}_{H{s-1}}+||v||^{2}_{H^{s-1}}||u||_{H^{s}}\right.\right.\\\nonumber \left.\left.+||\delta\rho||_{H^{s-1}}+||u||_{H^{s}}||\delta\rho||_{H^{s-1}}+||\delta\rho||_{H^{s-1}}||v||^{2}_{H^{s-1}}\right]||\delta\rho||^{2}_{H^{s-1}}+||v||_{H^{s-1}}\right.\\\nonumber 
\left.\left[||\partial_{t}\gamma||_{H^{s-1}}||v||_{H^{s-1}}+||v||^{2}_{H^{s-1}}+||\delta\rho||_{H^{s-1}}||\partial_{t}v||_{H^{s-2}}+\frac{1}{a}||k^{tr}||_{H^{s-1}}||v||_{H^{s-1}}\right.\right.\\\nonumber 
\left.\left.+\frac{1}{a}||X||_{H^{s+1}}||v||_{H^{s-1}}+\frac{1}{a}||v||^{2}_{H^{s-1}}+\frac{1}{a}||\omega||_{H^{s+1}}+\frac{1}{a}||\delta\rho||_{H^{s-1}}\right.\right.\\\nonumber 
\left.\left.+\frac{1}{a}||\delta\rho||_{H^{s-1}}||v||^{2}_{H^{s-1}}+\frac{1}{a}||\delta\rho||_{H^{s-1}}||u||_{H^{s}}+||\partial_{t}\gamma||_{H^{s-1}}||v||_{H^{s-1}}||\delta\rho||_{H^{s-1}}\right.\right.\\\nonumber 
\left.\left. +||\partial_{t}\gamma||_{H^{s-1}}||v||_{H^{s-1}}+\frac{1}{a}||v||_{H^{s-1}}||\delta\rho||_{H^{s-1}}||X||_{H^{s+1}}\right]\right.\\\nonumber 
\left. +\frac{\dot{a}}{a}||v||^{4}_{H^{s-1}}+||\delta\rho||_{H^{s-1}}\left[||\partial_{t}\gamma||_{H^{s-1}}||\delta\rho||_{H^{s-1}}+\frac{||v||_{H^{s-1}}}{a}(||\delta\rho||_{H^{s-1}}+||\omega||_{H^{s+1}}\right.\right.\\\nonumber 
\left.\left.+||v||^{2}_{H^{s-1}})+\frac{1}{a}||X||_{H^{s+1}}||\delta\rho||_{H^{s-1}}+||v||^{2}_{H^{s-1}}||\partial_{t}\gamma||_{H^{s-1}}\right.\right.\\\nonumber \left.\left.+||\partial_{t}v||_{H^{s-2}}(||\delta\rho||_{H^{s-1}}||v||_{H^{s-1}}+||u||_{H^{s}}||v||_{H^{s-1}}+||v||_{H^{s-1}})\right.\right.\\\nonumber \left.\left.+\frac{1}{a}||v||_{H^{s-1}}||\delta\rho||_{H^{s-1}}+\frac{1}{a}||X||_{H^{s+1}}||v||^{2}_{H^{s-1}}+\frac{1}{a}||k^{tr}||_{H^{s-1}}||v||^{2}_{H^{s-1}}+\frac{\dot{a}}{a}||v||^{2}_{H^{s-1}}\right.\right.\\\nonumber 
\left.\left.+\frac{1}{a}||v||_{H^{s-1}}||\omega||_{H^{s+1}}+\frac{\dot{a}}{a}||\omega||_{H^{s+1}}\right]\right.\\\nonumber 
\left.+[||\partial_{t}g||_{H^{s-1}}||v||^{2}_{H^{s-1}}+||v||_{H^{s-1}}||\partial_{t}v||_{H^{s-2}}+||\partial_{t}N||_{H^{s}}+||\partial_{t}\delta\rho||_{H^{s-2}}]||\delta\rho||^{2}_{H^{s-1}}\right.\\\nonumber 
\left. +[||\partial_{t}N||_{H^{s}}||v||_{H^{s-1}}+||v||^{2}_{H^{s-1}}||\partial_{t}g||_{H^{s-1}}+||v||_{H^{s-1}}||\partial_{t}v||_{H^{s-2}}]||\delta\rho||_{H^{s-1}}||v||_{H^{s-1}}\right.\\\nonumber 
\left.+||v||^{2}_{H^{s-1}}[||\partial_{t}\delta\rho||_{H^{s-2}}+||\partial_{t}g||_{H^{s-1}}||v||^{2}_{H^{s-1}}+||v||_{H^{s-1}}||\partial_{t}v||_{H^{s-2}}+||\partial_{t}N||_{H^{s}}]\right.\\\nonumber 
\left.+||\partial_{t}g||_{H^{s-1}}||v||^{2}_{H^{s-1}}+||v||^{2}_{H^{s-1}}[||\partial_{t}g||_{H^{s-1}}||v||_{H^{s-1}}+||v||_{H^{s-1}}||\partial_{t}v||_{H^{s-2}}]\right.\\\nonumber 
\left. +\frac{1}{a}\left[||v||_{H^{s-1}}||\delta\rho||^{3}_{H^{s-1}}+||v||_{H^{s-1}}(1+||u||_{H^{s}})||\delta\rho||^{2}_{H^{s-1}}+||v||^{3}_{H^{s-1}}(||\delta\rho||_{H^{s-1}}\right.\right.\\\nonumber 
\left.\left.+||u||_{H^{s}})\right]\right)
\end{eqnarray}
}\\
\textbf{Proof:} Recall from section 3, the expression (\ref{eq:reducedexpression}) for $\int_{\Sigma}\left(\partial_{t}\mathcal{C}^{0}+\frac{1}{a}\nabla_{i}\mathcal{C}^{i}\right)N\mu_{g}$
\begin{eqnarray}
\label{eq:principalfluid}
\int_{\Sigma}(\partial_{t}\mathcal{C}^{0}+\frac{1}{a}\nabla_{i}\mathcal{C}^{i})N\mu_{g}\\\nonumber
=\int_{\Sigma}\left(\underbrace{\frac{2(\gamma_{ad}-1)(1+g(v,v))^{1/2}}{a\gamma_{ad}\rho N}\dot{\rho}L_{X}\dot{\rho}}_{I}-\underbrace{\frac{2\gamma_{ad}\rho g(v,\dot{v})}{aN[1+g(v,v)]^{1/2}}g(v,L_{X}\dot{v})}_{II}\right.\\\nonumber 
\left.-\underbrace{\frac{2g(v,\dot{v})}{N[1+g(v,v)]^{1/2}}g(v,\frac{\mathcal{T}_{2}}{u^{0}})}_{IM_{2}}+\underbrace{\frac{2\gamma_{ad}\rho[1+g(v,v)]^{1/2}}{aN}g(\dot{v},L_{X}\dot{v})}_{III}+\underbrace{\frac{2[1+g(v,v)]^{1/2}}{N}g(\dot{v},\frac{\mathcal{T}_{2}}{u^{0}})}_{IM_{1}:=important~term}\right.\\\nonumber 
\left.-\frac{2(\gamma_{ad}-1)}{aN[1+g(v,v)]^{1/2}}\left(\underbrace{L_{X}(\dot{\rho}g(v,\dot{v}))}_{IV}+\dot{\rho}g(X^{k}\nabla_{k}v,\dot{v})+\dot{\rho}g(v,\dot{v}^{k}\nabla_{k}X)\right)\right.\\\nonumber 
\left.+\underbrace{\frac{2(\gamma_{ad}-1)[1+g(v,v)]^{1/2}}{\gamma_{ad}\rho N}\dot{\rho}\mathcal{T}_{1}}_{IM_{3}}\right.\\\nonumber 
\left.+\frac{\gamma_{ad}-1}{\gamma_{ad}}\left(\frac{[1+g(v,v)]^{-1/2}}{2\rho N}(\partial_{t}g(v,v)+2g(\partial_{t}v,v))-\frac{[1+g(v,v)]^{1/2}}{\rho^{2}N^{2}}[\rho\partial_{t}N+N\partial_{t}\rho]\right)\dot{\rho}^{2}\right.\\\nonumber 
\left.+2(\gamma_{ad}-1)\dot{\rho}\dot{v}^{j}\left(-\frac{\partial_{t}Ng_{ij}v^{i}}{N^{2}[1+g(v,v)]^{1/2}}-\frac{g_{ij}v^{i}}{N[1+g(v,v)]^{3/2}}[\partial_{t}g(v,v)+2g(\partial_{t}v,v)]\right.\right.\\\nonumber 
\left.\left.+\frac{1}{N[1+g(v,v)]^{1/2}}(\partial_{t}g_{ij}v^{i}+g_{ij}\partial_{t}v^{i})\right)+\gamma_{ad}\left[g(\dot{v},\dot{v})-\frac{g(v,\dot{v})^{2}}{1+g(v,v)}\right]\left(\frac{[1+g(v,v)]^{1/2}}{N}\partial_{t}\rho\right.\right.\\\nonumber 
\left.\left.+\frac{\rho[1+g(v,v)]^{-1/2}}{2N}(\partial_{t}g(v,v)+2g(\partial_{t}v,v))-\frac{\rho[1+g(v,v)]^{1/2}}{N^{2}}\partial_{t}N\right)\right.\\\nonumber 
\left.+\frac{\gamma_{ad}\rho[1+g(v,v)]^{1/2}}{N}\left[(\partial_{t}g_{ij})\dot{v}^{i}\dot{v}^{j}+\frac{g(v,\dot{v})^{2}}{[1+g(v,v)]^{2}}(\partial_{t}g(v,v)+2g(\partial_{t}v,v))\right.\right.\\\nonumber \left.\left. -\frac{g(v,\dot{v})}{1+g(v,v)}(\partial_{t}g_{ij}v^{i}\dot{v}^{j}+g_{ij}\dot{v}^{i}\partial_{t}v^{j})\right]+\frac{1}{a}\left(-\frac{\gamma_{ad}-1}{\gamma_{ad}}\frac{\dot{\rho}^{2}}{\rho^{2}}v^{i}\nabla_{i}\rho+\frac{\gamma_{ad}-1}{\gamma_{ad}}\frac{\dot{\rho}^{2}}{\rho}\nabla_{i}v^{i}\nonumber\right.\right.\\\nonumber 
\left.\left.+\gamma_{ad}[v^{i}\nabla_{i}\rho+\rho\nabla_{i}v^{i}] 
\left[-\frac{g(v,\dot{v})^{2}}{1+g(v,v)}+g(\dot{v},\dot{v})\right]+\gamma_{ad}\rho v^{i}\left(\frac{2g(v,\nabla_{i}v)g(v,\dot{v})^{2}}{[1+g(v,v)]^{2}}\right.\right.\right.\\\nonumber\left.\left.\left.-\frac{2g(v,\dot{v})g(\dot{v},\nabla_{i}v)}{1+g(v,v)}\right)\right)\right)N\mu_{g}.
\end{eqnarray}
As we mentioned previously, the terms $I$, $II$, $III$, and $IV$ contain derivatives of $\dot{\rho}$ and $\dot{v}$ in the direction of the shift vector field $X$. We will reduce these derivative terms to lower order terms by integration by parts. We proceed sequentially starting with the term $I$   
\begin{eqnarray}
I=\frac{2(\gamma_{ad}-1)}{a\gamma_{ad}}\int\frac{(1+g(v,v))^{1/2}}{\rho N}\dot{\rho}L_{X}\dot{\rho}N\mu_{g}\\\nonumber =\frac{\gamma_{ad}-1}{a\gamma_{ad}}\int\left(L_{X}(\frac{[1+g(v,v)]^{1/2}\dot{\rho}^{2}}{\rho})-\dot{\rho}^{2}L_{X}(\frac{[1+g(v,v)]^{1/2}}{\rho})\right)\mu_{g}\\\nonumber 
=\frac{\gamma_{ad}-1}{a\gamma_{ad}}\int\left(\nabla_{i}(\frac{\dot{\rho}^{2}[1+g(v,v)]^{1/2}}{\rho}X^{i})-\frac{\dot{\rho}^{2}[1+g(v,v)]^{1/2}}{\rho}\nabla_{i}X^{i}\right.\\\nonumber
\left.+\frac{\dot{\rho}^{2}[1+g(v,v)]^{1/2}}{\rho^{2}}L_{X}\rho-\frac{\dot{\rho}^{2}}{\rho}L_{X}[1+g(v,v)]^{1/2}\right)\mu_{g}\\\nonumber 
=\frac{\gamma_{ad}-1}{a\gamma_{ad}}\int\left(-\frac{\dot{\rho}^{2}[1+g(v,v)]^{1/2}}{\rho}\nabla_{i}X^{i}+\frac{\dot{\rho}^{2}[1+g(v,v)]^{1/2}}{\rho^{2}}L_{X}\rho\right.\\\nonumber
\left.-\frac{\dot{\rho}^{2}}{\rho}L_{X}[1+g(v,v)]^{1/2}\right)\mu_{g},
\end{eqnarray}
where we have used Stokes's theorem on the closed manifold $\Sigma$. Notice that the directional derivative of $\dot{\rho}$ has disappeared. Using the usual trick, we now express $\nabla_{i}X^{i}$ as follows 
\begin{eqnarray}
\label{eq:divergence}
\nabla_{i}X^{i}=\nabla[\gamma]_{i}X^{i}+(\Gamma[g]^{i}_{ik}-\Gamma[\gamma]^{i}_{ik})X^{k}\\\nonumber 
=\nabla[\gamma]_{i}X^{i}+\frac{1}{2}g^{il}\left(\nabla[\gamma]_{i}g_{lk}+\nabla[\gamma]_{k}g_{il}-\nabla[\gamma]_{l}g_{ik}\right)X^{k}\\\nonumber 
=\nabla[\gamma]_{i}X^{i}+\frac{1}{2}g^{il}\nabla[\gamma]_{k}g_{il}.
\end{eqnarray}
Substituting the expression (\ref{eq:divergence}) of $\nabla_{i}X^{i}$  together with elementary calculations yield the following 
\begin{eqnarray}
I=\frac{\gamma_{ad}-1}{a\gamma_{ad}}\int\left(-\frac{\dot{\rho}^{2}[1+g(v,v)]^{1/2}}{\rho}(\nabla[\gamma]_{i}X^{i}+\frac{1}{2}g^{il}\nabla[\gamma]_{k}g_{il})\right.\\\nonumber
\left.+\frac{\dot{\rho}^{2}[1+g(v,v)]^{1/2}}{\rho^{2}}L_{X}\rho-\frac{\dot{\rho}^{2}}{\rho}[1+g(v,v)]^{-1/2}g_{ij}X^{k}v^{j}\left(\nabla[\gamma]_{k}v^{i}\right.\right.\\\nonumber
\left.\left.+\frac{1}{2}g^{il}(\nabla[\gamma]_{k}g_{lm}+\nabla[\gamma]_{m}g_{kl}-\nabla[\gamma]_{l}g_{km})v^{m}\right)\right)\mu_{g}.
\end{eqnarray}
Utilizing inequalities (\ref{eq:product1}-\ref{eq:product}) and Sobolev inequalities for $s>\frac{n}{2}+2$, we estimate $I$ as follows 
\begin{eqnarray}
|I|\leq C\frac{\gamma_{ad}-1}{a\gamma_{ad}}\left(||\frac{[1+g(v,v)]^{1/2}}{C_{\rho}+\delta\rho}||_{L^{\infty}}(||\nabla[\gamma] X||_{L^{\infty}}+||X||_{L^{\infty}}||\nabla[\gamma] u||_{L^{\infty}})\right.\\\nonumber 
\left.+||\frac{[1+g(v,v)]^{1/2}}{(C_{\rho}+\delta\rho)^{2}}||_{L^{\infty}}||X||_{L^{\infty}}||\nabla\rho||_{L^{\infty}}+||\frac{[1+g(v,v)]^{-1/2}}{C_{\rho}+\delta\rho}||_{L^{\infty}}||X||_{L^{\infty}}||v||_{L^{\infty}}\right.\\\nonumber 
\left.(1+||u||_{L^{\infty}})(||\nabla[\gamma] v||_{L^{\infty}}+||\nabla[\gamma] u||_{L^{\infty}}||v||_{L^{\infty}}\right)E_{fluid}\\\nonumber 
\leq C\frac{\gamma_{ad}-1}{a\gamma_{ad}}\left((1+||v||^{2}_{L^{\infty}}+||u||_{L^{\infty}}||v||^{2}_{L^{\infty}})(C_{\rho}+||\delta\rho||_{L^{\infty}})(||\nabla[\gamma] X||_{L^{\infty}}\right.\\\nonumber 
\left.+||X||_{L^{\infty}}||\nabla[\gamma] u||_{L^{\infty}})+(1+||v||^{2}_{L^{\infty}}+||u||_{L^{\infty}}||v||^{2}_{L^{\infty}})(C^{2}_{\rho}+||\delta\rho||^{2}_{L^{\infty}})||X||_{L^{\infty}}||\nabla \delta\rho||_{L^{\infty }}\right.\\\nonumber 
\left.+||X||_{L^{\infty}}||v||_{L^{\infty}}(1+||u||_{L^{\infty}})(||\nabla[\gamma] v||_{L^{\infty}}+||\nabla[\gamma] u||_{L^{\infty}}||v||_{L^{\infty}})\right)E_{fluid}\\\nonumber
\leq \frac{C(\gamma_{ad}-1)}{a\gamma_{ad}}(1+||v||^{2}_{L^{\infty}}+||u||_{L^{\infty}}||v||^{2}_{L^{\infty}})\left((1+||\delta\rho||_{L^{\infty}})(||\nabla[\gamma] X||_{L^{\infty}}\right.\\\nonumber 
\left.+||X||_{L^{\infty}}||\nabla[\gamma] u||_{L^{\infty}})+(1+||\delta\rho||^{2}_{L^{\infty}})||X||_{L^{\infty}}||v||_{L^{\infty}}(1+||u||_{L^{\infty}})\right.\\\nonumber 
\left.(||\nabla[\gamma] v||_{L^{\infty}}+||\nabla[\gamma] u||_{L^{\infty}}||v||_{L^{\infty}})\right)E_{fluid}\\\nonumber
\leq\frac{C(\gamma_{ad}-1)}{a\gamma_{ad}}(1+||v||^{2}_{H^{s-1}}+||u||_{H^{s}}||v||^{2}_{H^{s-1}})\left((1+||\delta\rho||_{H^{s-1}})\right.\\\nonumber
\left.(||X||_{H^{s+1}}+||X||_{H^{s+1}}|| u||_{H^{s-1}})+(1+||\delta\rho||^{2}_{H^{s-1}})||X||_{H^{s+1}}||v||_{H^{s-1}}\right.\\\nonumber
\left.(1+||u||_{H^{s}})(||v||_{H^{s-1}}+|| u||_{H^{s}}||v||_{H^{s-1}})\right)E_{fluid}.
\end{eqnarray}
Now we need to estimate term $II$. In order to do so, first note the following calculations 
\begin{eqnarray}
g(v,\dot{v})g(v,L_{X}\dot{v})=g(v,\dot{v})g(v,X^{k}\nabla_{k}\dot{v}-\dot{v}^{k}\nabla_{k}X)\\\nonumber 
=g(v,\dot{v})X^{k}\nabla_{k}g(v,\dot{v})-g(v,\dot{v})g(v,\dot{v}^{k}\nabla_{k}X)\\\nonumber 
=\frac{1}{2}X^{k}\nabla_{k}\left((g(v,\dot{v}))^{2}\right)-g(v,\dot{v})g(v,\dot{v}^{k}\nabla_{k}X),
\end{eqnarray}
 substitution of which together with the usual integration by parts argument yield
 \begin{eqnarray}
 II=\frac{2\gamma_{ad}}{a}\int\frac{\rho g(v,\dot{v})}{[1+g(v,v)]^{1/2}}g(v,L_{X}\dot{v})\mu_{g}\\\nonumber 
=\frac{\gamma_{ad}}{a}\int \frac{X^{k}\nabla_{k}(g(v,\dot{v})^{2})}{[1+g(v,v)]^{1/2}}\mu_{g}-\frac{2\gamma_{ad}}{a}\int\frac{g(v,\dot{v})g(v,\dot{v}^{k}\nabla_{k}X)}{[1+g(v,v)]^{1/2}}\mu_{g}\\\nonumber 
=-\frac{\gamma_{ad}}{a}\int g(v,\dot{v})^{2}\nabla_{k}\left(\frac{X^{k}}{[1+g(v,v)]^{1/2}}\right)\mu_{g}-\frac{2\gamma_{ad}}{a}\int\frac{g(v,\dot{v})g(v,\dot{v}^{k}\nabla_{k}X)}{[1+g(v,v)]^{1/2}}\mu_{g}\\\nonumber
=-\frac{\gamma_{ad}}{a}\int\left( \frac{g(v,\dot{v})^{2}}{[1+g(v,v)]^{1/2}}\nabla_{k}X^{k}-\frac{g(v,\dot{v})^{2}X^{k}g(\nabla_{k}v,v)}{[1+g(v,v)]^{3/2}}\right)\mu_{g}\\\nonumber 
-\frac{2\gamma_{ad}}{a}\int\frac{g(v,\dot{v})g(v,\dot{v}^{k}\nabla_{k}X)}{[1+g(v,v)]^{1/2}}\mu_{g}\\\nonumber 
=-\frac{\gamma_{ad}}{a}\int\left( \frac{g(v,\dot{v})^{2}}{[1+g(v,v)]^{1/2}}(\nabla[\gamma]_{k}X^{k}+(\Gamma[g]^{k}_{kj}-\Gamma[\gamma]^{k}_{kj})X^{k})\right.\\\nonumber \left.-\frac{g(v,\dot{v})^{2}X^{k}g(\nabla_{k}v,v)}{[1+g(v,v)]^{3/2}}\right)\mu_{g}-\frac{2\gamma_{ad}}{a}\int\frac{g(v,\dot{v})g(v,\dot{v}^{k}\nabla_{k}X)}{[1+g(v,v)]^{1/2}}\mu_{g}.
 \end{eqnarray}
Therefore $II$ is estimated as follows 
\begin{eqnarray}
|II|\leq \frac{C\gamma_{ad}}{a}\left\{||v||^{2}_{L^{\infty}}(||\nabla[\gamma] X||_{L^{\infty}}+||\nabla[\gamma] u||_{L^{\infty}}||X||_{L^{\infty}})\nonumber+||X||_{L^{\infty}}||v||_{L^{\infty}}\right.\\\nonumber 
\left.(||\nabla[\gamma] v||_{L^{\infty}}+||\nabla[\gamma]u||_{L^{\infty}}||v||_{L^{\infty}})+||v||^{2}_{L^{\infty}}(||\nabla[\gamma]X||_{L^{\infty}}+||\nabla[\gamma]u||_{L^{\infty}}||X||_{L^{\infty}})\right\}E_{fluid}\\\nonumber 
\leq \frac{C\gamma_{ad}}{a}\left\{||v||^{2}_{H^{s-1}}(||X||_{H^{s+1}}+|| u||_{H^{s}}||X||_{X^{s+1}})\nonumber+||X||_{H^{s+1}}||v||_{H^{s-1}}\right.\\\nonumber 
\left.(|| v||_{H^{s-1}}+||u||_{H^{s-1}}||v||_{H^{s-1}})+||v||^{2}_{H^{s-1}}(||X||_{H^{s+1}}+||u||_{H^{s}}||X||_{H^{s+1}})\right\}E_{fluid}.
\end{eqnarray}
Expressing the covariant derivatives with respect to the metric $g$ in terms of those of the metric $\gamma$ i.e., $\nabla_{k}v^{i}=\nabla[\gamma]_{k}v^{i}+(\Gamma[g]^{i}_{kj}-\Gamma[\gamma]^{i}_{kj})v^{j}$, the term $III$ reads 
\begin{eqnarray}
III=-\frac{\gamma_{ad}}{a}\int_{\Sigma} g(\dot{v},\dot{v})\left(\rho\frac{X^{k}g_{ij}v^{i}(\nabla[\gamma]_{k}v^{j}\nonumber+(\Gamma[g]^{j}_{kl}-\Gamma[\gamma]^{j}_{jl})v^{l})}{[1+g(v,v)]^{1/2}}+\rho[1+g(v,v)]^{1/2}\right.\\\nonumber 
\left.(\nabla[\gamma]_{k}X^{k}+(\Gamma[g]^{k}_{kl}-\Gamma[\gamma]^{k}_{kl})X^{l})+[1+g(v,v)]^{1/2}X^{k}\nabla_{k}\rho\right)\mu_{g}\\\nonumber 
-\frac{\gamma_{ad}}{a}\int_{\Sigma}\rho[1+g(v,v)]^{1/2}g_{ij}\dot{v}^{i}\dot{v}^{k}(\nabla[\gamma]_{k}X^{j}+(\Gamma[g]^{j}_{kl}-\Gamma[\gamma]^{j}_{kl})X^{l})\mu_{g},
\end{eqnarray}
which is estimated as follows 
\begin{eqnarray}
|III|\leq \frac{C}{a}\left(||v||^{2}_{L^{\infty}}||\nabla[\gamma]u||_{L^{\infty}}||X||_{L^{\infty}}+||X||_{L^{\infty}}||v||_{L^{\infty}}||\nabla[\gamma]v||_{L^{\infty}}\right.\\\nonumber \left.+(1+||v||^{2}_{L^{\infty}}+||v||^{2}_{L^{\infty}}||u||_{L^{\infty}})(||\rho||_{L^{\infty}}(||\nabla[\gamma]X||_{L^{\infty}}+||X||_{L^{\infty}}||\nabla[\gamma]u||_{L^{\infty}})\right.\\\nonumber 
\left.+||X||_{L^{\infty}}||\nabla w||_{L^{\infty}})\right)E_{fluid}\\\nonumber 
\leq \frac{C}{a}\left(||v||^{2}_{H^{s-1}}||u||_{H^{s}}||X||_{H^{s+1}}+||X||_{H^{s+1}}||v||_{H^{s-1}}||v||_{H^{s-1}}\right.\\\nonumber \left.+(1+||v||^{2}_{H^{s-1}}+||v||^{2}_{H^{s-1}}||u||_{H^{s}})((1+||\delta\rho||_{L^{H^{s-1}}})(||X||_{H^{s+1}}+||X||_{H^{s+1}}||u||_{H^{s}})\right.\\\nonumber 
\left.+||X||_{H^{s+1}}||\delta\rho||_{H^{s-1}})\right)E_{fluid}.
\end{eqnarray}
Similarly, the term $IV$ 
\begin{eqnarray}
IV=-\int_{\Sigma}\frac{2(\gamma_{ad}-1)}{a[1+g(v,v)]^{1/2}}L_{X}(\dot{\rho}g(v,\dot{v}))\mu_{g}\\\nonumber 
=\frac{2(\gamma_{ad}-1)}{a}\int_{\Sigma}\dot{\rho}g(v,\dot{v})\nabla_{k}(\frac{X^{k}}{[1+g(v,v)]^{1/2}})\mu_{g}\\\nonumber 
=\frac{2(\gamma_{ad}-1)}{a}\int_{\Sigma}\dot{\rho}g(v,\dot{v})\left(\frac{\nabla[\gamma]_{k}X^{k}+(\Gamma[g]^{k}_{ki}-\Gamma[\gamma]^{k}_{ki})X^{i})}{[1+g(v,v)]^{1/2}}-\frac{X^{k}g(v,\nabla_{k}v)}{[1+g(v,v)]^{3/2}}\right)\mu_{g}
\end{eqnarray}
is estimated as 
\begin{eqnarray}
|IV|\leq \frac{C(\gamma_{ad}-1)}{a}||\left(\frac{\nabla[\gamma]_{k}X^{k}+(\Gamma[g]^{k}_{ki}-\Gamma[\gamma]^{k}_{ki})X^{i})}{[1+g(v,v)]^{1/2}}\nonumber-\frac{X^{k}g(v,\nabla_{k}v)}{[1+g(v,v)]^{3/2}}\right)||_{L^{\infty}}\\\nonumber \int_{\Sigma}(\dot{\rho}^{2}+g(v,\dot{v})^{2})\mu_{g}\\\nonumber 
\leq \frac{C(\gamma_{ad}-1)}{a}\left(||X||_{H^{s+1}}+||X||_{H^{s+1}}||u||_{H^{s}}+||X||_{H^{s+1}}||v||^{2}_{H^{s-1}}(1+||u||_{H^{s}})\right)\\\nonumber
(1+||v||^{2}_{H^{s-1}}(1+||u||_{H^{s}}))E_{fluid}.
\end{eqnarray}
Now we will control the remaining terms. One of the terms that is most important to us is the following 
\begin{eqnarray}
IM_{1}:=\int_{\Sigma}\frac{2[1+g(v,v)]^{1/2}}{N}g(\dot{v},\frac{\mathcal{T}_{2}}{u^{0}})N\mu_{g},
\end{eqnarray}
which upon using the normalization condition $Nu^{0}=[1+g(v,v)]^{1/2}$ yields 
\begin{eqnarray}
IM_{1}=\int_{\Sigma}\frac{2[1+g(v,v)]^{1/2}}{N}g(\dot{v},\frac{\mathcal{T}_{2}}{u^{0}})N\mu_{g}=2\int_{\Sigma}Ng(\dot{v},\mathcal{T}_{2})\mu_{g}.
\end{eqnarray}
Now recall the expression for $\mathcal{T}_{2}$ from equation (\ref{eq:T2})
\begin{eqnarray}
\mathcal{T}^{i}_{2}:=-\underbrace{\gamma_{ad}\rho u^{0}[2N-1-n(\gamma_{ad}-1)]\frac{\dot{a}}{a}\dot{v}^{i}}_{one~of~the~most~important~term:~generates~decay}\\\nonumber
+\underbrace{\gamma_{ad}[\mathcal{D}^{s-1},\rho u^{0}\partial_{t}]v^{i}}_{i_{1}}-\underbrace{\gamma_{ad}\left[\mathcal{D}^{s-1},\rho u^{0}[2N-1-n(\gamma_{ad}-1)]\right]v^{i}}_{i_{2}}+\underbrace{\frac{2\gamma_{ad}}{a}\mathcal{D}^{s-1}[\rho N u^{0}k^{tri}_{j}v^{j}]}_{i_{3}}\\\nonumber 
+\underbrace{\frac{\gamma_{ad}}{a}[\mathcal{D}^{s-1},\rho u^{0}L_{X}]v^{i}}_{i_{4}}-\underbrace{\frac{\gamma_{ad}}{a}[\mathcal{D}^{s-1},\rho v^{j}\nabla_{j}]v^{i}}_{i_{5}}+\underbrace{\frac{\gamma_{ad}}{a}\mathcal{D}^{s-1}\left[\rho[1+g(v,v)]\nabla^{i}N\right]}_{i_{6}}\\\nonumber 
+\frac{(\gamma_{ad}-1)}{a}\left\{\underbrace{[\mathcal{D}^{s-1},\nabla^{i}]\rho}_{i_{7}}+\underbrace{[\mathcal{D}^{s-1},v^{i}L_{v}]\rho}_{i_{8}}+\underbrace{a[\mathcal{D}^{s-1},u^{0}v^{i}\partial_{t}]\rho}_{i_{9}}-\underbrace{[\mathcal{D}^{s-1},u^{0}v^{i}L_{X}]\rho}_{i_{10}}\right\}\\
\label{eq:t2re}
=-\underbrace{\gamma_{ad}\rho u^{0}[2N-1-n(\gamma_{ad}-1)]\frac{\dot{a}}{a}\dot{v}^{i}}_{one~of~the~most~important~term:~generates~decay}+\mathcal{T}^{i}_{2remainder}
\end{eqnarray}
and substitute in the expression of $IM_{1}$ to yield
\begin{eqnarray}
\label{eq:firstpart}
IM_{1}=-2\int_{\Sigma}\gamma_{ad}\rho Nu^{0}[2N-1-n(\gamma_{ad}-1)]\frac{\dot{a}}{a}g(\dot{v},\dot{v})\mu_{g}\\\nonumber
+2\int_{\Sigma}Ng(\dot{v},\mathcal{T}_{2remainder})\mu_{g}.
\end{eqnarray}
This essentially yields the first part of the lemma. Now We will estimate the term $2\int_{\Sigma}Ng(\dot{v},\mathcal{T}_{2remainder})\mu_{g}$ as follows 
\begin{eqnarray}
|\int_{\Sigma}Ng(\dot{v},\mathcal{T}_{2remainder})\mu_{g}|\\\nonumber 
\leq (1+||u||_{L^{\infty}})||\dot{v}||_{L^{2}}||N\mathcal{T}_{2remainder}||_{L^{2}}\\\nonumber 
\leq C(1+||u||_{H^{s}})||\dot{v}||_{L^{2}}||N\mathcal{T}_{2remainder}||_{L^{2}}.
\end{eqnarray}
In order to obtain $||\mathcal{T}_{2remainder}||_{L^{2}}$, we use its expression (\ref{eq:t2re}) and estimate each term involved i.e., terms $i_{1}$ throuh $i_{10}$. In doing so we use the commutator estimates (\ref{eq:commutator}), the composition estimate (\ref{eq:composition}), and the Sobolev inequalities for $s>\frac{n}{2}+2$. 

\begin{eqnarray}
||i_{1}||_{L^{2}}\nonumber=||-N\gamma_{ad}[\mathcal{D}^{s-1},\rho u^{0}\partial_{t}]v^{i}||_{L^{2}}\\\nonumber
\leq ||N\rho u^{0}\sum_{l=1}^{s-1}\mathcal{D}^{s-1-l}((T_{\gamma}\mathcal{D}\cdot \partial_{t}\gamma)\mathcal{D}^{l-1}v^{i})||_{L^{2}}+||[\mathcal{D}^{s-1},(\rho u^{0}-C_{\rho})](\partial_{t}v^{i})||_{L^{2}}\\\nonumber 
\leq C\left((C_{\rho}+||\delta\rho||_{L^{\infty}})(||Nu^{0}-1||_{L^{\infty}}+1)||\partial_{t}\gamma||_{H^{s-1}}||v||_{H^{s-1}}\right.\\\nonumber \left.+\left(||\nabla[\gamma](N\rho u^{0}-C_{\rho})||_{L^{\infty}}||\partial_{t}v^{i}||_{H^{s-1}}+||N\rho u^{0}-C_{\rho}||_{H^{s-1}}||\partial_{t}v^{i}||_{L^{\infty}}\right)\right)
\end{eqnarray}
Now utilizing the Sobolev embedding and composition estimate (\ref{eq:composition}), we obtain for small data 
\begin{eqnarray}
||Nu^{0}-1||_{L^{\infty}}\leq C||Nu^{0}-1||_{H^{s-1}}\leq C||v||^{2}_{H^{s-1}}(1+||u||_{H^{s}}),
\end{eqnarray}
which yields 
\begin{eqnarray}
||i_{1}||_{L^{2}}\leq C\left((C_{\rho}+||\delta\rho||_{H^{s-1}})(||v||^{2}_{H^{s-1}}(1+||u||_{H^{s}})\nonumber+1)||\partial_{t}\gamma||_{H^{s-1}}||v||_{H^{s-1}}\right.\\\nonumber \left.+\left(||v||^{2}_{H^{s-1}}(1+||u||_{H^{s}})+||\delta\rho||_{H^{s-1}}(1+||v||^{2}_{H^{s-1}}(1+||u||_{H^{s}})) ||\partial_{t}v||_{H^{s-2}}\right)\right).
\end{eqnarray}
Similarly, we estimate
\begin{eqnarray}
\nonumber||i_{2}||_{L^{2}}\\\nonumber
=||\left([\mathcal{D}^{s-1},(\rho u^{0}-C_{\rho})\left(2(N-1)-n(\gamma_{ad}-\frac{n+1}{n})]\right)v^{i}\nonumber+[\mathcal{D}^{s-1},2C_{\rho}(N-1)]v^{i}\right)||_{L^{2}}\\\nonumber 
\leq C\left(||\rho u^{0}-C_{\rho}||_{H^{s-1}}(1+||N-1||_{H^{s-1}})+||N-1||_{H^{s-1}}\right)||v||_{H^{s-1}}\\\nonumber \leq C\left(||v||^{2}_{H^{s-1}}(1+||N-1||_{H^{s-1}})+||N-1||_{H^{s-1}}\right)||v||_{H^{s-1}},
\end{eqnarray}

\begin{eqnarray}
||i_{3}||_{L^{2}}\nonumber=\frac{1}{a}||\mathcal{D}^{s-1}(\rho(Nu^{0}-1)k^{tri}_{j}v^{j}+\rho k^{tri}_{j}v^{j})||_{L^{2}}\\\nonumber 
\leq \frac{C}{a}\left(||\rho-C_{\rho}||_{H^{s-1}}||Nu^{0}-1||_{H^{s-1}}+||\rho-C_{\rho}||_{H^{s-1}}+||Nu^{0}-1||_{H^{s-1}}+1\right)\\\nonumber(1+||u||_{H^{s-1}})||k^{tr}||_{H^{s-1}}||v||_{H^{s-1}}\\\nonumber 
\leq \frac{C}{a}\left(||\delta\rho||_{H^{s-1}}||v||^{2}_{H^{s-1}}(1+||u||_{H^{s-1}})+||\delta\rho||_{H^{s-1}}+||v||^{2}_{H^{s-1}}(1+||u||_{H^{s-1}})+1\right)\\\nonumber(1+||u||_{H^{s-1}})||k^{tr}||_{H^{s-1}}||v||_{H^{s-1}},
\end{eqnarray}

\begin{eqnarray}
||Ni_{4}||_{L^{2}}\nonumber=\frac{1}{a}||N[\mathcal{D}^{s-1},\rho u^{0}L_{X}]v^{i}||_{L^{2}}\\\nonumber 
\leq \frac{C}{a}\left(||N\rho u^{0}||_{L^{\infty}}||X^{k}[\mathcal{D}^{s-1},\nabla[\gamma]_{k}]v^{i}+[\mathcal{D}^{s-1},X^{k}]\nabla[\gamma]_{k}v^{i}-v^{k}[\mathcal{D}^{s-1},\nabla[\gamma]_{k}]X^{i}\right.\\\nonumber
\left.-[\mathcal{D}^{s-1},v^{k}]\nabla[\gamma]_{k}X^{i}||_{L^{2}}+||([\mathcal{D}^{s-1},(N\rho u^{0}-C_{\rho})](X^{k}\nabla[\gamma]_{k}v^{i}-v^{k}\nabla[\gamma]_{k}X^{i}))||_{L^{2}}\right)\\\nonumber 
\leq \frac{C}{a}\left((1+||\delta\rho||_{L^{\infty}}(1+||Nu^{0}-1||_{L^{\infty}})+||Nu^{0}-1||_{L^{\infty}})\right.\\\nonumber 
\left.
(||X||_{L^{\infty}}||v||_{H^{s-1}}+||\nabla[\gamma]X||_{L^{\infty}}||\nabla[\gamma]v||_{H^{s-2}}+||v||_{L^{\infty}}||X||_{H^{s-1}}\right.\\\nonumber 
\left.+||X||_{H^{s-1}}||\nabla[\gamma]v||_{L^{\infty}}+||v||_{H^{s-1}}||\nabla[\gamma] X||_{L^{\infty}})+\{||\nabla(N\rho u^{0}-C_{\rho})||_{L^{\infty}}||X^{k}\nabla[\gamma]_{k}v^{i}\right.\\\nonumber \left. -v^{k}\nabla[\gamma]_{k}X^{i}||_{H^{s-1}}+||N\rho u^{0}-C_{\rho}||_{H^{s-1}}||X^{k}\nabla[\gamma]_{k}v^{i}-v^{k}\nabla[\gamma]_{k}X^{i}||_{L^{\infty}}\}\right)\\\nonumber 
\leq \frac{C}{a}\left((1+||\delta\rho||_{H^{s-1}}(1+||v||^{2}_{H^{s-1}}(1+||u||_{H^{s-1}}))+||v||^{2}_{H^{s-1}}(1+||u||_{H^{s-1}}))\right.\\\nonumber \left.||X||_{H^{s+1}}||v||_{H^{s-1}}+||\delta\rho||_{H^{s-1}}(1+||v||^{2}_{H^{s-1}}(1+||u||_{H^{s-1}})+||v||^{2}_{H^{s-1}}(1+||u||_{H^{s-1}}))\right.\\\nonumber 
\left.||X||_{H^{s+1}}||v||_{H^{s-1}}\right)
\end{eqnarray}

\begin{eqnarray}
||i_{5}||_{L^{2}}\nonumber=\frac{1}{a}||[\mathcal{D}^{s-1},\rho v^{k}\nabla_{k}]v^{i}||_{L^{2}}\\\nonumber
\leq \frac{C}{a}||\left(\rho v^{k}[\mathcal{D}^{s-1},\nabla[\gamma]_{k}]v^{i}+\rho v^{k}[\mathcal{D}^{s-1},(\Gamma[g]-\Gamma[\gamma])_{k}]v^{i}+[\mathcal{D}^{s-1},\rho v^{k}]\right.\\\nonumber \left.(\nabla[\gamma]_{k}v^{i}+(\Gamma[g]^{i}_{kl}-\Gamma[\gamma]^{i}_{kl})v^{l})\right)||_{L^{2}}\\\nonumber 
\leq \frac{C}{a}\left(||\rho v||_{L^{\infty}}||v||_{H^{s-1}}+||\nabla[\gamma]\nabla[\gamma]u||_{L^{\infty}}||v||^{2}_{H^{s-1}}+||\nabla[\gamma]u||_{H^{s-1}}||v||^{2}_{L^{\infty}}\right.\\\nonumber 
\left.+||\nabla[\gamma](\rho v)||_{L^{\infty}}||\nabla[\gamma]v||_{H^{s-2}}+||\rho v||_{H^{s-1}}||\nabla v||_{L^{\infty}}+||\nabla[\gamma](\rho v)||_{L^{\infty}}||\nabla[\gamma]u||_{H^{s-2}}||v||_{L^{\infty}}\right.\\\nonumber 
\left.+||\rho v||_{H^{s-1}}||\nabla[\gamma]u||_{L^{\infty}}||v||_{L^{\infty}}\right)\\\nonumber 
\leq \frac{C}{a}\left((1+||\delta\rho||_{H^{s-1}})||v||^{2}_{H^{s-1}}+||v||^{2}_{H^{s-1}}||u|||_{H^{s}}+(1+||\delta\rho||_{H^{s-1}})||v||^{2}_{H^{s-1}}||u||_{H^{s-1}}\right),
\end{eqnarray}
\begin{eqnarray}
||i_{6}||_{L^{2}}\nonumber=\frac{1}{a}||\mathcal{D}^{s-1}\left(\rho(1+g(v,v))g^{ij}\nabla_{j}N\right)||_{L^{2}}\\\nonumber\leq \frac{C}{a}\left((1+||\delta\rho||_{H^{s-1}})(1+||v||^{2}_{H^{s-1}}(1+||u||_{H^{s-1}}))||N-1||_{H^{s}}\right),
\end{eqnarray}
\begin{eqnarray}
||i_{7}||_{L^{2}}\nonumber=\frac{1}{a}||[\mathcal{D}^{s-1},g^{ij}\nabla[\gamma]_{j}]\rho||_{L^{2}}\\\nonumber 
=\frac{1}{a}||[\mathcal{D}^{s-1},g^{ij}\nabla[\gamma]_{j}]\delta\rho||_{L^{2}}\\\nonumber 
\leq \frac{C}{a}\left(||\nabla\delta\rho||_{L^{\infty}}+||\nabla[\gamma] u||_{L^{\infty}}||\nabla\delta\rho||_{H^{s-1}}+||u||_{H^{s-1}}||\nabla\delta\rho||_{L^{\infty}}+||u||_{L^{\infty}}||\delta\rho||_{H^{s-2}}\right)\\\nonumber 
\leq \frac{C}{a}\left(||\delta\rho||_{H^{s-1}}+||u||_{H^{s}}||\delta\rho||_{H^{s-1}}\right),
\end{eqnarray}
\begin{eqnarray}
||i_{8}||_{L^{2}}\nonumber=\frac{1}{a}||[\mathcal{D}^{s-1},v^{i}L_{v}]\rho||_{L^{2}}\\\nonumber
=\frac{1}{a}||[\mathcal{D}^{s-1},v^{i}L_{v}]\delta\rho||_{L^{2}}\\\nonumber 
\leq \frac{C}{a}\left(||v||^{2}_{L^{\infty}}||\delta\rho||_{H^{s-1}}+||v\nabla[\gamma]v||_{L^{\infty}}||\nabla\delta\rho||_{H^{s-2}}+||v||^{2}_{H^{s-1}}||\nabla\delta\rho||_{L^{\infty}}\right)\\\nonumber 
\leq \frac{C}{a}||v||^{2}_{H^{s-1}}||\delta\rho||_{H^{s-1}},
\end{eqnarray}
\begin{eqnarray}
||Ni_{9}||_{L^{2}}\nonumber=||N[\mathcal{D}^{s-1},u^{0}v^{i}\partial_{t}]\delta\rho||_{L^{2}}\\\nonumber 
\leq C\left(||Nu^{0}v^{i}\sum_{l=1}^{s-1}\mathcal{D}^{s-1-l}((T_{\gamma}\mathcal{D}\cdot \partial_{t}\gamma)\mathcal{D}^{l-1}\delta\rho)||_{L^{2}}+||N[\mathcal{D}^{s-1},u^{0}v^{i}]\partial_{t}\rho||_{L^{2}}\right)\\\nonumber 
\leq C\left(||v||_{L^{\infty}}(1+||Nu^{0}-1||_{L^{\infty}})||\partial_{t}\gamma||_{H^{s-1}}||\delta \rho||_{H^{s-1}}+(1+||N-1||_{L^{\infty}})\right.\\\nonumber\left. (||\nabla[\gamma](Nu^{0}v^{i})||_{L^{\infty}}||\partial_{t}\rho||_{H^{s-2}}+||Nu^{0}v^{i}||_{H^{s-1}}||\partial_{t}\rho||_{L^{\infty}})\right)\\\nonumber 
\leq C\left(||v||_{H^{s-1}}(1+||v||^{2}_{H^{s-1}}(1+||u||_{H^{s-1}})))||\partial_{t}\gamma||_{H^{s-1}}||\delta\rho||_{H^{s-1}}\right.\\\nonumber
\left.+(1+||N-1||_{H^{s+1}})||v||_{H^{s-1}}(1+||v||^{2}_{H^{s-1}}(1+||u||_{H^{s-1}}))||\partial_{t}\rho||_{H^{s-2}}\right),
\end{eqnarray}
\begin{eqnarray}
||Ni_{10}||_{L^{2}}\nonumber=\frac{1}{a}||N[\mathcal{D}^{s-1},u^{0}v^{i}L_{X}]\rho||_{L^{2}}\\\nonumber
=\frac{1}{a}||N[\mathcal{D}^{s-1},u^{0}v^{i}L_{X}]\delta\rho||_{L^{2}}\\\nonumber 
\leq \frac{C}{a}\left(||Nu^{0}v^{i}X^{k}[\mathcal{D}^{s-1},\nabla[\gamma]_{k}]\rho||_{L^{2}}+(1+||N-1||_{L^{\infty}})||[\mathcal{D}^{s-1},u^{0}v^{i}X^{k}]\nabla[\gamma]_{k}\rho||_{L^{2}}\right)\\\nonumber 
\leq \frac{C}{a}\left(||v||_{L^{\infty}}(1+||Nu^{0}-1||_{L^{\infty}})||X||_{L^{\infty}}||\delta\rho||_{H^{s-1}}+(1+||N-1||_{L^{\infty}})\right.\\\nonumber 
\left.(||\nabla[\gamma](Nu^{0}vX)||_{L^{\infty}}||\nabla\rho||_{H^{s-2}}+||Nu^{0}vX||_{H^{s-1}}||\nabla[\gamma]\rho]||_{L^{\infty}})\right)\\\nonumber 
\leq\frac{C}{a}(1+||v||^{2}_{H^{s-1}}(1+||u||_{H^{s-1}}))||X||_{H^{s+1}}||\delta\rho||_{H^{s-1}}||v||_{H^{s-1}}.
\end{eqnarray}
The estimates of the terms $i_{1}-i_{10}$ yield the required $L^{2}$ estimates for $\mathcal{T}_{2remainder}$. Using these estimates, we may also estimate the following term that appears in the energy integral 
\begin{eqnarray}
IM_{2}=|\int_{\Sigma}\frac{2g(v,\dot{v})}{[1+g(v,v)]^{1/2}}g(v,\frac{\mathcal{T}_{2}}{u^{0}})\mu_{g}|\\\nonumber
=|\int_{\Sigma}\frac{2\gamma(v,\dot{v})\nonumber+2u(v,\dot{v})}{[1+g(v,v)]}N\left(\gamma(v,\mathcal{T}_{2})+u(v,\mathcal{T}_{2})\right)\mu_{g}|\\\nonumber 
\lesssim||v||^{2}_{L^{\infty}}\left(||\dot{v}||_{L^{2}}||N\mathcal{T}_{2}||_{L^{2}}\right)+||u||_{L^{\infty}}||v||^{2}_{L^{\infty}}\left(||\dot{v}||_{L^{2}}||N\mathcal{T}_{2}||_{L^{2}}\right)\\\nonumber
+||u||^{2}_{L^{\infty}}||v||^{2}_{L^{\infty}}\left(||\dot{v}||_{L^{2}}||N\mathcal{T}_{2}||_{L^{2}}\right)
\end{eqnarray}
and since $\mathcal{T}^{i}_{2}=-\gamma_{ad}\rho u^{0}[2N-1-n(\gamma_{ad}-1)]\frac{\dot{a}}{a}\dot{v}^{i}+\mathcal{T}^{i}_{2remainder}$, 
\begin{eqnarray}
||N\mathcal{T}_{2}||_{L^{2}}\\\nonumber 
\leq C\frac{\dot{a}}{a}(||v||_{H^{s-1}}(1+||u||_{H^{s-1}})(1+||\delta\rho||_{H^{s-1}})(1+||N-1||_{H^{s+1}})\\\nonumber +||\mathcal{T}_{2remainder}||_{L^{2}}.
\end{eqnarray}

The next term reads 
\begin{eqnarray}
\label{eq:estimateIM3}
IM_{3}=|\int_{\Sigma}\frac{[1+g(v,v)]^{1/2}}{\rho N}\dot{\rho}\nonumber\mathcal{T}_{1}N\mu_{g}|\leq ||[1+g(v,v)]^{1/2}||_{L^{\infty}}||\dot{\rho}||_{L^{2}}||\frac{\mathcal{T}_{1}}{\rho}||_{L^{2}}\\\nonumber 
=||[1+g(v,v)]^{1/2}-1+1||_{L^{\infty}}||\dot{\rho}||_{L^{2}}||\frac{\mathcal{T}_{1}}{\rho}||_{L^{2}}\\ 
\leq (1+||[1+g(v,v)]^{1/2}-1||_{L^{\infty}})||\dot{\rho}||_{L^{2}}||\frac{\mathcal{T}_{1}}{\rho}||_{L^{2}}.
\end{eqnarray}
In order to estimate $IM_{3}$, we recall the expression of $\mathcal{T}_{1}$ from equation (\ref{eq:T1})
\begin{eqnarray}
\frac{\mathcal{T}_{1}}{\rho}:=\frac{1}{\rho}\left(\underbrace{-[\mathcal{D}^{s-1},\partial_{t}]\rho}_{ii_{1}}-\underbrace{[\mathcal{D}^{s-1},\frac{\gamma_{ad}\rho N\nabla_{i}}{a[1+g(v,v)]^{1/2}}]v^{i}}_{ii_{2}}+\underbrace{\frac{1}{a}[\mathcal{D}^{s-1},L_{X}]\rho}_{ii3}\right.\\\nonumber
\left.\underbrace{-[\mathcal{D}^{s-1},\frac{\gamma_{ad}\rho g_{ij}v^{j}\partial_{t}}{1+g(v,v)}]v^{i}}_{ii4}-\underbrace{\frac{1}{a}[\mathcal{D}^{s-1},\frac{NL_{v}}{[1+g(v,v)]^{1/2}}]\rho}_{ii5}+\underbrace{\gamma_{ad}[\mathcal{D}^{s-1},\frac{\rho v_{i}L_{X}}{a[1+g(v,v)]}]v^{i}}_{ii6}\right.\\\nonumber 
\left.+\underbrace{\mathcal{D}^{s-1}\left[\frac{\gamma_{ad}\rho N}{1+g(v,v)}\left(\frac{1}{a}k^{tr}(v,v)-\frac{\dot{a}}{a}g(v,v)\right)\right]}_{ii7}-\underbrace{\mathcal{D}^{s-1}\left[\frac{\gamma_{ad}N\rho L_{v}N}{a[1+g(v,v)]^{1/2}}\right]}_{ii8}\right.\\\nonumber 
\left.\underbrace{-\mathcal{D}^{s-1}\left[\frac{(N-1)n\gamma_{ad}\dot{a}}{a}\rho\right]}_{ii9}\right).
\end{eqnarray}
We proceed to estimate $\mathcal{T}_{1}/\rho$ exactly the same way, that is, term by term
\begin{eqnarray}
||ii_{1}||_{L^{2}}\nonumber=||\frac{-[\mathcal{D}^{s-1},\partial_{t}]\rho}{\rho}||_{L^{2}}\\\nonumber 
\leq ||\frac{1}{\rho}||_{L^{\infty}}||\sum_{l=1}^{s-1}\mathcal{D}^{s-1-l}((T_{\gamma}\mathcal{D}\cdot \partial_{t}\gamma)\mathcal{D}^{l-1}\rho)||_{L^{2}}\\\nonumber 
\leq C||\frac{1}{\rho}||_{L^{\infty}}||\partial_{t}\gamma||_{H^{s-1}}||\rho||_{H^{s-1}},
\end{eqnarray}
\begin{eqnarray}
\label{eq:ii2}
ii_{2}=-\frac{1}{\rho}[\mathcal{D}^{s-1},\frac{\gamma_{ad}\rho N\nabla_{i}}{a[1+g(v,v)]^{1/2}}]v^{i}\\\nonumber 
=\frac{1}{\rho}[\mathcal{D}^{s-1},\frac{\gamma_{ad}\rho N}{a[1+g(v,v)]^{1/2}}](\nabla[\gamma]_{i}v^{i}+\Gamma[g]^{i}_{ik}-\Gamma[\gamma]^{i}_{ik})v^{k}\\\nonumber+\frac{\gamma_{ad} N}{a[1+g(v,v)]^{1/2}}\left([\mathcal{D}^{s-1},\nabla[\gamma]_{i}]v^{i}+[\mathcal{D}^{s-1},(\Gamma[g]^{\cdot}_{i\cdot}-\Gamma[\gamma]^{\cdot}_{i\cdot})]v^{i}\right)
\end{eqnarray}
We write $\frac{\rho N}{[1+g(v,v)]^{1/2}}$ as follows 
\begin{eqnarray}
\frac{\rho N}{[1+g(v,v)]^{1/2}}\nonumber=(\rho-C_{\rho}+C_{\rho})(N-1+1)\left(\frac{1}{[1+g(v,v)]^{1/2}}-1+1\right)\\\nonumber 
=\left((\rho-C_{\rho})(N-1)+(\rho-C_{\rho})+C_{\rho}(N-1)+C_{\rho}\right)\left(\frac{1}{[1+g(v,v)]^{1/2}}-1\right)\\\nonumber 
+\left((\rho-C_{\rho})(N-1)+(\rho-C_{\rho})+C_{\rho}(N-1)+C_{\rho}\right)
\end{eqnarray}
Therefore the commutator reads 
\begin{eqnarray}
\nonumber[\mathcal{D}^{s-1},\frac{\rho N}{[1+g(v,v)]^{1/2}}]\\\nonumber 
=[\mathcal{D}^{s-1},\left((\rho-C_{\rho})(N-1)+(\rho-C_{\rho})+C_{\rho}(N-1)+C_{\rho}\right)\left(\frac{1}{[1+g(v,v)]^{1/2}}-1\right)\\\nonumber 
+\left((\rho-C_{\rho})(N-1)+(\rho-C_{\rho})+C_{\rho}(N-1)+C_{\rho}\right)]\\\nonumber 
=[\mathcal{D}^{s-1},\left(\delta\rho \omega+\delta\rho+C_{\rho}\omega+C_{\rho}\right)\left(\frac{1}{[1+g(v,v)]^{1/2}}-1\right)\\\nonumber 
+\left(\delta\rho \omega+\delta\rho+C_{\rho}\omega\right)]
\end{eqnarray}
which after substituting to the expression (\ref{eq:ii2}) of $ii_{2}$ yields
\begin{eqnarray}
||ii_{2}||_{L^{2}}=\nonumber
\frac{1}{a}||\frac{1}{\rho}[\mathcal{D}^{s-1},\frac{\rho N}{[1+g(v,v)]^{1/2}}](\nabla[\gamma]_{i}v^{i}+\Gamma[g]^{i}_{ik}-\Gamma[\gamma]^{i}_{ik})v^{k}||_{L^{2}}\\\nonumber 
\leq \frac{C}{a}||\frac{1}{\rho}||_{L^{\infty}}\left(||\nabla\left(\left(\delta\rho\omega+\delta\rho+C_{\rho}\omega+C_{\rho}\right)\left(\frac{1}{[1+g(v,v)]^{1/2}}-1\right)\right.\right.\\\nonumber
\left.\left.+\left(\delta\rho\omega+\delta\rho+C_{\rho}\omega\right)\right)||_{L^{\infty}}\left(||\nabla[\gamma] v||_{H^{s-2}}+||\nabla[\gamma]u v||_{H^{s-2}}\right)\right.\\\nonumber 
\left.+||\left(\delta\rho\omega+\delta\rho+C_{\rho}\omega+C_{\rho}\right)\left(\frac{1}{[1+g(v,v)]^{1/2}}-1\right)\right.\\\nonumber 
\left.+\left(\delta\rho\omega+\delta\rho+C_{\rho}\omega\right)||_{H^{s-1}}\left(||\nabla v||_{L^{\infty}}+||\nabla u v||_{L^{\infty}}\right)\right)\leq\\\nonumber 
\frac{C}{a}||\frac{1}{\rho}||_{L^{\infty}}\left[(||\delta\rho||_{H^{s-1}}||\omega||_{H^{s+1}}+||\delta\rho||_{H^{s-}}+||\omega||_{H^{s+1}})(1+||v||^{2}_{H^{s-1}})+||v||^{2}_{H^{s-1}}\right]\\\nonumber 
\left(||v||_{H^{s-1}}+||u||_{H^{s}}||v||_{H^{s-1}}\right),
\end{eqnarray}
where we have used the commutator, composition, and product estimates (\ref{eq:commutator}-\ref{eq:product}). We continue to estimate the remaining terms
\begin{eqnarray}
||ii_{3}||_{L^{2}}\nonumber=\frac{1}{a}||\frac{[\mathcal{D}^{s-1},L_{X}]\rho}{\rho}||_{L^{2}}=\frac{1}{a}||\frac{[\mathcal{D}^{s-1},L_{X}]\delta\rho}{\rho}||_{L^{2}}\\\nonumber 
\leq \frac{1}{a}||\frac{1}{u^{0}}||_{L^{\infty}}||X^{k}[\mathcal{D}^{s-1},\nabla[\gamma]_{k}]\delta\rho\nonumber+[\mathcal{D},X^{k}]\nabla[\gamma]_{k}\delta\rho||_{L^{2}}\\\nonumber
\leq \frac{C}{a}||\frac{1}{\rho}||_{L^{\infty}}\left(||X||_{L^{\infty}}||\delta\rho||_{H^{s-1}}\nonumber+||\nabla[\gamma]X||_{L^{\infty}}||\nabla[\gamma]\delta\rho||_{H^{s-2}}+||X||_{H^{s-1}}||\nabla[\gamma]\delta\rho||_{L^{\infty}}\right),
\end{eqnarray}

\begin{eqnarray}
\label{eq:ii4}
||ii_{4}||_{L^{2}}=||\frac{[\mathcal{D}^{s-1},\frac{\gamma_{ad}\rho g_{ij}v^{j}}{1+g(v,v)}\partial_{t}]v^{i}}{\rho}||_{L^{2}}\\\nonumber 
\leq ||g_{ij}v^{j}\sum_{l=1}^{s-1}\mathcal{D}^{s-1-l}\left((T_{\gamma}\mathcal{D}\cdot \partial_{t}\gamma)\mathcal{D}^{l-1}v^{i}\right)||_{L^{2}}+||\frac{1}{\rho}||_{L^{\infty}}||[\mathcal{D}^{s-1},\frac{\gamma_{ad}\rho g_{ij}v^{j}}{1+g(v,v)}]\partial_{t}v^{i}||_{L^{2}}\\\nonumber 
\leq C||v||_{L^{\infty}}(1+||u||_{L^{\infty}})||\partial_{t}\gamma||_{C^{\infty}}||v||_{H^{s-2}}+||\frac{1}{\rho}||_{L^{\infty}}||[\mathcal{D}^{s-1},\frac{\gamma_{ad}\rho g_{ij}v^{j}}{1+g(v,v)}]\partial_{t}v^{i}||_{L^{2}}.
\end{eqnarray}
In order to estimate the last term in the previous inequality (\ref{eq:ii4}), we perform the following calculations and use the commutator, composition, and product estimates (\ref{eq:commutator}-\ref{eq:product})
\begin{eqnarray}
||[\mathcal{D}^{s-1},\frac{\gamma_{ad}\rho g_{ij}v^{j}}{1+g(v,v)}]\partial_{t}v^{i}||_{L^{2}}\\\nonumber 
=||[\mathcal{D}^{s-1},(\rho-C_{\rho})g_{ij}v^{j}(\frac{1}{1+g(v,v)}-1)\\\nonumber +(\rho-C_{\rho})g_{ij}v^{j}+C_{\rho}g_{ij}v^{j}(\frac{1}{1+g(v,v)}-1)
+C_{\rho}g_{ij}v^{j}]\partial_{t}v^{i}||_{L^{2}}\\\nonumber 
\leq ||[\mathcal{D}^{s-1},\delta\rho g_{ij}v^{j}(\frac{1}{1+g(v,v)}-1)]\partial_{t}v^{i}||_{L^{2}}+||[\mathcal{D}^{s-1},\delta\rho g_{ij}v^{j}]\partial_{t}v^{i}||_{L^{2}}\\\nonumber 
+||[\mathcal{D}^{s-1},C_{\rho}g_{ij}v^{j}(\frac{1}{1+g(v,v)}-1)]\partial_{t}v^{i}||_{L^{2}}+||[\mathcal{D}^{s-1},C_{\rho}g_{ij}v^{j}]\partial_{t}v^{i}||_{L^{2}}\\\nonumber 
\leq C\left(||\nabla[\gamma](\delta\rho g_{ij}v^{j}(\frac{1}{1+g(v,v)}-1))||_{L^{\infty}}||\partial_{t}v^{i}||_{H^{s-2}}\right.\\\nonumber\left.+||\delta\rho g_{ij}v^{j}(\frac{1}{1+g(v,v)}-1)||_{H^{s-1}}||\partial_{t}v^{i}||_{L^{\infty}}+||\nabla[\gamma](\delta\rho g_{ij}v^{j})||_{L^{\infty}}||\partial_{t}v^{i}||_{H^{s-2}}\right.\\\nonumber 
\left.+||\delta\rho g_{ij}v^{j}||_{H^{s-1}}||\partial_{t}v^{i}||_{L^{\infty}}+||\nabla[\gamma](C_{\rho}g_{ij}v^{j}(\frac{1}{1+g(v,v)}-1))||_{L^{\infty}}||\partial_{t}v^{i}||_{H^{s-2}}\right.\\\nonumber 
\left.+||C_{\rho}g_{ij}v^{j}(\frac{1}{1+g(v,v)}-1)||_{H^{I}}||\partial_{t}v^{i}||_{L^{\infty}}+||\nabla[\gamma](g_{ij}v^{j})||_{L^{\infty}}||\partial_{t}v^{i}||_{H^{s-2}}\right.\\\nonumber \left.+||g_{ij}v^{j}||_{H^{s-1}}||\partial_{t}v^{i}||_{L^{\infty}}\right)\\\nonumber 
\leq C\left[||\delta\rho||_{H^{s-1}}||v||^{3}_{H^{s-1}}||u||_{H^{s-1}}||\partial_{t}v||_{H^{s-2}}\right.\\\nonumber 
\left.+||\delta\rho||_{H^{s-1}}||v||_{H^{s-1}}(1+||u||_{H^{s-1}})||\partial_{t}v||_{L^{\infty}}+||u||_{H^{s-1}}||v||_{H^{s-1}}||v||^{2}_{H^{s-1}}\right.\\\nonumber 
\left.||\partial_{t}v||_{H^{s-2}}+||u||_{H^{s-1}}||v||_{H^{s-1}}||\partial_{t}v||_{H^{s-2}}+||v||_{H^{s-1}}(1+||u||_{H^{s-1}})||\partial_{t}v||_{L^{\infty}}\right].
\end{eqnarray}
Note that in the view of Sobolev inequalities, the term $||\partial_{t}v||_{L^{\infty}}$ is once again dominated by $||\partial_{t}v||_{H^{s-2}}$ since $s>\frac{n}{2}+2$. This completes the estimation of the term $ii_{4}$.
The last two terms $ii_{5}$ and $ii_{6}$ obey
\begin{eqnarray}
||ii_{5}||_{L^{2}}=\frac{1}{a}||\frac{[\mathcal{D}^{s-1},\frac{NL_{v}}{[1+g(v,v)]^{1/2}}]\rho}{\rho}||_{L^{2}}\\\nonumber
\leq \frac{1}{a}||\frac{1}{\rho}||_{L^{\infty}}||\left([\mathcal{D}^{s-1},Nv^{k}(\frac{1}{[1+g(v,v)]^{1/2}}-1)\nonumber+Nv^{k}]\nabla[\gamma]_{k}\delta\rho\right.\\\nonumber 
\left.\nonumber+\frac{Nv^{k}}{[1+g(v,v)]^{1/2}}[\mathcal{D}^{s-1},\nabla[\gamma]_{k}]\delta\rho\right)||_{L^{2}}\\\nonumber 
\leq \frac{C}{a}||\frac{1}{\rho}||_{L^{\infty}}\left(||\nabla[\gamma](Nv(\frac{1}{[1+g(v,v)]^{1/2}}-1))||_{L^{\infty}}||\nabla\delta\rho||_{H^{s-2}}\right.\\\nonumber 
\left.+||(Nv(\frac{1}{[1+g(v,v)]^{1/2}}-1))||_{H^{s-1}}||\nabla\delta\rho||_{L^{\infty}}+||\nabla[\gamma]((N-1)v)||_{L^{\infty}}||\nabla\delta\rho||_{H^{s-2}}\right.\\\nonumber 
\left.+||(N-1)v||_{H^{s-1}}||\nabla\delta\rho||_{L^{\infty}}+||\nabla[\gamma]v||_{L^{\infty}}||\nabla\delta\rho||_{H^{s-2}}+||v||_{H^{s-1}}||\nabla\delta\rho||_{L^{\infty}}\right.\\\nonumber 
\left.+||N||_{L^{\infty}}||v||_{L^{\infty}}||\delta\rho||_{H^{s-1}}\right)\\\nonumber 
\leq \frac{C}{a}||\frac{1}{\rho}||_{L^{\infty}}\left(||\omega||_{H^{s-1}}||v||^{3}_{H^{s-1}}||\delta\rho||_{H^{s-1}}+||v||^{3}_{H^{s-1}}(1+||\omega||_{H^{s-1}})||\delta\rho||_{H^{s-1}}\right.\\\nonumber 
\left.+||\omega||_{H^{s-1}}||v||_{H^{s-1}}||\delta\rho||_{H^{s-1}}+||v||_{H^{s-1}}||\delta\rho||_{H^{s-1}}\right)
\end{eqnarray}

\begin{eqnarray}
||ii_{6}||_{L^{2}}=\frac{1}{a}||\frac{[\mathcal{D}^{s-1},\frac{\rho g_{ij}v^{j}L_{X}}{[1+g(v,v)]}]v^{i}}{\rho}||_{L^{2}}\\\nonumber
\leq \frac{1}{a}\nonumber||\frac{g_{ij}v^{j}}{1+g(v,v)}\left\{[\mathcal{D}^{s-1},X^{k}\nabla[\gamma]_{k}]v^{i}-[\mathcal{D}^{s-1},v^{k}\nabla[\gamma]_{k}]X^{i}\right\}||_{L^{2}}\\\nonumber 
+||\frac{1}{\rho}||_{L^{\infty}}||[\mathcal{D}^{s-1},\frac{\rho g_{ij}v^{j}}{1+g(v,v)}]L_{X}v^{i}||_{L^{2}}\\\nonumber 
\leq \frac{C}{a}||v||_{L^{\infty}}(1+||u||_{L^{\infty}})\left(||X||_{L^{\infty}}||v||_{H^{s-1}}+||\nabla[\gamma]X||_{L^{\infty}}||v||_{H^{s-1}}\right.\\\nonumber 
\left.+||X||_{H^{s-1}}||\nabla[\gamma] v||_{L^{\infty}}+||v||_{L^{\infty}}||X||_{H^{s-1}}+||\nabla[\gamma]v||_{L^{\infty}}||X||_{H^{s-1}}\right.\\\nonumber 
\left.+||v||_{H^{s-1}}||\nabla[\gamma] X||_{L^{\infty}}\right)\\\nonumber 
+\frac{C}{a}||\frac{1}{\rho}||_{L^{\infty}}\left[||\nabla[\gamma](\delta\rho g_{ij}v^{j}(\frac{1}{1+g(v,v)}-1))||_{L^{\infty}}(||X||_{H^{s-2}}||v||_{H^{s-1}}\right.\\\nonumber 
\left.+||v||_{H^{s-2}}||X||_{H^{s-1}})+||(\delta\rho g_{ij}v^{j}(\frac{1}{1+g(v,v)}-1))||_{H^{s-1}}\left(||X||_{L^{\infty}}||\nabla[\gamma]v||_{L^{\infty}}\right.\right.\\\nonumber 
\left.\left.+||v||_{L^{\infty}}||\nabla[\gamma]X||_{L^{\infty}}\right)\right]\\\nonumber 
+\frac{C}{a}||\frac{1}{\rho}||_{L^{\infty}}\left(||\nabla[\gamma](\delta\rho g_{ij}v^{j})||_{L^{\infty}}(||X||_{H^{s-2}}||v||_{H^{s-1}}+||v||_{H^{s-2}}||X||_{H^{s-1}})\right.\\\nonumber 
\left.+||(\delta\rho g_{ij}v^{j})||_{H^{s-1}}\left(||X||_{L^{\infty}}||\nabla[\gamma]v||_{L^{\infty}}+||v||_{L^{\infty}}||\nabla[\gamma]X||_{L^{\infty}}\right)\right)\\\nonumber
+\frac{C}{a}||\frac{1}{\rho}||_{L^{\infty}}\left(||\nabla[\gamma](g_{ij}v^{j}(\frac{1}{1+g(v,v)}-1))||_{L^{\infty}}[||X||_{H^{s-2}}||v||_{H^{s-1}}\right.\\\nonumber 
\left.+||v||_{H^{s-2}}||X||_{H^{s-1}}]\right.\\\nonumber 
\left.+||(g_{ij}v^{j}(\frac{1}{1+g(v,v)}-1))||_{H^{s-1}}\left(||X||_{L^{\infty}}||\nabla[\gamma]v||_{L^{\infty}}+||v||_{L^{\infty}}||\nabla[\gamma]X||_{L^{\infty}}\right)\right)\\\nonumber 
+\frac{C}{a}||\frac{1}{\rho}||_{L^{\infty}}\left(||\nabla[\gamma](g_{ij}v^{j})||_{L^{\infty}}(||X||_{H^{s-2}}||v||_{H^{s-1}}+||v||_{H^{s-2}}||X||_{H^{s-1}})\right.\\\nonumber 
\left.+||(g_{ij}v^{j}||_{H^{s-1}}\left(||X||_{L^{\infty}}||\nabla[\gamma]v||_{L^{\infty}}+||v||_{L^{\infty}}||\nabla[\gamma]X||_{L^{\infty}}\right)\right)
\end{eqnarray}
which yields after simplification and application of the Sobolev inequality for $s>\frac{n}{2}+2$  
\begin{eqnarray}
||ii_{6}||_{L^{2}}\leq \frac{C}{a}||\frac{1}{\rho}||_{L^{\infty}}\left(||X||_{H^{s+1}}||v||^{2}_{H^{s-1}}(1+||v||^{2}_{H^{s-1}}))(1+||\delta\rho||_{H^{s-1}})\right.\\\nonumber 
\left.+||v||^{3}_{H^{s-1}}||X||_{H^{s+1}}\right)(1+||u||_{H^{s}}).
\end{eqnarray}
Estimating $ii_{7},ii_{8},$ and $ii_{9}$ is straightforward. Using product estimates, the Sobolev inequality, and $H^{s_{1}}\hookrightarrow H^{s_{2}}, s_{1}>s_{2}$, we obtain 
\begin{eqnarray}
||ii_{7}||_{L^{2}}=||\frac{\mathcal{D}^{s-1}\left[\frac{\gamma_{ad}\rho N}{1+g(v,v)}\left(\frac{1}{a}k^{tr}(v,v)-\frac{\dot{a}}{a}g(v,v)\right)\right]}{\rho}||_{L^{2}}\\\nonumber 
\leq C||\frac{1}{\rho}||_{L^{\infty}}\left(\frac{1}{a}||k^{tr}||_{H^{s-1}}||v||^{2}_{H^{s-1}}+\frac{\dot{a}}{a}||v||^{2}_{H^{s-1}}(1+||u||_{H^{s}})\right)\\\nonumber 
(1+||\delta\rho||_{H^{s-1}})(1+||\omega||_{H^{s+1}})(1+||v||^{2}_{H^{s-1}}(1+||u||_{H^{s}})),
\end{eqnarray}

\begin{eqnarray}
||ii_{8}||_{L^{2}}=||\frac{1}{\rho}\mathcal{D}^{s-1}\left[\frac{\gamma_{ad}N\rho L_{v}N}{a[1+g(v,v)]^{1/2}}\right]||_{L^{2}}\\\nonumber 
\leq \frac{C}{a}||\frac{1}{\rho}||_{L^{\infty}}||v||_{H^{s-1}}||\omega||_{H^{s+1}}(1+||v||^{2}_{H^{s-1}}(1+||u||_{H^{s}}))(1+||\omega||_{H^{s+1}})\\\nonumber (1+||\delta\rho||_{H^{s-1}}),
\end{eqnarray}

\begin{eqnarray}
||ii_{9}||_{L^{2}}=||-\frac{1}{\rho}\mathcal{D}^{s-1}\left[\frac{(N-1)n\gamma_{ad}\dot{a}}{a}\rho\right]||_{L^{2}}\\\nonumber 
\leq C\frac{\dot{a}}{a}||\frac{1}{\rho}||_{L^{\infty}}||\omega||_{H^{s+1}}(1+||\delta\rho||_{H^{s-1}}) 
\end{eqnarray}
Estimates of $ii_{1}-ii_{9}$ determine $IM_{3}$ in (\ref{eq:estimateIM3}). The remaining terms (apart from $I,II,III,IV,IM_{1},IM_{2},$ and $IM_{3}$) in the expression (\ref{eq:principalfluid}) of $\int_{\Sigma}(\partial_{t}\mathcal{C}^{0}+\frac{1}{q}\nabla_{i}\mathcal{C}^{i})N\mu_{g}$ are straightforward to estimate. The estimates read
\begin{eqnarray}
\frac{\gamma_{ad}-1}{\gamma_{ad}}|\int_{\Sigma}\left(\frac{[1+g(v,v)]^{-1/2}}{2\rho N}(\partial_{t}g(v,v)\nonumber+2g(\partial_{t}v,v))]\right)\dot{\rho}^{2}N\mu_{g}|\\\nonumber 
\leq C||\frac{1}{\rho}||_{L^{\infty}}\left(||\partial_{t}g(v,v)||_{L^{\infty}}+2||g(\partial_{t}v,v)||_{L^{\infty}}\right)||\delta\rho||^{2}_{H^{s-1}},
\end{eqnarray}

\begin{eqnarray}
|\int_{\Sigma}\frac{[1+g(v,v)]^{1/2}}{N\rho^{2}}\left(\rho\partial_{t}N+N\partial_{t}\rho\right)\dot{\rho}^{2}\mu_{g}|\\\nonumber 
\leq C||\sqrt{1+g(v,v)}||_{L^{\infty}}||\frac{1}{N}||_{L^{\infty}}\left(||\frac{1}{\rho}||_{L^{\infty}}||\partial_{t}N||_{L^{\infty}}+||\frac{1}{\rho^{2}}||_{L^{\infty}}||\partial_{t}\rho||_{L^{\infty}}(1+||\omega||_{L^{\infty}})\right)\\\nonumber
||\delta\rho||^{2}_{H^{s-1}},
\end{eqnarray}

\begin{eqnarray}
|\int_{\Sigma}2(\gamma_{ad}-1)\dot{\rho}\dot{v}^{j}\left(-\frac{\partial_{t}Ng_{ij}v^{i}}{N[1+g(v,v)]^{1/2}}-\frac{g_{ij}v^{i}}{[1+g(v,v)]^{3/2}}[\partial_{t}g(v,v)\nonumber+2g(\partial_{t}v,v)]\right.\\\nonumber 
\left.+\frac{1}{[1+g(v,v)]^{1/2}}(\partial_{t}g_{ij}v^{i}+g_{ij}\partial_{t}v^{i})\right)\mu_{g}|\\\nonumber 
\leq C\left(||\frac{1}{N}||_{L^{\infty}}||\partial_{t}N||_{L^{\infty}}||v||_{L^{\infty}}(1+||u||_{L^{\infty}})+||v||_{L^{\infty}}(1+||u||_{L^{\infty}})(||\partial_{t}g(v,v)||_{L^{\infty}}\right.\\\nonumber
\left.+||g(v,\partial_{t}v)||_{L^{\infty}})+(||\partial_{t}g(\cdot,v)||_{L^{\infty}}+||\partial_{t}v||_{L^{\infty}}(1+||u||_{L^{\infty}}))||v||_{H^{s-1}}||\delta\rho||_{H^{s-1}}\right),
\end{eqnarray}

\begin{eqnarray}
|\int_{\Sigma}\gamma_{ad}\left[g(\dot{v},\dot{v})-\frac{g(v,\dot{v})^{2}}{1+g(v,v)}\right]\left(\frac{[1+g(v,v)]^{1/2}}{N}\partial_{t}\rho\right.\\\nonumber 
\left.+\frac{\rho[1+g(v,v)]^{-1/2}}{2N}(\partial_{t}g(v,v)+2g(\partial_{t}v,v))-\frac{\rho[1+g(v,v)]^{1/2}}{N^{2}}\partial_{t}N\right)N\mu_{g}|\\\nonumber 
\leq C\left(||\sqrt{1+g(v,v)}||_{L^{\infty}}||\partial_{t}\rho||_{L^{\infty}}+||\rho||_{L^{\infty}}(||\partial_{t}g(v,v)||_{L^{\infty}}+||g(v,\partial_{t}v)||_{L^{\infty}})\right.\\\nonumber 
\left.+||\frac{1}{N}||_{L^{\infty}}||\rho||_{L^{\infty}}||\sqrt{1+g(v,v)}||_{L^{\infty}}||\partial_{t}N||_{L^{\infty}}\right)\left(||v||^{2}_{H^{s-1}}(1+||u||_{L^{\infty}})+||v||^{2}_{L^{\infty}}||v||^{2}_{H^{s-1}}\right.\\\nonumber 
\left.(1+||u||^{2}_{L^{\infty}})\right),
\end{eqnarray}

\begin{eqnarray}
|\int_{\Sigma}\frac{\gamma_{ad}\rho[1+g(v,v)]^{1/2}}{N}\left[(\partial_{t}g_{ij})\dot{v}^{i}\dot{v}^{j}+\frac{g(v,\dot{v})^{2}}{[1+g(v,v)]^{2}}(\partial_{t}g(v,v)\nonumber+2g(\partial_{t}v,v))\right.\\\nonumber \left. -\frac{g(v,\dot{v})}{1+g(v,v)}(\partial_{t}g_{ij}v^{i}\dot{v}^{j}+g_{ij}\dot{v}^{i}\partial_{t}v^{j})\right]N\mu_{g}|\\\nonumber 
\leq C||\sqrt{1+g(v,v)}||_{L^{\infty}}\left(||\partial_{t}g||_{L^{\infty}}+||v||^{2}_{L^{\infty}}(1+||u||^{2}_{L^{\infty}})(||\partial_{t}g(v,v)||_{L^{\infty}}+||g(\partial_{t}v,v)||_{L^{\infty}})\right.\\\nonumber 
\left.+||\partial_{t}g||_{L^{\infty}}||v||^{2}_{L^{\infty}}(1+||u||_{L^{\infty}})+||\partial_{t}v||_{L^{\infty}}(1+||u||_{L^{\infty}})||v||_{L^{\infty}}\right)||v||^{2}_{H^{s-1}},
\end{eqnarray}

\begin{eqnarray}
\frac{1}{a}|\int_{\Sigma}\left(-\frac{\gamma_{ad}-1}{\gamma_{ad}}\frac{\dot{\rho}^{2}}{\rho^{2}}v^{i}\nabla_{i}\rho+\frac{\gamma_{ad}-1}{\gamma_{ad}}\frac{\dot{\rho}^{2}}{\rho}\nabla_{i}v^{i}\nonumber+\gamma_{ad}[v^{i}\nabla_{i}\rho\right.\\\nonumber\left.+\rho\nabla_{i}v^{i}] 
\left[-\frac{g(v,\dot{v})^{2}}{1+g(v,v)}+g(\dot{v},\dot{v})\right]+\gamma_{ad}\rho v^{i}\left(\frac{2g(v,\nabla_{i}v)g(v,\dot{v})^{2}}{[1+g(v,v)]^{2}}-\frac{2g(v,\dot{v})g(\dot{v},\nabla_{i}v)}{1+g(v,v)}\right)\right)\\\nonumber N\mu_{g}|\\\nonumber 
\leq \frac{C}{a}||N||_{L^{\infty}}\left(||\frac{1}{\rho^{2}}||_{L^{\infty}}||v||_{L^{\infty}}||\nabla\delta\rho||_{L^{\infty}}+||\frac{1}{\rho}||_{L^{\infty}}(||\nabla v||_{L^{\infty}}\right.\\\nonumber 
\left.+||\nabla u||_{L^{\infty}}(1+||u||_{L^{\infty}})||v||_{L^{\infty}})\right)||\delta\rho||^{2}_{H^{s-1}}+C||N||_{L^{\infty}}\left(||v||_{L^{\infty}}||\nabla\delta\rho||_{L^{\infty}}\right.\\\nonumber \left.+||\rho||_{L^{\infty}}(||\nabla v||_{L^{\infty}}+||\nabla u||_{L^{\infty}}(1+||u||_{L^{\infty}})||v||_{L^{\infty}})\right)\\\nonumber
\left(||v||^{2}_{L^{\infty}}(1+||u||^{2}_{L^{\infty}})+(1+||u||_{L^{\infty}})\right)||v||^{2}_{H^{s-1}}\\\nonumber 
+C||N||_{L^{\infty}}||\rho||_{L^{\infty}}||v||_{L^{\infty}}\left(||v||^{2}_{L^{\infty}}(1+||u||_{L^{\infty}})\right.\\\nonumber 
\left.(||\nabla[\gamma]v||_{L^{\infty}}+||\nabla[\gamma]u||_{L^{\infty}}||v||_{L^{\infty}}(1+||u||_{L^{\infty}})+||v||_{L^{\infty}}(1+||u||_{L^{\infty}})(||\nabla[\gamma]v||_{L^{\infty}}\right.\\\nonumber
\left.+||\nabla[\gamma]u||_{L^{\infty}}||v||_{L^{\infty}}(1+||u||_{L^{\infty}}))\right)||v||^{2}_{H^{s-1}}.
\end{eqnarray}
In view of the Sobolev embedding with $s>\frac{n}{2}+2$, all of these terms are controlled by the maximum available Sobolev norms of the perturbed fields ($u,k^{tr},\delta\rho,v,\omega:=N-1,X$). In addition, assuming small data, composition and product estimates (\ref{eq:composition}-\ref{eq:product}), and Sobolev embedding, we may write the following 
\begin{eqnarray}
||\frac{1}{\rho}||_{L^{\infty}}=||\frac{1}{\rho}-\frac{1}{C_{\rho}}+\frac{1}{C_{\rho}}||_{L^{\infty}}\leq C||\rho-C_{\rho}||_{H^{s-1}}\nonumber+\frac{1}{C_{\rho}}\leq C(1+||\delta\rho||_{H^{s-1}}),\\
||\frac{1}{N}||_{L^{\infty}}=||\frac{1}{N}-1+1||_{L^{\infty}}\leq 1+C||N-1||_{H^{s+1}}=1+C||\omega||_{H^{s+1}},\\
||\sqrt{1+g(v,v)}||_{L^{\infty}}\leq 1+C||v||^{2}_{H^{s-1}}(1+||u||_{H^{s}})
\end{eqnarray}
which take care of the terms $||\frac{1}{\rho}||_{L^{\infty}},||\frac{1}{N}||_{L^{\infty}},$ and $||\sqrt{1+g(v,v)}||_{L^{\infty}}$. In addition, we also note the following estimate using the evolution equation of the metric $g$ (\ref{eq:fixed_new}) in terms of the maximum Sobolev norms of the perturbed fields 
\begin{eqnarray}
||\partial_{t}g||_{H^{s-1}}\leq C\left(\frac{\dot{a}}{a}||\omega||_{H^{s+1}}(1+||u||_{H^{s}})\nonumber+\frac{1}{a}(1+||\omega||_{H^{s+1}})||k^{tr}||_{H^{s-1}}\right.\\\nonumber+\left.\frac{1}{a}(||X||_{H^{s+1}}||u||_{H^{s}}+||X||_{H^{s+1}})\right).
\end{eqnarray}
Collecting all the terms and carefully keeping track of the scale factor $a$, the estimate (\ref{eq:estimateerror1}) follows. This completes the proof of the lemma.~~~~~~~~~~~~~~~~~~~~~$\Box$

Now we obtained all the necessary estimates we require in order to obtain an energy inequality. The next section contains the energy inequality, the boundedness of the total energy and decay of the spatial velocity field (of the fluid), and the trace-less second fundamental form. Utilizing these estimates, we will establish the geodesic completeness of the perturbed spacetimes as well. 
\subsection{\textbf{Lyapunov stability and geodesic completeness}}
In the last section, we have obtained all the necessary estimates. Using these estimates, we will obtain an inequality for the total energy. As we have mentioned earlier, due to the non-autonomous character of Einstein's equations, there are terms such as $\frac{\dot{a}}{a}$ and $\frac{1}{a}$ that appear explicitly in the estimates obtained so far ($a$ is the time-dependent scale factor). Therefore, we will find that the global existence will be critically dependent on a precise integrability property of the scale factor $a$. The following lemma provides the desired energy inequality.

\textbf{Lemma 15:} \textit{Let $s>\frac{n}{2}+2$ and $B_{\delta}(\mathbf{0})\subset H^{s}\times H^{s-1}\times H^{s-1}\times H^{s-1}$ be a ball for sufficiently small $\delta>0$. Also let $(u,k^{tr},\delta\rho,v)\in B_{\delta}(\mathbf{0})$ solve the re-scaled Einstein-Euler-$\Lambda$ evolution equations. Then the total energy (\ref{eq:totalenergy}) satisfies the following inequality 
\begin{eqnarray}
\label{eq:energyinequalityfinal}
d_{t}E_{total}\leq -\frac{\dot{a}}{a}\left[(n-1)-C(||u||_{H^{s}}+||k^{tr}||_{H^{s-1}}+||u||^{2}_{H^{s-1}}+||k^{tr}||^{2}_{H^{s-1}}\right.\\\nonumber 
\left.+||u||^{3}_{H^{s}}+||k^{tr}||^{3}_{H^{s-1}}+||\delta\rho||_{H^{s-1}}+||v||_{H^{s-1}}+||\delta\rho||^{2}_{H^{s-1}}+||v||^{2}_{H^{s-1}}\right.\\\nonumber 
\left.+||v||^{3}_{H^{s-1}})\right]\langle k^{tr}|\mathcal{L}_{g,\gamma}k^{tr}\rangle_{L^{2}}-\frac{\dot{a}}{a}\left[2\gamma_{ad}\left\{(n+1)-n\gamma_{ad}\right\}-C(||\delta\rho||_{H^{s-1}}+||v||_{H^{s-1}}\right.\\\nonumber 
\left.+||v||^{2}_{H^{s-1}}+||\delta\rho||^{2}_{H^{s-1}}+||u||^{3}_{H^{s}}+||v||^{3}_{H^{s-1}}+||\delta\rho||^{3}_{H^{s-1}})\right]||v||^{2}_{H^{s-1}}\\\nonumber 
+\frac{\dot{a}}{a}a^{2-n\gamma_{ad}}||\delta\rho||^{3}_{H^{s-1}}+\frac{\dot{a}}{a}a^{2-n\gamma_{ad}}||\delta\rho||^{2}_{H^{s-1}}||v||^{2}_{H^{s-1}}+\frac{\dot{a}}{a}a^{2-n\gamma_{ad}}||\delta\rho||_{H^{s-1}}||u||^{2}_{H^{s}}\\\nonumber
+\frac{\dot{a}}{a}a^{2-n\gamma_{ad}}||\delta\rho||_{H^{s-1}}||k^{tr}||^{2}_{H^{s-1}}+\frac{\dot{a}}{a}a^{2-n\gamma_{ad}}||v||^{2}_{H^{s-1}}||u||^{2}_{H^{s}}+\frac{\dot{a}}{a}a^{2-n\gamma_{ad}}||v||^{2}_{H^{s-1}}\\\nonumber||k^{tr}||^{2}_{H^{s-1}}
+\frac{\dot{a}}{a}a^{2-n\gamma_{ad}}||v||^{2}_{H^{s-1}}||\delta\rho||_{H^{s-1}}+\frac{\dot{a}}{a}a^{2-n\gamma_{ad}}||v||^{4}_{H^{s-1}}+a^{1-n\gamma_{ad}}||u||_{H^{s}}||v||_{H^{s-1}}\\\nonumber+a^{1-n\gamma_{ad}}||v||^{2}_{H^{s-1}}||k^{tr}||_{H^{s-1}}
||\delta\rho||^{2}_{H^{s-1}}+a^{1-n\gamma_{ad}}||v||_{H^{s-1}}||\delta\rho||^{2}_{H^{s-1}}\\\nonumber+a^{1-n\gamma_{ad}}||v||^{3}_{H^{s-1}}||\delta\rho||^{2}_{H^{s-1}}+a^{1-n\gamma_{ad}}||k^{tr}||_{H^{s-1}}||\delta\rho||_{H^{s-1}}||u||^{2}_{H^{s}}\\\nonumber +a^{1-n\gamma_{ad}}||k^{tr}||_{H^{s-1}}||v||^{2}_{H^{s-1}}||u||^{2}_{H^{s}}+\frac{1}{a}||\delta\rho||^{2}_{H^{s-1}}||k^{tr}||_{H^{s-1}}||u||_{H^{s}}+\frac{1}{a}||\delta\rho||^{2}_{H^{s-1}}\\\nonumber||k^{tr}||_{H^{s-1}}+\frac{1}{a}||k^{tr}||_{H^{s-1}}||u||^{2}_{H^{s}}+\frac{1}{a}||k^{tr}||^{3}_{H^{s-1}}+\frac{1}{a}||\delta\rho||^{2}_{H^{s-1}}||v||_{H^{s-1}}\\\nonumber+\frac{1}{a}||\delta\rho||_{H^{s-1}}||v||^{2}_{H^{s-1}}+\frac{1}{a}||v||^{2}_{H^{s-1}}||k^{tr}||^{2}_{H^{s-1}}+\frac{1}{a}||v||^{4}_{H^{s-1}}.
\end{eqnarray}
}\\
\textbf{Proof:} The proof of this energy inequality is straightforward. Recall from equation (\ref{eq:energyevolution}) of section 3 that we have
\begin{eqnarray}
\frac{dE_{total}}{dt}=\frac{dE_{Ein}}{dt}+\frac{dE_{fluid}}{dt}.
\end{eqnarray}
Now using the lemmas (4-14), we may substitute the expressions for $\frac{dE_{Ein}}{dt}$ and $\frac{dE_{fluid}}{dt}$. However, as we have discussed earlier, the presence of non-autonomous terms such as the scale factor $a(t)$ makes the analysis a bit more subtle. This is accomplished in four steps. We notice that there are four types of non-autonomous terms that arise. The first one contains the term $\frac{\dot{a}}{a}$ and is most dangerous. We collect these terms together. Secondly, the terms of type $\frac{\dot{a}}{a}a^{2-n\gamma_{ad}}$ appear, which seem to be innocuous given that certain integrability criteria hold. Next, we encounter terms of the type $a^{1-n\gamma_{ad}}$ and $\frac{1}{a}$, which once again could be innocuous provided that the scale factor satisfies suitable integrability criteria. Collecting all the terms together and using the elliptic estimates for $\omega=N-1$, $X$, $Y^{||}+\frac{1}{a}X$, $\partial_{t}N$ from lemmas (5-9), and estimates for $\partial_{t}\delta\rho$ and $\partial_{t}v$ from lemma (10), we obtain the desired differential inequality for $E_{total}$. ~~~~~~~~~~~~~~~~~~~~~~~~~~~~~~~~~~~~~~~~~~~~~~~~~~$\Box$     

Note that we have kept all the higher-order terms as well. This is done simply because the explicitly time-dependent terms such as $a$ and $\dot{a}$ appear which may cause trouble. However, the structure of Einstein's equations is such that the potentially problematic terms containing the factor $\frac{\dot{a}}{a}$ either contribute a decay factor or are of higher orders. This enables us to control these terms in the small data limit. Now let us obtain a global existence result. The following lemma provides the boundedness of the total energy.

\textbf{Lemma 16:}  \textit{Let $s>\frac{n}{2}+2$ and $B_{\epsilon}(\mathbf{0})\subset H^{s}\times H^{s-1}\times H^{s-1}\times H^{s-1}$ be a ball for sufficiently small $\epsilon>0$. Also let $(u_{0},k^{tr}_{0},\delta\rho_{0},v_{0})\in B_{\epsilon}(\mathbf{0})$ be the initial data at time $t=t_{0}\in (0,\infty)$ for the re-scaled Einstein-Euler-$\Lambda$ evolution equations. Then there exists a small $\delta=\delta(\epsilon)>0$ such that the following boundedness holds for all $t\in (t_{0},\infty)$
\begin{eqnarray}
\label{eq:bounded}
||u(t)||_{H^{s}}+||k^{tr}(t)||_{H^{s-1}}+||\delta\rho(t)||_{H^{s-1}}+||v(t)||_{H^{s-1}}< \delta
\end{eqnarray}
}
\textbf{Proof:} Recall the energy inequality
\begin{eqnarray}
d_{t}E_{total}\leq -\frac{\dot{a}}{a}\left[(n-1)-C(||u||_{H^{s}}+||k^{tr}||_{H^{s-1}}\nonumber+||u||^{2}_{H^{s-1}}+||k^{tr}||^{2}_{H^{s-1}}\right.\\\nonumber 
\left.+||u||^{3}_{H^{s}}+||k^{tr}||^{3}_{H^{s-1}}+||\delta\rho||_{H^{s-1}}+||v||_{H^{s-1}}+||\delta\rho||^{2}_{H^{s-1}}+||v||^{2}_{H^{s-1}}\right.\\\nonumber 
\left.+||v||^{3}_{H^{s-1}})\right]
\langle k^{tr}|\mathcal{L}_{g,\gamma}k^{tr}\rangle_{L^{2}}-\frac{\dot{a}}{a}\left[2\gamma_{ad}\left\{(n+1)-n\gamma_{ad}\right\}-C(||\delta\rho||_{H^{s-1}}\right.\\\nonumber 
\left.+||v||_{H^{s-1}}+||v||^{2}_{H^{s-1}}+||\delta\rho||^{2}_{H^{s-1}}+||u||^{3}_{H^{s}}+||v||^{3}_{H^{s-1}}+||\delta\rho||^{3}_{H^{s-1}})\right]||v||^{2}_{H^{s-1}}\\\nonumber 
+\dot{a}a^{1-n\gamma_{ad}}\mathcal{HE}_{1}+a^{1-n\gamma_{ad}}\mathcal{HE}_{2}+\frac{1}{a}\mathcal{HE}_{3},
\end{eqnarray}
where the terms $\mathcal{HE}_{1}$, $\mathcal{HE}_{2}$, and $\mathcal{HE}_{3}$ may be bounded by the total energy in the small data limit. Since we are assuming $\gamma_{ad}<\frac{n+1}{n}$, we set $\left\{(n+1)-n\gamma_{ad}\right\}=\beta>0$ fixed. Let $\delta>0$ be such that it satisfies $\delta<\min (\frac{(n-1)}{2C},\frac{\gamma_{ad}\beta}{C})$. Now, if we assume the small data limit i.e., 
$E_{total}< \delta^{2}$ (essentially a boot-strap assumption), then the potentially problematic terms  $-\frac{\dot{a}}{a}\left[(n-1)-C(||u||_{H^{s}}+||k^{tr}||_{H^{s-1}}\nonumber+||u||^{2}_{H^{s-1}}+||k^{tr}||^{2}_{H^{s-1}}\right.\\\nonumber 
\left.+||u||^{3}_{H^{s}}+||k^{tr}||^{3}_{H^{s-1}}+||\delta\rho||_{H^{s-1}}+||v||_{H^{s-1}}+||\delta\rho||^{2}_{H^{s-1}}+||v||^{2}_{H^{s-1}}+||v||^{3}_{H^{s-1}})\right]$ and\\ $-\frac{\dot{a}}{a}\left[2\gamma_{ad}\left\{(n+1)-n\gamma_{ad}\right\}-C(||\delta\rho||_{H^{s-1}}+||v||_{H^{s-1}}\right.\\\nonumber 
\left.+||v||^{2}_{H^{s-1}}+||\delta\rho||^{2}_{H^{s-1}}+||u||^{3}_{H^{s}}+||v||^{3}_{H^{s-1}}+||\delta\rho||^{3}_{H^{s-1}})\right]$ are both negative i.e., 
\begin{eqnarray}
-\frac{\dot{a}}{a}\left[(n-1)-C(||u||_{H^{s}}+||k^{tr}||_{H^{s-1}}\nonumber+||u||^{2}_{H^{s-1}}+||k^{tr}||^{2}_{H^{s-1}}\right.\\\nonumber 
\left.+||u||^{3}_{H^{s}}+||k^{tr}||^{3}_{H^{s-1}}+||\delta\rho||_{H^{s-1}}+||v||_{H^{s-1}}+||\delta\rho||^{2}_{H^{s-1}}+||v||^{2}_{H^{s-1}}+||v||^{3}_{H^{s-1}})\right]\\\nonumber 
<-\frac{\dot{a}}{a}\frac{n-1}{2},\\
-\frac{\dot{a}}{a}\left[2\gamma_{ad}\left\{(n+1)-n\gamma_{ad}\right\}-C(||\delta\rho||_{H^{s-1}}+||v||_{H^{s-1}}\right.\\\nonumber 
\left.+||v||^{2}_{H^{s-1}}+||\delta\rho||^{2}_{H^{s-1}}+||u||^{3}_{H^{s}}+||v||^{3}_{H^{s-1}}+||\delta\rho||^{3}_{H^{s-1}})\right]<-\frac{\dot{a}\gamma_{ad}\beta}{a}
\end{eqnarray}
Therefore we immediately obtain the following energy inequality
\begin{eqnarray}
 \frac{dE_{total}}{dt}\leq C(\dot{a}a^{1-n\gamma_{ad}}+a^{1-n\gamma_{ad}}+a^{-1})E_{total}.
\end{eqnarray}
Integration of this energy inequality yields 
 \begin{eqnarray}
E_{total}(t)\leq C E_{total}(t_{0})e^{C\left(-\frac{1}{n\gamma_{ad}-2}a^{2-n\gamma_{ad}}+\int_{t_{0}}^{t}a^{1-n\gamma_{ad}}dt^{'}+\int_{t_{0}}^{t}a^{-1}dt^{'}\right)}.
 \end{eqnarray}
 Noting that $n\gamma_{ad}-2>0$ ($n\geq 3$ and $\gamma_{ad}\in (1,\frac{n+1}{n})$), the integrals in the exponential are finite if the following integrability criteria hold
\begin{eqnarray} 
\label{eq:integrability}
\int_{t_{0}}^{\infty}a^{1-n\gamma_{ad}}dt<\infty,~\int_{t_{0}}^{\infty}a^{-1}dt<\infty.
\end{eqnarray}
In our cosmological model where a positive cosmological constant is included (\ref{eq:expansion}), both of these integrability criteria hold yielding 
\begin{eqnarray}
E_{total}(t)\leq CE_{total}(t_{0}).
\end{eqnarray}
Now in order to close the bootstrap assumption, the following must hold 
\begin{eqnarray}
E_{total}(t)\leq CE_{total}(t_{0})< \delta^{2}
\end{eqnarray}
which can be satisfied by a $\delta=\delta(\epsilon)>0$ if the initial data $(u_{0},k^{tr}_{0},\delta\rho_{0},v_{0})\in B_{\epsilon}(\mathbf{0})$ is chosen sufficiently small i.e., $\epsilon>0$ is chosen sufficiently small. This yields
\begin{eqnarray}
||u(t)||_{H^{s}}+||k^{tr}(t)||_{H^{s-1}}+||\delta\rho(t)||_{H^{s-1}}\nonumber+||v(t)||_{H^{s-1}} < \delta~\forall t\in (t_{0},\infty).~\Box
\end{eqnarray}

Once we have proven the lemma 16, we apply the local existence theorem of section 2.2 to yield a global existence result. Note that we have proved that the solution remains uniformly bounded in terms of initial data for all time or we have proved a \textit{Lyapunov} stability of FLRW cosmological models in the presence of a positive cosmological constant. On the other hand, the expectation would be that a decay estimate holds since there is a positive cosmological constant involved, which causes accelerated expansion. Since we only have decay terms that involve $k^{tr}$ and $v$ in the energy inequality (\ref{eq:energyinequalityfinal}), a natural attempt would be to construct a modified energy similar to the one defined by Andersson and Moncrief \cite{andersson2011einstein}. However, that does not seem to work in the current context due to the presence of the factor $1/a$ with the higher-order terms. As such we will not be able to obtain decay to zero behavior of all the fields involved. The following lemma provides the necessary decay estimates.

\textbf{Lemma 17:} \textit{Let $s>\frac{n}{2}+2$ and $B_{\epsilon}(\mathbf{0})\subset H^{s}\times H^{s-1}\times H^{s-1}\times H^{s-1}$ be a ball for sufficiently small $\epsilon>0$. Also let $(u_{0},k^{tr}_{0},\delta\rho_{0},v_{0})\in B_{\epsilon}(\mathbf{0})$ be the initial data at time $t=t_{0}\in (0,\infty)$ for the re-scaled Einstein-Euler-$\Lambda$ evolution equations, $\alpha:=\sqrt{\frac{2\Lambda}{n(n-1)}}$, and $\beta:=\left\{(n+1)-n\gamma_{ad}\right\}>0~fixed$. Then there exists a small $\delta=\delta(\epsilon)>0$ such that the following decay estimates hold 
\begin{eqnarray}
||g-g^{\dag}||_{H^{s-1}}\lesssim \delta e^{-2\alpha t},~||k^{tr}||_{H^{s-2}}\lesssim \delta e^{-\alpha t},~||v||_{H^{s-2}}\lesssim \delta e^{-\frac{\zeta}{2} t},\\\nonumber 
~||\rho-\rho^{'}||_{H^{s-2}}\lesssim \delta e^{-(\alpha+\frac{\zeta}{2})t}~as~t~\to\infty,
\end{eqnarray}
where $\zeta:=\min(\alpha,\beta)$, $R(g^{\dag})=-1$, and $\partial_{i}\rho^{'}\neq 0$ in general.
}\\
\textbf{Proof:} In order to proof these decay estimates, we will need to consider time evolution of the following two entities separately and use the boundedness result of the previous lemma (lemma 16) 
\begin{eqnarray}
\label{eq:EK}
E_{K}:=\frac{1}{2}\sum_{I=1}^{s-1}\langle k^{tr}|\mathcal{L}^{I-1}_{g,\gamma}k^{tr}\rangle_{L^{2}}
\end{eqnarray}
and 
\begin{eqnarray}
E_{v}:=\frac{1}{2}|| \mathcal{D}^{s-2}v||^{2}_{L^{2}}.
\end{eqnarray}
Notice that we have lowered the regularity level by one, the reason of which will be clear soon. Explicit calculations yield 
\begin{eqnarray}
\label{eq:EKdiff}
\frac{dE_{K}}{dt}\\\nonumber 
\leq -\frac{\dot{a}}{a}\left[(n-1)-C(||k^{tr}||_{H^{s-2}}\nonumber+||k^{tr}||^{2}_{H^{s-2}})\right]\sum_{I=1}^{s-1}\langle k^{tr}|\mathcal{L}^{I-1}_{g,\gamma}k^{tr}\rangle_{L^{2}}\\\nonumber+\frac{C}{a}||k^{tr}||_{H^{s-2}}(||u||_{H^{s}}+||u||^{2}_{H^{s}}+||k^{tr}||^{2}_{H^{s-2}}+a^{2-n\gamma_{ad}}||v||^{2}_{H^{s-2}}\\\nonumber+a^{2-n\gamma_{ad}}||\delta\rho||_{H^{s-2}}), 
\end{eqnarray}
where we notice that the term $\frac{1}{a}\langle k^{tr}|\mathcal{L}^{s-1}_{g,\gamma}u\rangle$ does not cancel out due to the absence of the term $\langle u|\mathcal{L}^{s-1}_{g,\gamma}u\rangle$ in the expression of $E_{k}$ (\ref{eq:EK}). Therefore we need to estimate the term $\frac{1}{a}\langle k^{tr}|\mathcal{L}^{s-1}_{g,\gamma}u\rangle$ in a brute force way. We distribute $2(s-1)$ derivatives among $u$ and $k^{tr}$ by integration by parts in the following way: put $s-2$ derivatives on $k^{tr}$ and so that the remaining $s$ derivatives are left to act on $u$. Therefore this term is dominated by $\frac{1}{a}||k^{tr}||_{H^{s-2}}||u||_{H^{s}}$ allowing us to control the time derivative of $E_{K}$. In view of the boundedness of the total energy (lemma 16), and $a\sim e^{\alpha t},t\to\infty,~\alpha:=\sqrt{\frac{2\Lambda}{n(n-1)}}$ (equation (\ref{eq:expansion}) of section 2.2), the differential inequality for $E_{K}$ (\ref{eq:EKdiff}) may be written as follows 
\begin{eqnarray}
\frac{dE_{K}}{dt}\leq -2\alpha [(n-1)-C\delta]E_{K}+Ce^{-\alpha t}\delta^{2}.
\end{eqnarray}
Now in the view of the smallness of $\delta$ i.e., $\delta<\frac{n-1}{2C}$, the previous inequality takes the following form 
\begin{eqnarray}
\frac{dE_{K}}{dt}\leq -(n-1)\alpha E_{K}+Ce^{-\alpha t}\delta^{2}
\end{eqnarray}
integration of which yields 
\begin{eqnarray}
E_{k}\lesssim \delta^{2}e^{-\alpha t}
\end{eqnarray}
since $n-1\geq 2$. This yields the following estimate for $||k^{tr}||_{H^{s-2}}$
\begin{eqnarray}
\label{eq:rough}
||k^{tr}||_{H^{s-2}}\lesssim \delta e^{-\frac{\alpha t}{2}}.
\end{eqnarray}
In fact we may improve this estimate by substituting (\ref{eq:rough}) in the inequality (\ref{eq:EKdiff})
\begin{eqnarray}
\frac{dE_{K}}{dt}\leq -2\alpha (n-1)E_{K}+C\delta^{2}e^{-\alpha(1+\frac{1}{2})t}
\end{eqnarray}
integration of which yields 
\begin{eqnarray}
E_{K}\lesssim \delta^{2}e^{-\alpha(1+\frac{1}{2})t}.
\end{eqnarray}
Continuing in the similar fashion (estimate $||k^{tr}||_{H^{s-2}}$ from $E_{K}$ after each step and substitute it in the inequality (\ref{eq:EKdiff})) we obtain the following final (saturated) decay estimate of $E_{K}$
\begin{eqnarray}
E_{k}\lesssim \delta^{2}e^{-\alpha\left(1+\frac{1}{2}(1+\frac{1}{2}(1+\cdot\cdot\cdot\cdot\right)t}=\delta^{2}e^{-2\alpha t}
\end{eqnarray}
which yields 
\begin{eqnarray}
||k^{tr}||_{H^{s-2}}\lesssim \delta e^{-\alpha t}.
\end{eqnarray}
Utilizing this decay estimate for $||k^{tr}||_{H^{s-2}}$, the boundedness result (\ref{eq:bounded}) and the elliptic estimates (\ref{eq:lapsepert},\ref{eq:shiftpert}), we obtain the following decay estimates for the perturbations to the lapse function and the shift vector field
\begin{eqnarray}
||\omega||_{H^{s}}\lesssim \delta^{2}e^{-2\alpha t}+\delta e^{-(n\gamma_{ad}-2)\alpha t},\\
||X||\lesssim \delta e^{-\alpha t}+\delta e^{-(n\gamma_{ad}-2)\alpha t}~as~t\to\infty.
\end{eqnarray}
Integration of the evolution equation for the metric 
\begin{eqnarray}
\frac{\partial g_{ij}}{\partial t}&=&2\frac{\dot{a}}{a}\left(N-1\right)g_{ij}-\frac{2}{a}Nk^{tr}_{ij}+\frac{1}{a}(L_{X}g)_{ij}
\end{eqnarray}
yields 
\begin{eqnarray}
||g^{\dag}-g(t)||_{H^{s-1}}\leq \int_{t}^{\infty}||\partial_{t^{'}}g||_{H^{s-1}}dt^{'}\lesssim \delta e^{-2\alpha t},
\end{eqnarray}
since $(n\gamma_{ad}-2)\alpha>\alpha$ for $\gamma_{ad}\in (1,\frac{n+1}{n})$ and $g^{\dag}:=g(t=\infty)$. In order to show that the limit metric $g^{\dag}$ satisfies $R(g^{\dag})=-1$, we use the Hamiltonian constraint 
\begin{eqnarray}
R(g)-|k^{tr}|^{2}-2a^{2-n\gamma_{ad}}\rho\{1+\gamma_{ad} g(v,v)\}+1+2C_{\rho}a^{2-n\gamma}=0.
\end{eqnarray}
The decay estimates obtained so far yields 
\begin{eqnarray}
\lim_{t\to\infty}R(g(t))=-1,
\end{eqnarray}
which makes sense since $s>\frac{n}{2}+2$ and since $g\in H^{s}$. This is due to the fact that $H^{s}\hookrightarrow H^{s-1}$ compactly for the compact manifold $\Sigma$ and therefore the limit metric $g^{\dag}$ actually lies in $H^{s}$. Therefore $g^{\dag}\in \mathcal{M}^{\epsilon}_{-1}$, where $\mathcal{M}^{\epsilon}_{-1}$ is a sufficiently small neighbourhood of the space of Einstein metrics $\mathcal{N}$ (the center manifold defined in section 2.3) in the space of metrics of constant negative scalar curvature $-1$. This completes the proof of the first part of the lemma. 

For the second part, we perform similar calculations with $E_{v}:=\frac{1}{2}|| \mathcal{D}^{s-2}v||^{2}_{L^{2}}$. Using the commutation equations for $v$ and $\delta\rho$ (equations \ref{eq:commuted1}-\ref{eq:commuted2} but we commute by $ \mathcal{D}^{s-2}$ instead of $\mathcal{D}^{s-1}$; this is done so simply because of the regularity issue since the evolution equation for $(\delta\rho,v)$ contain spatial derivatives of $(\delta\rho,v)$) we obtain 
\begin{eqnarray}
\frac{dE_{v}}{dt}\leq-2[\left\{(n+1)-n\gamma_{ad}\right\}-C(||v||^{2}_{H^{s-2}}+||k^{tr}||^{2}_{H^{s-2}})]E_{v}\\\nonumber +\frac{C}{a}||v||_{H^{s-2}}(||\delta\rho||_{H^{s-1}}+||\delta\rho||^{2}_{H^{s-1}}+||v||^{2}_{H^{s-1}}\nonumber+||\omega||_{H^{s-1}}
\\\nonumber 
+||X||_{H^{s-2}}||\delta\rho||_{H^{s-1}}||v||_{H^{s-2}}+||X||_{H^{s-1}}||v||_{H^{s-1}}+||k^{tr}||_{H^{s-1}}||v||_{H^{s-1}}\\\nonumber +a^{2-n\gamma_{ad}}||\delta\rho||_{H^{s-1}}+a^{2-n\gamma_{ad}}||v||^{2}_{H^{s-1}}+\dot{a}a^{2-n\gamma_{ad}}||\delta\rho||_{H^{s-1}}\\\nonumber +\dot{a}a^{2-n\gamma_{ad}}||\delta\rho||^{2}_{H^{s-1}}).
\end{eqnarray}
Using the boundedness property from lemma 16 and the smallness of $\delta$ we may obtain as $t\to\infty$
\begin{eqnarray}
\frac{dE_{v}}{dt}\leq -\beta E_{v}+C\delta^{2} e^{-\alpha t},~\beta:=(n+1)-n\gamma_{ad}>0~fixed,
\end{eqnarray}
integration of which yields 
\begin{eqnarray}
E_{v}\lesssim \delta^{2}e^{-\zeta t},
\end{eqnarray}
where $\zeta:=\min(\beta,\alpha)$. This yields \begin{eqnarray}
||v||_{H^{s-2}}\lesssim\delta e^{-\frac{\zeta}{2}t}. 
\end{eqnarray}
Using the estimate (\ref{eq:rhoestimate}) of $\partial_{t}\delta\rho$ (or of $\partial_{t}\rho$), we obtain 
\begin{eqnarray}
||\rho^{'}-\rho(t)||_{H^{s-2}}\leq\int_{t}^{\infty}||\partial_{t^{'}}\rho||_{H^{s-3}}dt^{'}\lesssim \delta e^{-(\alpha+\frac{\zeta}{2})t},
\end{eqnarray}
where $\rho^{'}:=\rho(t=\infty)$. In general, $\partial_{i}\rho^{'}\neq 0$. This concludes the proof of the lemma.~~~~~~~~~~~~~~~~~~~~~~~~~~~~~~~~~~$\Box$ 

Now we establish the geodesic completeness of the perturbed spacetimes using the estimates obtained so far. To be precise, we show that if we perturb the FLRW background, then this perturbed spacetime is causally complete. In order to establish this future completeness of the perturbed spacetime, we need to show that the solutions of the geodesic equation must exist for an infinite interval of the associated affine parameter. We provide a rough sketch of the proof in this context following \cite{andersson2004future}. 
Let's designate a timelike geodesic in the homotopy class of a family of timelike curves by $\mathcal{\chi}$. The tangent vector $\xi=\frac{d\mathcal{\chi}}{d\lambda}=\xi^{\mu}\partial_{\mu}$ to $\chi$ for the affine parameter $\lambda$ satisfies $\hat{g}(\xi,\xi)=-1$, where $\hat{g}$ is the spacetime metric. As $\mathcal{\chi}$ is causal, we may parametrize it as $(t,\mathcal{\chi}^{i})$, $i=1,2,3$. We must show that $\lim_{t\to\infty}\lambda=+\infty$, that is, 
\begin{eqnarray}
\lim_{t\to\infty}\int_{t_{0}}^{t}\frac{d\lambda}{dt^{'}}dt^{'}=+\infty.
\end{eqnarray}
Noting that $\xi^{0}=\frac{dt}{d\lambda}$, we must show 
\begin{eqnarray}
\lim_{t\to\infty}\int_{t_{0}}^{t}\frac{1}{\xi^{0}}dt^{'}=+\infty.
\end{eqnarray}
Showing that $|\tilde{N}\xi^{0}|$ is bounded and therefore $\lim_{t\to\infty}\int_{t_{0}}^{t}\tilde{N}dt^{'}=+\infty$ is enough to ensure the geodesic completeness. We first show that $|\tilde{N}\xi^{0}|$ is bounded. Proceeding the same way as that of \cite{mondal2019attractors} together with the estimates obtained from lemma 17, we obtain
\begin{eqnarray}
\tilde{N}^{2}(\xi^{0})^{2}\leq C
\end{eqnarray}
as $t\to\infty$ for some $C<\infty$. Therefore, we only need to show that the following holds 
\begin{eqnarray}
\lim_{t\to\infty}\int_{t_{0}}^{t}\tilde{N}dt=\infty
\end{eqnarray}
in order to finish the proof of timelike geodesic completeness. Notice that $\tilde{N}=N=1+\omega$. Now $|\omega(t)|\leq  C^{'}(\delta^{2}e^{-2\alpha t}+\delta e^{-\alpha(n\gamma_{ad}-2)t}), 0<C^{'}<\infty$ as $t\to\infty$ yielding 
\begin{eqnarray}
\lim_{t\to\infty}\left(\int_{t_{0}}^{t}dt-C^{'}\alpha(2(e^{-2\alpha t_{0}}-e^{-2\alpha t})+(n\gamma-2)(e^{-\alpha(n\gamma_{ad}-2)t_{0}}\nonumber-e^{-\alpha(n\gamma_{ad}-2)t}))\right)\\\nonumber 
\leq \lim_{t\to\infty}\int_{t_{0}}^{t}\tilde{N}dt\\\nonumber 
\leq \lim_{t\to\infty}\left(\int_{t_{0}}^{t}dt+C^{'}\alpha(2(e^{-2\alpha t_{0}}-e^{-2\alpha t})+(n\gamma-2)(e^{-\alpha(n\gamma_{ad}-2)t_{0}}\nonumber-e^{-\alpha(n\gamma_{ad}-2)t}))\right)\\\nonumber 
=>\lim_{t\to\infty}\int_{t_{0}}^{t}\tilde{N}dt=\infty.
\end{eqnarray}
This completes the proof of the geodesic completeness of the spacetimes of interest under fully nonlinear perturbations. Collecting the results of lemma 16, lemma 17, and the geodesic completeness, we obtain the following main theorem.

\textbf{Main Theorem:} \textit{Let $(g_{0},k^{tr}_{0},\rho_{0},v_{0})\in B_{\epsilon}(\gamma,0,C_{\rho},0)\subset H^{s}\times H^{s-1}\times H^{s-1}\times H^{s-1}$, where $s>\frac{n}{2}+2$ and $\epsilon>0$ is sufficiently small. Also consider that $\mathcal{N}$ is the integrable deformation space of $\gamma$ and $\Lambda>0$ is the cosmological constant. Assume that the adiabatic index $\gamma_{ad}$ lies in the interval $(1,\frac{n+1}{n})$. Let $t \mapsto (g(t), k^{tr}(t), \rho(t), v(t))$ be the maximal development of the Cauchy problem for the re-scaled Einstein-Euler-$\Lambda$ system (\ref{eq:metricevol}-\ref{eq:mc}) in constant mean extrinsic curvature spatial harmonic gauge (CMCSH) with initial data $(g_{0},k^{tr}_{0},\rho_{0},v_{0})$. Then there exists a $\gamma^{\dag}\in \mathcal{M}^{\epsilon}_{-1}\cap \mathcal{S}_{\gamma}$ and $\gamma^{*}\in \mathcal{N}$ such that $(g,\gamma, k^{tr}, \rho, v)$ flows toward $(\gamma^{\dag},\gamma^{*},0, \rho^{'}, 0)$ in the limit of infinite time, that is, 
\begin{eqnarray}
\lim_{t\to\infty}(g(t,x),\gamma(t,x),k^{tr}(t,x),\rho(t,x),v(t,x))\nonumber=(\gamma^{\dag}(x),\gamma^{*}(x),0,\rho^{'}(x),0),
\end{eqnarray}
where $\gamma^{\dag}=\gamma^{*}$ or  $\gamma^{*}$ is the shadow of $\gamma^{\dag}$ and the convergence is understood in the strong sense i.e., with respect to the available Sobolev norms..\\
In addition, this Cauchy problem for the re-scaled Einstein-Euler system in constant mean extrinsic curvature (CMC) and spatial harmonic (SH) gauge is globally well posed to the future and the space-time is future complete.
Here $\mathcal{M}^{\epsilon}_{-1}$ denotes the space of metrics of constant negative ($-1$) scalar curvature sufficiently close to and containing the deformation space $\mathcal{N}$. $\mathcal{S}_{\gamma}$ denotes a harmonic slice through the metric $\gamma$. $\rho^{'}$ is dependent on the initial density and $\partial_{i}\rho^{'}\neq 0$ in general. 
}

Since the perturbed solutions do not, in general, return to the background solutions (\ref{eq:model}) but rather stay in a small neighbourhood of the later for all time, we designate this stability as \textit{Lyapunov} stability. 

\begin{center}
\begin{figure}
\begin{center}
\includegraphics[width=13cm,height=60cm,keepaspectratio,keepaspectratio]{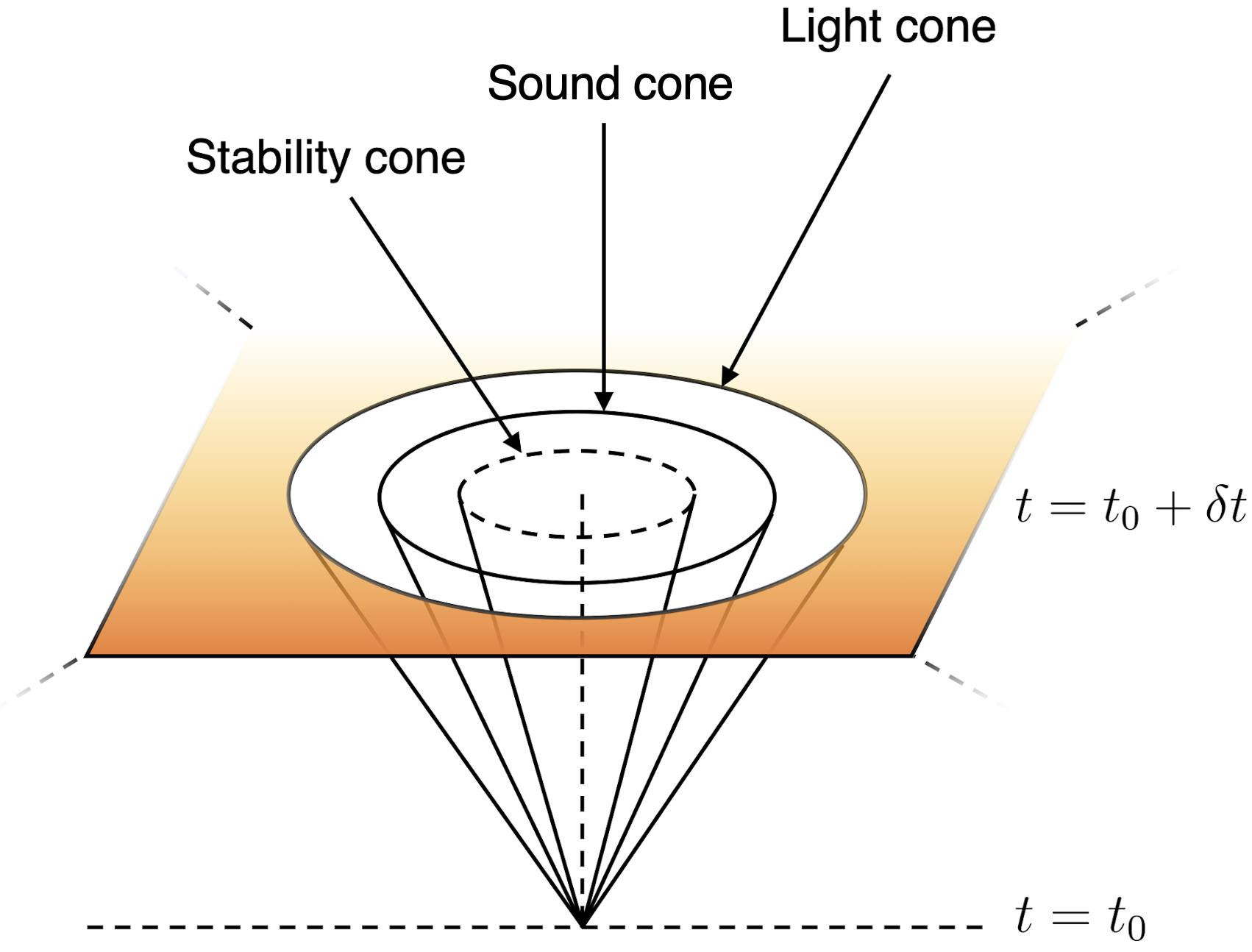}
\end{center}
\begin{center}
\caption{The tangent space picture of null-geometry associated with the Einstein-Euler equations. The nonlinear stability criteria is essentially defined by the fact that the perturbations to the matter part do not travel at a speed more than $\sqrt{1/n}$ i.e., they are restricted within the innermost cone.}
\label{fig:pdf}
\end{center}
\end{figure}
\end{center}

\section{Concluding Remarks}
We have proven a fully non-linear \textit{Lyapunov} stability of the spatially compact negative spatial sectional curvature FLRW solutions (and their higher-dimensional generalizations). A serious difficulty that arises in the context of the Einstein-Euler system is the formation of shocks in addition to concentration and blow-up of the gravitational energy. The inclusion of a positive cosmological constant removes such possibility at the level of small data since the rapid expansion of the space does not allow for energy concentration. If, on the other hand, one were to turn off the positive cosmological constant, then one would expect formation of shocks (or gravitational collapse) since the integrability condition $\int_{t_{0}}^{\infty}\frac{1}{a(t)}dt$ is no longer satisfied as $a(t)\sim t,~t\to\infty$ when $\Lambda=0$. Currently, we cannot prove a precise statement about the shock formation (or gravitational singularity). There is however a recent study by \cite{fajman2021slowly}, which proves the asymptotic stability of the dust-filled Milne universe (isometric to negative curvature FLRW model with vanishing energy density) in the absence of a positive cosmological constant. An advantage of dust matter sources is the absence of sound phenomena (responsible for shock formation) since the pressure vanishes for such sources and the individual fluid particles propagate along their respective geodesics. It is however currently unknown whether such a result holds true for the FLRW spacetimes (of direct physical interest). 

Since Euler's equations are not of a manifestly diagonal nature, it is not trivial to prove their hyperbolic character. To that end, we have constructed an energy current motivated by the energy current first constructed by Christodoulou \cite{christodoulou2007formation}. Prior to us, Speck \cite{speck2012nonlinear}, used the exact form of Christodoulou's energy current to establish non-linear Lyapunov stability of the spatially compact flat FLRW model. However, as we mentioned previously, our framework is different from that of Speck's due to the choice of gauge and the geometry and the topology of the background solutions. We have constructed the necessary energy current by hand, which satisfies all the required properties including positivity (for arbitrarily large data), coercivity, and convexity (for small data) in addition to the cancellation of principal terms in its divergence expression that is essential to close the energy argument. Application of CMCSH gauge turns the Einstein's equations into an elliptic-hyperbolic system and together with the aforementioned energy current for the fluid degrees of freedom, we obtain a coupled elliptic-hyperbolic system. This is an advantage of using this particular choice of gauge, where the analysis of Einstein's equations simplifies substantially.

In addition, the re-scaled Einstein-Euler system considered in the current context is an example of a non-autonomous (infinite-dimensional) dynamical system since the explicitly time-dependent scale factor $a$ appears in the equations. This makes the analysis non-trivial since there are terms in the equations of motion which involve factors such as $\frac{\dot{a}}{a}$, $\frac{1}{a}$, $\dot{a}a^{1-n\gamma_{ad}}$, and $a^{1-n\gamma_{ad}}$. The decay estimates that we obtained in the last section heavily depended on the fact that these factors exhibit a particular pattern in their appearances such that $\frac{\dot{a}}{a}$ should only multiply the term that generates decay or higher-order terms. Remarkably, the structure of Einstein's equations is such that all these factors appear in the field equations in an innocuous way. The stability is then found to be dependent on an integrability criteria of this scale factor namely $\int_{t_{0}}^{\infty}a^{1-n\gamma_{ad}}dt<\infty,~\int_{t_{0}}^{\infty}a^{-1}dt<\infty$. Notice that the crucial integrability criterion $\int_{t_{0}}^{\infty}a^{-1}dt<\infty$ does not appear at the linear level (in the study of $\cite{mondal2021linear}$, this integrability criterion arises due to the consideration of the motion tangential to the center manifold, and as such is absent for the $n=3$ case). It is therefore tempting to conjecture these integrability criteria to be the conditions to avoid shock and naked singularity formation. However, there is much left to be done. It is worth mentioning that there are studies in the literature \cite{fajman2021stabilizing, speck2013stabilizing} that deal with such issue of shock formation on a background expanding spacetimes. 

Our results only provide a notion of the Lyapunov stability of the FLRW solutions considered here. In other words, the perturbed solutions to the Einstein-Euler field equations around this class of spatially compact FLRW metrics (spatial slices are compact negative Einstein spaces in general and hyperbolic for the physically relevant $n=3$ case) arising from regular Cauchy data remain uniformly bounded and decay to a family of metrics with constant negative spatial scalar curvature. Since the metrics with constant scalar curvature do not imply constant sectional curvature for dimensions greater than 2 and the final mass-energy density is not spatially constant, the asymptotic state is essentially described by an inhomogeneous and anisotropic physical universe filled with dispersed fluid. This may be slightly unsatisfactory since one would expect the perturbed solutions to return to the FLRW family. However, since the physical universe does consist of large-scale structures, it is tempting to conjecture that our result indicates a possibility of structure formation through Einstein flow. In addition, it provides yet another reason to believe that FLRW models are \textit{physical} contrary to their isometric cousin Milne model (which is devoid of background energy density) which does not exhibit a possibility of an inhomogeneous and anisotropic final state in the absence of a positive cosmological constant. This notion of structure formation through perturbing the FLRW solution was noted in several heuristic studies (e.g., \cite{percival2005cosmological, macpherson2019einstein, roy2011global,raychaudhuri2003general}). It would be worth investigating this issue further in future studies due to its tremendous physical implications.  

Lastly, it is worth pointing out that our stability result heavily relies on the assumption on the allowed range of the adiabatic index $\gamma_{ad}$ in the chosen barotropic equation of state ($P=(\gamma_{ad}-1)\rho$). The non-linear stability result holds if $\gamma_{ad}$ lies in the range $(1,\frac{n+1}{n})$ ($(1,4/3)$ for the physically relevant $n=3$ case). We recall that the same condition was obtained in the linear stability analysis by \cite{mondal2021linear}. Prior to our study, Rodnianski and Speck \cite{rodnianski2009stability} and Speck \cite{speck2012nonlinear} obtained the same condition for nonlinear stability of small data on $\mathbb{R}\times \mathbb{T}^{3}$. In addition to \cite{rodnianski2009stability} and \cite{speck2012nonlinear}, avoidance of shock formation by exponential expansion was also studied by \cite{lubbe2013conformal} for a radiation fluid (on the $\mathbb{S}^{3}$ spatial topology), by \cite{oliynyk2021future} for $\gamma_{ad}\in (4/3,3/2)$ (with $\mathbb{T}^{3}$ spatial topology), and \cite{hadvzic2015global} for dust (with $\mathbb{T}^{3}$ spatial topology). It would be interesting to extend our study to these special cases and include different types of equations of state as well.      

\section{Acknowledgement}
P.M would like to thank Prof. Vincent Moncrief for numerous useful discussions related to this project and for his help improving the manuscript. This work was supported by CMSA at Harvard University.

\author{Puskar Mondal$^{\dag}$}
\address{$^{\dag}$ Center of Mathematical Sciences and Applications,\\
Department of Mathematics,\\
Harvard University, Cambridge, MA02138, USA}
\ead{puskar$\_$mondal@fas.harvard.edu}

\end{document}